\documentclass[]{spie}  %>>> use for US letter paper
\usepackage[T1]{fontenc} % For correct coding of accentuated chars.
\usepackage{graphicx}
\usepackage{transparent}
\usepackage{xkeyval,xcolor}% http://ctan.org/pkg/{xkeyval,xcolor}
\makeatletter
\newlength{\sfp@hseplen}\newlength{\sfp@vseplen}
\define@cmdkey{subfigpos}[sfp@]{pos}[ul]{}% \sfp@pos
\define@cmdkey{subfigpos}[sfp@]{font}[\small]{}% \sfp@font
\define@cmdkey{subfigpos}[sfp@]{vsep}[0.75\baselineskip]{\setlength{\sfp@vseplen}{\sfp@vsep}}% \sfp@vsep
\define@cmdkey{subfigpos}[sfp@]{hsep}[3.5pt]{\setlength{\sfp@hseplen}{\sfp@hsep}}% \sfp@hsep
\newcommand{\subfigimg}[4][,]{%
        \setkeys{Gin,subfigpos}{pos,font,vsep,hsep,#1}% Set (default) keys
        \setbox1=\hbox{\includegraphics{#4}}% Store image in box
        \ifnum\pdfstrcmp{\sfp@pos}{ul}=0% UPPER LEFT placement of subfig label
                \leavevmode\rlap{\usebox1}% Print image
                \rlap{\hspace*{\sfp@hsep}\raisebox{\dimexpr\ht1-\sfp@vsep}{\transparent{#3}{\setlength{\fboxsep}{1pt}\colorbox{white}{%
\transparent{1}\sfp@font{#2}}}%
}}% Print label
                \phantom{\usebox1}% Insert appropriate spacing
        \else\ifnum\pdfstrcmp{\sfp@pos}{ur}=0% UPPER RIGHT placement of subfig label
                \leavevmode\usebox1% Print image
                \llap{\raisebox{\dimexpr\ht1-\sfp@vsep}{\sfp@font{#2}}\hspace*{\sfp@hsep}}% Print label
        \else\ifnum\pdfstrcmp{\sfp@pos}{lr}=0% LOWER RIGHT placement of subfig label
                \leavevmode\usebox1% Print image
                \llap{\raisebox{\sfp@vsep}{\sfp@font{#2}}\hspace*{\sfp@hsep}}% Print label
        \else% Assume LOWER LEFT placement of subfig label
                \leavevmode\rlap{\usebox1}% Print image
                \rlap{\hspace*{\sfp@hseplen}\raisebox{\sfp@vsep}{\sfp@font{#2}}}% Print label
                \phantom{\usebox1}% Insert appropriate spacing
        \fi\fi\fi
}
\newcommand{\fontfig}[1]{\tiny$\!\!$\color{#1}\textbf}
\newcommand{\AspectRatio}[1]{\dimexpr 1pt * \wd#1 / \ht#1 \relax} % Aspect ratio of a subfigure

\usepackage{amsmath} % Table of equations
\usepackage{array}
\usepackage{adjustbox}
\usepackage{multirow}
\usepackage{makecell}
\newcolumntype{C}[1]{>{\centering\arraybackslash}p{#1}} % Centered column with cell aligned at the bottom
\newcolumntype{L}[1]{>{\raggedright\arraybackslash}m{#1}} % Left-flushed column with cell aligned at the middle
\newcolumntype{M}[1]{>{\centering\arraybackslash}m{#1}} % Centered column with text aligned at the middle

\newcommand{\eq}[1]{Eq.~\eqref{#1}\xspace}
\newcommand{\eqs}[2]{Eqs.~(\ref{#1},\ref{#2})\xspace}
\newcommand{\eqfull}[1]{Equation~\eqref{#1}\xspace} % Full expansion of the name

\newcommand{\subfigref}[1]{(#1)} % How to refere to a subfigure.
\newcommand{\fig}[1]{Fig.~\ref{#1}\xspace}
\newcommand{\figs}[2]{Figs.~\ref{#1} and~\ref{#2}\xspace} % 2 figures
\newcommand{\figfull}[1]{Figure~\ref{#1}\xspace} % Full expansion of the name
\newcommand{\subfig}[2]{Fig.~\ref{#1}\subfigref{#2}\xspace}
\newcommand{\subfigfull}[2]{Figure~\ref{#1}\subfigref{#2}\xspace} % Full expansion of the name
\newcommand{\subfigs}[2]{Figs.~\ref{#1}\subfigref{#2}\xspace}
\newcommand{\subfigsfull}[2]{Figures~\ref{#1}\subfigref{#2}\xspace} % Full expansion of the name

\newcommand{\refsec}[1]{Section~\ref{#1}\xspace} % Section and subsection
\newcommand{\refsecs}[2]{Sections~\ref{#1} and~\ref{#2}\xspace} % Section and subsection
\newcommand{\refapp}[1]{Appendix~\ref{#1}\xspace} % Appendix
\newcommand{\refline}[1]{Line~\ref{#1}\xspace} % Line of an algorithm

\usepackage{algorithm}
\usepackage[noend]{algpseudocode}
\newcommand{\commentalgo}[1]{\Comment{{\tiny #1}}} % for the comments in an algorithm
\usepackage{amssymb}
\usepackage{mathtools}
\usepackage{mathrsfs}
\usepackage{nicefrac}
\usepackage{stmaryrd} % For the \llbracket and \rrbracket commands.
\SetSymbolFont{stmry}{bold}{U}{stmry}{m}{n} % To remove the warning "LaTeX Font Warning: Font shape `U/stmry/b/n' undefined"

\newcommand{\Tag}[1]{\text{#1}}   

\newcommand{\V}[1]{{\boldsymbol{#1}}}                 % vector (with amsmath)
\newcommand{\M}[1]{{\mathbf{#1}}}                     % matrix
\newcommand{\T}{^{\mathrm{T}}}                        % transpose
\newcommand{\Inv}{^{-1}}                              % inverse
\newcommand{\I}{\mathrm{i}}                           % sqrt(-1)
\newcommand{\E}[1]{\mathrm{e}^{#1}}                   % e^{...}
\newcommand{\conj}[1]{#1^{\star}}                     % c^{\star} Complex conjugate

\newcommand{\D}[1]{\Tag{d}#1}
\newcommand{\Dx}{{\D \Vx}}            % integrant dx
\newcommand{\Dxp}{{\D \Vxp}}          % integrant dx'
\newcommand{\Dk}{{\D \Vk}}            % integrant dk
\newcommand{\Dkp}{{\D \Vkp}}          % integrant dk'

\newcommand{\pupil}{P}                       % Pupil of the telescope 
\newcommand{\PSF}{h}                         % PSF
\newcommand{\FTPSF}{\FT{h}}
\newcommand{\PSFturb}{h_{\pha}}              % PSF of the turbulence
\newcommand{\FTPSFturb}{\FT{h}_{\pha}} 
\newcommand{\PSFtel}{h_{\Tag{tel}}}          % PSF of the telescope
\newcommand{\FTPSFtel}{\FT{h}_{\Tag{tel}}} 
\newcommand{\PSFres}{h_{\epsilon}}           % PSF of the residuals

\newcommand{\strehl}{\gamma_{\Tag{Strehl}}}  % Strehl ratio
\DeclarePairedDelimiterX{\paren}[1]{(}{)}{#1}
\newcommand{\Paren}[1]{\paren*{#1}}
\let\brace=\undefined % Redifine \brace
\DeclarePairedDelimiterX{\brace}[1]{\{}{\}}{#1}
\newcommand{\Brace}[1]{\brace*{#1}}
\let\brack=\undefined % Redifine \brack
\DeclarePairedDelimiterX{\brack}[1]{[}{]}{#1}
\newcommand{\Brack}[1]{\brack*{#1}}
\DeclarePairedDelimiterX{\bbrack}[1]{\llbracket}{\rrbracket}{#1}%  [| ... |]

\DeclarePairedDelimiterX{\abs}[1]{\rvert}{\lvert}{#1}     %  | ... |
\newcommand{\Abs}[1]{\abs*{#1}}
\DeclarePairedDelimiterX{\norm}[1]{\lVert}{\rVert}{#1}    % || ... ||
\newcommand{\Norm}[1]{\norm*{#1}}

\DeclarePairedDelimiterX{\avg}[1]{\langle}{\rangle}{#1}   %  < ... >

\DeclarePairedDelimiterX{\ceil}[1]{\lceil}{\rceil}{#1}     % ceil operator

\DeclarePairedDelimiterX{\floor}[1]{\lfloor}{\rfloor}{#1}  % floor operator

\newcommand{\Set}[1]{\mathbb{#1}}
\newcommand{\Reals}{\Set{R}}
\newcommand{\PosIntegers}{\Set{N}}

\newcommand{\FTnot}{\mathscr{F}}
\newcommand{\FTnotinv}{\mathscr{F}^{-1}}
\newcommand{\FTfull}[2]{\FTnot\left[#1\right]\left(#2\right)}
\newcommand{\FTfullinv}[2]{\FTnotinv\left[#1\right]\left(#2\right)}
\newcommand{\FT}[1]{\tilde{#1}}

\newcommand{\Dsim}{D_{\Tag{sim}}}                     % The simulation diameter
\newcommand{\Dsub}{D_{\Tag{sub}}}                     % The sub-aperture diameter

\newcommand{\Vc}{{\V{c}}}                   % command vector a
\newcommand{\ap}{{a^{\prime}}}              % actuator a'
\newcommand{\Vx}{{\V{x}}}                   % vector x
\newcommand{\Vxp}{{\V{x^{\prime}}}}         % vector x'
\newcommand{\Vk}{{\V{k}}}                   % vector k
\newcommand{\Vkp}{{\V{k^{\prime}}}}         % vector k'

\newcommand{\setactu}{\mathcal{A}}    % Set of actuators
\newcommand{\pitchactu}{\Delta}       % Pitch of the actuators
\newcommand{\Vpitch}{\V{\pitchactu}}  % Pitch of the actuators (vector)
\newcommand{\Pa}{\V{p}_{a}}           % Position of the acuator a
\newcommand{\Pap}{\V{p}_{\ap}}        % Position of the acuator a'

\newcommand{\EWprod}{\otimes}         % Element-wise product

\newcommand{\pha}{w}                  % The phase in radian
\newcommand{\phaAO}{\pha_{\epsilon}}        % The AO corrected phase in radian
\newcommand{\rFried}{r_{0}}             % Fried's parameter
\newcommand{\nPha}{n_{\pha}}            % Number of phase screen

\newcommand{\IFref}{\varphi_{0}}               % Reference influence function
\newcommand{\FTIFref}{\FT{\pha}_{0}}        % The FT of \IFref
\newcommand{\IFperp}{\psi_{\perp}}          % Orthogonal influence function
\newcommand{\FTIFperp}{\FT{\psi}_{\perp}}   % The FT of \IFperp
\newcommand{\AvgT}[1]{\avg*{#1}}          % Temporally expected
\newcommand{\PSD}[1]{\Phi_{#1}}               % Power spectrum density
\newcommand{\PSDres}{\Phi_{\epsilon}}          % Power spectrum density of the fitting error
\newcommand{\PSDperp}{\Phi_\perp}             % Power spectrum density of the orthogonal reference function
\newcommand{\PSDturb}{\Phi_\pha}           % Power spectrum density of the turbulence

\newcommand{\SF}[1]{D_{#1}}         % Structure function
\newcommand{\SFturb}{\SF{\pha}}   	% Turbulent structure function
\newcommand{\fLF}{f_{\Tag{LF}}}           % Low frequency filter

\newcommand{\Strehl}{\gamma_{\mathrm{Strehl}}}
\newcommand{\ColorOne}[1]{\textcolor{blue}{#1}}
\newcommand{\ColorTwo}[1]{\textcolor{red}{#1}}
\newcommand{\COne}{C_{1.5\lD}}
\newcommand{\CTwo}{C_{4\lD}}
\newcommand{\sigRes}{\sigma_{\Tag{res}}}
\newcommand{\sigWF}{\sigma_{\Tag{WF}}}

\newcommand{\Vw}{\V{\pha}}
\newcommand{\Vd}{\V{d}}
\newcommand{\Vn}{\V{n}}

\newcommand{\Wpp}{\Vw_{\Tag{PP}}}
\newcommand{\Wdm}{\Vw_{\Tag{DM}}}
\newcommand{\Wao}{\Vw_{\Tag{AO}}}
\newcommand{\Wwfs}{\Vw_{\Tag{WFS}}}

\newcommand{\Pis}{\mathrm{P}}
\newcommand{\Pup}{\mathrm{\Sigma}}

\newcommand{\MS}{\M{S}}
\newcommand{\MSsy}{\MS_{\Tag{sy}}}
\newcommand{\MSfo}{\MS_{\Tag{FO}}}
\newcommand{\MM}{\M{M}}
\newcommand{\MMpis}{\bar{\MM}}
\newcommand{\MG}{\M{G}}
\newcommand{\MGsy}{\MG_{\Tag{sy}}}
\newcommand{\MGfo}{\MG_{\Tag{FO}}}
\newcommand{\MC}{\M{H}}
\newcommand{\MCLS}{\MC_{\Tag{LS}}}
\newcommand{\MCMV}{\MC_{\Tag{MV}}}
\newcommand{\MPis}{\M{P}}
\newcommand{\MPup}{\M{\Sigma}}

\newcommand{\MCov}[1]{\M{C}_{#1}}

\newcommand{\ndat}{n_{\Tag{dat}}}
\newcommand{\nw}{n_{\Tag{\pha}}}
\newcommand{\nact}{n_{\Tag{act}}}
\newcommand{\nsvd}{n_{\Tag{SVD}}}

\newcommand{\argmin}[1]{\text{arg min}_{#1} \;}

\newcommand{\yfo}{y_{\Tag{spot}}^{\Tag{FO}}}
\newcommand{\xfo}{x_{\Tag{spot}}^{\Tag{FO}}}
\newcommand{\ysy}{y_{\Tag{spot}}^{\Tag{sy}}}
\newcommand{\xsy}{x_{\Tag{spot}}^{\Tag{sy}}}

\newcommand{\npix}[1]{N_{\Tag{#1}}} % Number of pixel
\newcommand{\Sap}{\mathcal{S}_{A}} % Surface of the aperture
\newcommand{\pitch}[1]{\D{x}_{\Tag{#1}}} % Pixel pitch
\newcommand{\Is}[1]{I_{#1}} % Intensity simulated for the ratio s
\newcommand{\Iss}[1]{I_{\cdot, #1}} % Analytical intensity for the ratio s (sampled)
\newcommand{\Iis}[1]{I_{\circ, #1}} % Analytical intensity for the ratio s (integrated)
\newcommand{\Iips}[2]{I_{\Tag{#1}, #2}} % Intensity simulated for the strategy I or P for the ratio s
\newcommand{\angunit}{u_{\alpha}}

\newcommand{\lmin}{\lambda_{\Tag{min}}} % minimal wavelength
\newcommand{\lmax}{\lambda_{\Tag{max}}} % maximal wavelength

 \newcommand{\etal}{\textit{et~al.}\xspace}
\newcommand{\vs}{v.s.\xspace}
\usepackage{siunitx}
\DeclareSIUnit\angstrom{\text {Å}}
\newcommand{\arcsec}[1]{\SI{#1}{''}} % arcsecond
\newcommand{\diam}{\varnothing} % diameter
\newcommand{\lD}{\lambda/D}
\newcommand{\lDsub}{\lambda/\Dsub}
\usepackage[colorlinks=true, allcolors=blue]{hyperref}
\usepackage{xspace}
 
\title{Inverse problem approach in Extreme Adaptive Optics -- Analytical model of the fitting error and lowering of the aliasing}

\author[a,b]{Anthony Berdeu}
\author[c]{Michel Tallon}
\author[c]{Éric Thiébaut}
\author[a]{Mary Angelie Alagao}
\author[a]{Sitthichat Sukpholtham}
\author[c]{Maud Langlois}
\author[a]{Adithep Kawinkij}
\author[a]{Puttiwat Kongkaew}

\affil[a]{National Astronomical Research Institute of Thailand, Center for Optics and Photonics, 260 Moo 4, T. Donkaew, A. Maerim, Chiang Mai 50180, Thailand}
\affil[b]{Department of Physics, Faculty of Science, Chulalongkorn University, 254 Phayathai Road, Pathumwan, Bangkok 10330, Thailand}
\affil[c]{Univ Lyon, Univ Lyon1, Ens de Lyon, Centre de Recherche Astrophysique de Lyon, UMR 5574, F-69230, Saint-Genis-Laval, France}

\authorinfo{
Send correspondence to Anthony Berdeu: \href{mailto:anthony@narit.or.th}{anthony@narit.or.th}
}

\begin{document}
\maketitle

\begin{abstract}
We present the results obtained with an end-to-end simulator of an Extreme Adaptive Optics (XAO) system control loop. It is used to predict its on-sky performances and to optimise the AO loop algorithms. It was first used to validate a novel analytical model of the fitting error, a limit due to the Deformable Mirror (DM) shape. Standard analytical models assume a sharp correction under the DM cutoff frequency, disregarding the transition between the AO corrected and turbulence dominated domains. Our model account for the influence function shape in this smooth transition. Then, it is well-known that Shack-Hartmann wavefront sensors (SH-WFS) have a limited spatial bandwidth, the high frequencies of the wavefront being seen as low frequencies. We show that this aliasing error can be partially compensated (both in terms of Strehl ratio and contrast) by adding priors on the turbulence statistics in the framework of an inverse problem approach. This represents an alternative to the standard additional optical filter used in XAO systems. %This simulator is also used to investigate the optimal way to tackle the noise covariance in the SH-WFS slope measurements for the optimal command prediction. 
In parallel to this numerical work, a bench was aligned to experimentally test the AO system and these new algorithms comprising a DM192 ALPAO deformable mirror and a 15$\times$15 SH-WFS. We present the predicted performances of the AO loop based on end-to-end simulations.
\end{abstract}

% Include a list of keywords after the abstract 

\section{INTRODUCTION}
\label{sec:intro}

The Evanescent Wave Coronagraph (EvWaCo\cite{Buisset:17_EvWaCo}) is an on-going project at the National Astronomical Research Institute of Thailand (NARIT). Based on the frustration of the total internal reflection (FTIR\cite{Zhu:86_FTIR}) between a prism and a lens put in contact, the star light is transmitted through the contact area while the light of the stellar vicinity is internally reflected in the prism\cite{Alagao:21_EvWaCo_exp_contrast}. As the FTIR is chromatic, the mask adapts itself with the wavelength, providing almost achromatic contrast performances over the spectral domain~$\Brack{\SI{600}{\nano\meter},\SI{900}{\nano\meter}}$. In addition, the mask shape and size can be adjusted by tuning the pressure between the two pieces of glass. An on-sky demonstrator is currently under development at NARIT\cite{Buisset:18_EvWaCo_spec} to be installed on the \SI{2.4}{\meter} Thai National Telescope (TNT), using an elliptical unobstructed pupil of $1.17\times\SI{0.83}{\square\meter}$.

As a coronagraph dedicated to high contrast and high resolution imaging, EvWaCo implies the development of a dedicated extreme adaptive optics (AO) system. The description of its AO system and its associated characterisation bench aligned in the NARIT cleanroom is the subject of another paper by Berdeu~\etal\cite{Berdeu:22_AO_bench}. We remind here only its main features.

The EvWaCo AO system is based on a $\nact=192$ actuator deformable (DM) from ALPAO\cite{LeBouquin:18_charac_ALPAO}\footnote{\url{http://www.alpao.com}}, the DM192. As schemed in \subfig{fig:Fitting_error}{a}, the EvWaCo elliptical pupil spans over $16\times12$ actuators, or equivalently in the Fried's geometry\cite{Fried:77, Southwell:80}, over $15\times11$ sub-apertures of $7.8\times\SI{7.8}{\square\cm}$ with the actuators placed on their corners.

The wavefront sensor (WFS) of EvWaCo is a Shack-Hartmann\cite{Shack:71_SHWFS} (SH-WFS). Its working wavelength range is $\Brack{400, 600}\SI{}{\nano\meter}$. Its custom-designed lenslet array was manufactured by Smart MicroOptical Solution company\cite{Bahr:15_lenslet}. Its camera is a Nüvü~$128^\Tag{AO}$ from Nüvü Cam$\bar{\text{e}}$ras\footnote{\url{https://www.nuvucameras.com}}. A $8 \times 8$ pixels box is attributed to each sub-aperture with a field of view of $6.4 \times \arcsec{6.4}$.

The performances of an AO loop are mainly limited by four kinds of errors\cite{Rigaut:98_SH_error}. The \textbf{fitting error} is induced by the DM which cannot correct wavefront errors at a scale smaller than its actuator pitch. A SH-WFS is a low-pass sensor that produces \textbf{aliasing error}, by wrapping the unmeasured high spatial frequencies. Limited to faint targets and running at a high framerates, a WFS works in photon-starved conditions with measurements corrupted by readout noise and photon noise, impacting the estimation of the optimal command to send to the DM via a \textbf{noise error}. Finally, due the exposure time and the time needed to process the data, the AO loop is always delayed compared to the turbulence, inducing a \textbf{servo-lag error}.

This paper focuses on the two first errors via an end-to-end model (E2E) that was developed along with the bench design and alignment\cite{Berdeu:22_AO_bench}. The main objective of this tool is to have a numerical model as close as possible from the real bench to develop and test new AO algorithms or predict on-sky raw contrast\cite{Ridsdill:22}. Studying the discrepancies between the numerical predictions and the experimental data can be of great help to improve the models and develop more robust algorithms. Thus, our E2E model integrates the real influence function of the DM measured on the bench as well as the turbulence injected in the bench via a custom-designed phase plate\cite{Ebstein:02_Lexitek,Berdeu:22_AO_bench}.

First, in \refsec{sec:fitting_error}, we present a new analytical model of the fitting error. Up-to-date, the only way to study the impact of the shape of the DM influence function in the fitting error is to perform extensive Monte Carlo simulations\cite{Flicker:08_Keck_AO}. Analytical models\cite{Veran:97_PSF_AO_telemetry, Rigaut:98_SH_error, Tokovinin:00_MCAO_isoplanatism, Ellerbroek:05_PSD_AO_model, Jolissaint:06_AO_analytical, Correia:14_anti-aliasing} simply use a binary approximation to model the frequency correction induced by the DM, which leads to optimistic results. We show that an analytical solution of this problem exists.

Second, in \refsec{sec:aliasing_error}, we study the possibility to partially correct the aliasing error in close loop via an inverse problem approach that performs a super-resolved reconstruction of the incident wavefront. Up-to-date, the best solution to tackle this problem is to insert a pinhole to optically filter out the high-spatial frequencies\cite{Poyneer:04_SFWFS}. This concept is effective in the context of XAO\cite{Sauvage:14_SPHERE_SF}, but it cannot really be extended for more conventional systems such as the use of a laser guide star\cite{Fusco:04_SH_noise} or solar astronomy\cite{Tallon:19_THEMIS}. Finding a numerical solution to this problem implies to only change the AO control loop and prevents the need to modify and complicate the optical setups.

\section{ANALYTICAL MODEL OF THE FITTING ERROR}
\label{sec:fitting_error}

The atmosphere turbulence is a random process\cite{Tyson:15_principles_of_AO} that distorts the incoming wavefront~$\pha$. Thus, the instantaneous point spread function (PSF) of an instrument cannot be predicted in advance. The long-exposure PSF can nonetheless still be estimated via the knowledge of the statistical features of the turbulence\cite{Roddier:81} such as its structure function~$\SF{\pha}$ (SF, assumed stationary in the following) or equivalently its power spectrum density~$\PSD{\pha}$ (PSD)
\begin{equation} % eq:SF_PSD_def
	\label{eq:SF_PSD_def}
	\SF{\pha}\Paren{\Vx} \triangleq \AvgT{\Paren{\pha\Paren{\Vx,t} - \pha\Paren{\V{0},t}}^{2}} = 2\int{\PSD{\pha}\Paren{\Vk}\Dk} - 2\FTfullinv{\PSD{\pha}\Paren{\Vk}}{\Vx}
	\text{ with }
	\PSD{\pha}\Paren{\Vk} \triangleq \AvgT{\Abs{\FT{\pha}\Paren{\Vk,t}}^{2}}
	\,,
\end{equation}
where $\AvgT{.}$ denotes the time averaged value and where the 2D Fourier transform of~$\pha$ is defined as
\begin{equation} % eq:FT
	\label{eq:FT}
	\FT{\pha}\Paren{\Vk}
	{}\triangleq{} \FTfull{\pha}{\Vk}
	{}={} \int{\pha\Paren{\Vx}\E{-2\I\pi\Vx\T\Vk}\Dx}
	\text{ and }
	\pha\Paren{\Vx}
	{}\triangleq{} \FTfullinv{\FT{\pha}}{\Vx}
	{}={} \int{\FT{\pha}\Paren{\Vk}\E{2\I\pi\Vx\T\Vk}\Dk}
	\,.
\end{equation}

Doing so, the long-exposure PSF can be expressed as the convolution between the diffraction limited PSF of the telescope~$\PSFtel$ and an equivalent PSF induced by the incident turbulent wavefront~$\PSFturb$. This yields the following product between the corresponding optical transfer functions $\FTPSFturb$ and $\FTPSFtel$:
\begin{equation}
	\label{eq:FTlongPSF_SF_stat}
	\FTPSF\Paren{\Vx} =\FTPSFturb\Paren{\Vx} \FTPSFtel\Paren{\Vx}
	\text{ with }
	\begin{cases}
		\FTPSFturb\Paren{\Vx} \triangleq \E{-\frac{1}{2}\SF{\pha}\Paren{\Vx}}
		\\
		\FTPSFtel\Paren{\Vx} \triangleq  \int\pupil\Paren{\Vxp}\conj{\pupil}\Paren{\Vxp+\Vx}\Dxp
	\end{cases}
	\,,
\end{equation}
where $\pupil$ is the pupil of the instrument. The SF and the PSD can be parametrised by a few number of variables. For example, in the case of a Kolomogorov statistic, as assumed in the following, one gets\cite{Jolissaint:06_AO_analytical}
\begin{equation}
	\label{eq:Kolmogorov}
	\PSDturb\Paren{\Vk} \triangleq 0.023\rFried^{-5/3}\Norm{\Vk}^{-11/3}
	\Leftrightarrow
	\SFturb\Paren{\Vx} \triangleq 6.88\Paren{\Norm{\Vx}/\rFried}^{5/3}
	\,,
\end{equation}
where the Fried's parameter~$\rFried$ is the turbulence coherence length that describes its strength. 

In analytical studies\cite{Veran:97_PSF_AO_telemetry, Rigaut:98_SH_error, Tokovinin:00_MCAO_isoplanatism, Ellerbroek:05_PSD_AO_model, Jolissaint:06_AO_analytical, Correia:14_anti-aliasing}, a binary mask is assumed for the PSD~$\PSDres$ of the fitting error $\phaAO$ after an optimal AO correction\footnote{That is to say when applying the command on the DM that minimises the variance in the pupil, without any noise, aliasing, nor lag.}
\begin{equation} % eq:binary_mask
	\label{eq:binary_mask}
	\PSDres\Paren{\Vk} = \Paren{1-\fLF\Paren{\Vk}}\PSDturb\Paren{\Vk}
	\text{ with } \fLF\Paren{\Vk} \triangleq 
	\begin{cases}
	1 \text{ if } \Abs{\Vk}<\Paren{2\Vpitch}^{\EWprod-1}
	\\
	0 \text{ otherwise}	
	\end{cases}
	\hspace{-0.5cm}
	\,.
\end{equation}
\eqfull{eq:binary_mask} implies a perfect correction below the cut-off frequency of the DM and no correction beyond. This approximation is based on the fact than the DM cannot compensate for spatial frequencies higher than the pitch~$\Vpitch$ of its actuators.

\begin{figure}[ht!]
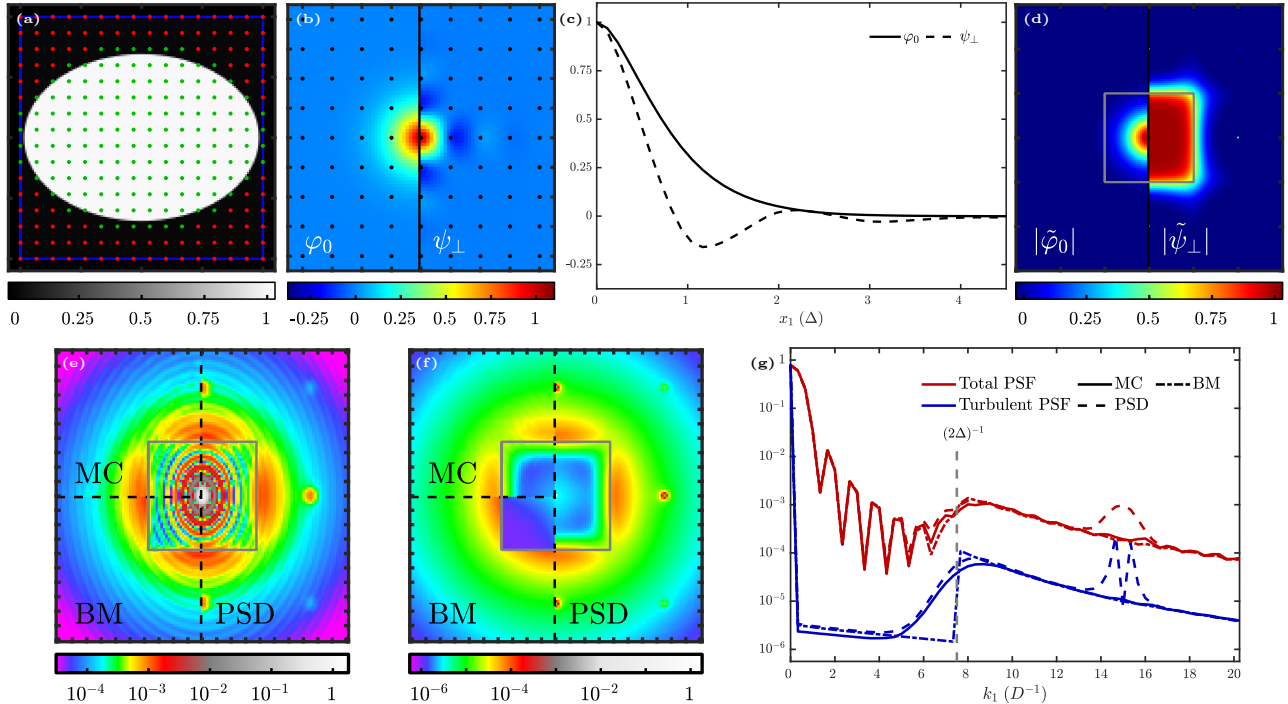
 % fig:Fitting_error
        \centering
        
        % Internal command of the figure for the automatic sizing
        % Line ratio
        \newcommand{\LineRatio}{0.99}
        \newcommand{\PathFig}{figures_Fitting_error_}
        \newcommand{\FlagAperture}{EvWaCo_97p5}
        \newcommand{\FlagIF}{ALPAO_repair_true}
        
        % First line
        \newcommand{\FigOne}{\PathFig aperture_\FlagAperture }
        \newcommand{\FigTwo}{\PathFig IF_\FlagIF }
        \newcommand{\FigThree}{\PathFig IF_prof}
        \newcommand{\FigFour}{\PathFig IF_\FlagIF _f}
        
        \newcommand{\subfigColor}{black}        
        
        % Getting the size of the boxes
        \sbox1{\includegraphics{\FigOne}}               % 1st column
        \sbox2{\includegraphics{\FigTwo}}               % 2nd column
        \sbox3{\includegraphics{\FigThree}}     % 3rd column
        \sbox4{\includegraphics{\FigFour}}     % 4th column
        % Defining column width command
        \newcommand{\ColumnWidth}[1]
                {\dimexpr \LineRatio \linewidth * \AspectRatio{#1} / (\AspectRatio{1} + \AspectRatio{2} + \AspectRatio{3} + \AspectRatio{4}) \relax
                }
        \newcommand{\ColumnGap}{\hspace {\dimexpr \linewidth /5 - \LineRatio\linewidth /5 }}

        % Figure table
        \begin{tabular}{
                @{\ColumnGap}
                M{\ColumnWidth{1}}
                @{\ColumnGap}
                M{\ColumnWidth{2}}
                @{\ColumnGap}
                M{\ColumnWidth{3}}
                @{\ColumnGap}
                M{\ColumnWidth{4}}
                @{\ColumnGap}
                }
                \subfigimg[width=\linewidth,pos=ul,font=\fontfig{white}]{$\;$(a)}{0.0}{\FigOne .pdf} &
                \subfigimg[width=\linewidth,pos=ul,font=\fontfig{white}]{$\;$(b)}{0.0}{\FigTwo .pdf} &
                \subfigimg[width=\linewidth,pos=ul,font=\fontfig{\subfigColor}]{\hspace{-4pt}(c)}{0.0}{\FigThree .pdf} &
                \subfigimg[width=\linewidth,pos=ul,font=\fontfig{white}]{$\;$(d)}{0.0}{\FigFour .pdf}
        \end{tabular}

        % Second line
        \renewcommand{\LineRatio}{0.85}
        \renewcommand{\FigOne}{\PathFig PSF_\FlagAperture _\FlagIF }
        \renewcommand{\FigTwo}{\PathFig PSF_PSD_\FlagAperture _\FlagIF }
        \renewcommand{\FigThree}{\PathFig PSF_comb_\FlagAperture _profile}
        
        % Getting the size of the boxes
        \sbox1{\includegraphics{\FigOne}}               % 1st column
        \sbox2{\includegraphics{\FigTwo}}               % 2nd column
        \sbox3{\includegraphics{\FigThree}}     % 3rd column
        \sbox4{\includegraphics{\FigFour}}     % 4th column
        % Defining column width command
        \renewcommand{\ColumnWidth}[1]
                {\dimexpr \LineRatio \linewidth * \AspectRatio{#1} / (\AspectRatio{1} + \AspectRatio{2} + \AspectRatio{3}) \relax
                }
        \renewcommand{\ColumnGap}{\hspace {\dimexpr \linewidth /4 - \LineRatio\linewidth /4 }}

        % Figure table
        \begin{tabular}{
                @{\ColumnGap}
                M{\ColumnWidth{1}}
                @{\ColumnGap}
                M{\ColumnWidth{2}}
                @{\ColumnGap}
                M{\ColumnWidth{3}}
                @{\ColumnGap}
                }
                \subfigimg[width=\linewidth,pos=ul,font=\fontfig{white}]{$\;$(e)}{0.0}{\FigOne .pdf} &
                \subfigimg[width=\linewidth,pos=ul,font=\fontfig{white}]{$\;$(f)}{0.0}{\FigTwo .pdf} &
                \subfigimg[width=\linewidth,pos=ul,font=\fontfig{\subfigColor}]{\hspace{-5pt}(g)}{0.0}{\FigThree .pdf}
        \end{tabular}

        \caption{\label{fig:Fitting_error} Analytical model of the fitting error. (a)~Pupil of EvWaCo. The actuators (dots) are on a Cartesian grid in a Fried geometry. The active actuators are in green. The blue square is the simulation diameter $\Dsim$. (b)~Influence function of the ALPAO DM192 before ($\IFref$, left) and after ($\IFperp$, right) its orthogonalisation. (c)~Cut-profile of (b) along the $x$-axis. (d)~Visualisation of the modulus of the Fourier transform of the influence functions before ($\Abs{\FTIFref}$, left) and after ($\abs{\FTIFperp}$, right) orthogonalisation. (e,f)~Simulated long-exposure PSF~$\PSF$~(e) and PSF of the turbulence residuals~$\PSFres$~(f) via the Monte Carlo simulation (MC, upper left corner), via the analytical power spectrum density model (PSD, right) and via the standard binary mask (BM, lower left corner). (g)~Cut profiles of~(e) and~(f). (d,e,f)~The gray squares represent the cut-off frequency~$\abs{\Vk}<\Paren{2\Vpitch}^{\EWprod-1}$ due to the actuator pitch.}
\end{figure}

Nonetheless, it is known that this approximation is optimistic. In fact, the fitting error depends on the influence function profile $\IFref\Paren{\Vx}$ of the DM, as shown in E2E simulations\cite{Flicker:08_Keck_AO}. We prove, see Berdeu~\etal\cite{Berdeu:22_IF}, that there is an analytical expression for the fitting error PSD $\PSDres\Paren{\Vk}$ in terms of $\IFref\Paren{\Vx}$ that can be written
\begin{equation} % eq:PSDres
	\label{eq:PSDres}
	\PSDres\Paren{\Vk} =
	\Paren{1 - 2 \nact\PSDperp\Paren{\Vk}}\PSDturb\Paren{\Vk} 
	+
	\sum_{a\in\setactu} \sum_{\ap\in\setactu} \PSDperp\Paren{\Vk}\E{-2\I\pi\Paren{\Pa-\Pap}\T\Vk}
	\times \int{\PSDturb\Paren{\Vkp}\PSDperp\Paren{\Vkp}\E{-2\I\pi\Paren{\Pa-\Pap}\T\Vkp}\Dkp}
	\,,
\end{equation}
where~$\nact$ is the total number of actuators, $\setactu$ is the set of actuators, $\Pa$ is the position of actuator $a$, and~$\PSDperp$ is the PSD of~$\IFperp$ which is the profile obtained from~$\IFref$ so that the basis defined by~$\Brace{\IFperp\Paren{\Vx - \Pa}}_{a\in\setactu}$ is orthonormal. \subfigsfull{fig:Fitting_error}{b,c} presents the influence function $\IFref$ of the DM192 from ALPAO that is used in the AO system of EvWaCo as well as its orthonormalised counterpart $\IFperp$. This is worth noticing that $\IFperp$ resembles a sinc profile whose Fourier transform, displayed in \subfig{fig:Fitting_error}{d}, gives some hints of the PSD area that can be corrected by the DM.

Using \eq{eq:PSDres} in \eq{eq:SF_PSD_def} to get the structure function after AO correction, it is possible to get $\PSF$ and the long-exposure PSF induced by the residual fitting error, $\PSFres$, via \eq{eq:FTlongPSF_SF_stat}. The smaller the wavefront residuals, the closer $\PSFres$ will be to a delta function and the closer the long exposure PSF, $\PSF$, will be to the diffraction limited PSF, $\PSFtel$.  The long-exposure PSF predicted by the analytical PSD is compared with Monte Carlo (MC) simulations in \subfig{fig:Fitting_error}{e}, averaging $\nPha = 10000$ random phase screens propagated through the EvWaCo aperture of \subfig{fig:Fitting_error}{a} with $\rFried = \Dsim/15$. The Kolmogorov screens are generated using the method described by Lane~\etal\cite{Lane:92_Subharmonics} with $16$ sub-harmonics to inject the low spatial frequencies of the wavefront. For information, the PSF obtained when assuming a binary mask (BM) via \eq{eq:binary_mask} is given. Except for the diffraction order secondary spots, the MC simulations and the theoretical PSD predictions are undistinguishable. No significant deviation can be noticed from the cut-profiles (red) of \subfig{fig:Fitting_error}{g}, but a close look at \subfig{fig:Fitting_error}{e} shows that the diffraction rings with the BM hypothesis are optimistically deeper.

The main results come from the analysis of the PSF~$\PSFres$ of the turbulence residuals after a perfect AO correction, in \subfig{fig:Fitting_error}{f}. Indeed, if one removes the central peak of this PSF, that gives the Strehl ratio\cite{Tyson:15_principles_of_AO} of the long-exposure PSF, one gets the 2D map of the best achievable contrast with a perfect coronagraph that would perfectly remove the on-axis light while leaving the off-axis light undisturbed\cite{Cavarroc:06_perfect_coro, Sauvage:10_AO_long_exp_perfect_coro}. Once again, the MC simulations and the analytical PSD model give the same results. The cut-profiles (blue) of \subfig{fig:Fitting_error}{g} show that the analytical model correctly describes the transition between the AO corrected area inside the DM cut-off frequency and the uncorrected area dominated by the turbulence. These figures also emphasise how the BM model is an optimistic approximation. As expected from its definition, it produces a sharp transition at the border of the AO corrected area, predicting a contrast that is almost two orders of magnitude better than the MC simulations.

Having such an analytical model can be very useful when it comes to perform analytical predictions of on-sky performances based on a limited number of meaningful parameters. This avoids to run extensive and time consuming MC simulations. It can also be used to optimise the influence function profile according to the needs and scientific targets of the instrument.

\section{LOWERING THE ALIASING ERROR VIA AN INVERSE PROBLEM APPROACH}
\label{sec:aliasing_error}

In this section, we focus on the possibility to tackle the aliasing error in a Shack-Hartmann wavefront sensor (SH-WFS) via an inverse problem approach based on a minimal variance estimator without the need to change the optical components. We describe its proof of concept and compare it with standard command estimators based on least-squares fit coupled or not with an optical low-pass filter.

\subsection{General Idea: Super-Resolved Wavefront Reconstruction}

For linear WFS (such as here a SH-WFS) the measurement model can be expressed as follows\cite{Thibaut:10_FRIM}
\begin{equation}
	\label{eq:data}
	\Vd = \MSsy \cdot \Vw + \Vn
	\,,
\end{equation}
where $\Vd\in\Reals^{\ndat}$ is the vector of the $\ndat\in\PosIntegers$ data (here the wavefront slopes in each sub-aperture), $\MSsy\in\Reals^{\ndat \times \nw}$ is the WFS synthetic model matrix (here the average phase derivative on each sub-aperture\cite{Rigaut:98_SH_error}), $\Vw\in\Reals^{\nw}$ is the vector of the wavefront described on $\nw\in\PosIntegers$ spatial knots (it can be infinite for a continuous description of the wavefront), and $\Vn$ is the noise on the measurements. Let us also introduce $\MM\in\Reals^{\nw \times \nact}$, the matrix of the mirror influence function \cite{Berdeu:22_AO_bench} of the $\nact\in\PosIntegers$ actuators of the DM, and $\MGsy = \MSsy \cdot \MM \in \Reals^{\ndat \times \nact}$, the synthetic interaction matrix of the AO system that links the commands $\Vc\in\Reals^{\nact}$ applied on the DM to their equivalent synthetic slopes.

Let us emphasise here that $\MSsy$ and $\MGsy$ represent a synthetic linear model of the SH-WFS, but not necessarily the `truth'. As further detailed in \refsec{sec:E2E}, the `truth' is given by the E2E model that accounts for the Fourier optics propagation through the system combined with centroiding algorithm to extract the slopes from the spots location. These `true' operators will be noted $\MSfo$ and $\MGfo$ in following.

In AO loop control, the objective is to find the optimal set of commands $\tilde{\Vc}$ to apply on the mirror based on the slopes measurements. As discussed in the following, different approaches can be used.

\subsubsection{Least-squares minimisation}
\label{sec:LS}

The least-squares (LS) method is the standard approach\cite{Rousset:99, Thibaut:10_FRIM} in AO. The commands $\tilde{\Vc}$ are considered to be the ones that produce slopes according to the interaction matrix $\MG$ as close as possible to the measured slopes. Let us notice here that depending on the context, as described in the following sections, $\MG$ can be either the synthetic model $\MGsy$ or the `true' model $\MGfo$. It consists in minimizing
\begin{equation}
	\tilde{\Vc}
	= \argmin{\Vc} \Norm{\Vd-\MG\cdot\Vc}^{2}_{\MCov{\Vn}}
	=
	\Paren{\MG\T\cdot\MCov{\Vn}\cdot\MG}\Inv\cdot
	\MG\T\cdot\MCov{\Vn} \cdot \Vd
	\,,
\end{equation}
where $\MCov{\Vn}=\AvgT{\Vd\cdot\Vd\T}$ is the symmetric covariance matrix of the measurement noise $\Vn$. To reduce the computational burden of estimating the covariance matrix $\MCov{\Vn}$ and computing the inverse, the noise is generally assumed independent and identically distributed and the equation becomes
\begin{equation}
	\label{eq:LS}
	\tilde{\Vc}
	= \Paren{\MG\T\cdot\MG}\Inv\cdot
	\MG\T \cdot \Vd
	= \MCLS \cdot \Vd
	\,,
\end{equation}
that is the standard pseudo-inverse approach. It can be solved either with Tikhonov regularisation when computing the inverse\cite{Bonnefond:16_FrIM_fragmented_pupil} or via singular value decomposition (SVD) by removing the last $\nsvd$ modes of $\MG$ to avoid noise amplification\cite{Boyer:90, Tyson:15_principles_of_AO}.

The LS method is fast and easy to implement as it only implies the experimental measurement of the interaction matrix $\MG$. Nonetheless, it does not account for the noise of the measurements and does not add any prior on the incoming wavefront that could help correcting the aliasing errors induced by the SH-WFS.

\subsubsection{Minimum variance estimator for optimal and super-resolved wavefront reconstruction}
\label{sec:MV}

The Maréchal's approximation\cite{Tyson:15_principles_of_AO} links the Strehl ratio $\strehl$ of the PSF with the phase variance of the wavefront $\Vw$ on the pupil 
\begin{equation} % eq:Strehl_Marechal
	\label{eq:Strehl_Marechal}
	\strehl \simeq \E{-\Norm{\MPis \cdot \Vw}_{\MPup}^2}
	\,,
\end{equation}
where $\MPis$ is the piston-removal operator on the pupil and $\MPup$ is the diagonal operator with the pupil weights $\V{m}$ defined as in \eq{eq:W_pupil}. Introducing the Kronecker symbol $\delta$, their expression is\cite{Wallner:83}
\begin{equation}
	\Pup_{i,j} =
	%\underset{\eq{eq:W_pupil}}{=}
	\delta_{i,j} m_{i}
	\text{ and }
	\Pis_{i,j} = 
	\delta_{i,j}-m_{j}/\sum_{k=1}^{\nw}m_{k}
	\,.
\end{equation}

As presented in \refsec{sec:fitting_error}, turbulence is a random process that can be statistically described by a limited number of parameters. When questing an optimal command estimator, one could think of finding the operator $\MCMV$ that minimises, in average, the variance on the pupil
\begin{equation}
	\MCMV
	= \argmin{\MC} \AvgT{\Norm{\MPis\cdot\Paren{\Vw - \MM \cdot \MC \cdot \Vd}}_{\MPup}^{2}}
	\,.
\end{equation}
The problem is quadratic and convex in $\MC$ and its solution is given by nulling its first derivative\cite{Petersen:12_MatrixCookbook}
\begin{equation}
	0 = \left.\frac{\partial \AvgT{\Norm{\MPis\cdot\Paren{\Vw - \MM \cdot \MC \cdot \Vd}}_{\MPup}}}{\partial \MC}\right|_{\MC = \MCMV}
	\Rightarrow
	\MM\T\cdot\MPis\T\cdot\MPup\cdot\MPis\cdot\MM\cdot\MCMV\cdot\AvgT{\Vd\cdot\Vd\T}
	= \MM\T\cdot\MPis\T\cdot\MPup\cdot\MPis\cdot\AvgT{\Vw\cdot\Vd\T}
	\,.
\end{equation}
And as $\Vn$ and $\Vw$ are independent, it comes from \eq{eq:data}
\begin{equation}
	\AvgT{\Vd\cdot\Vd\T} = \MSsy\cdot\MCov{\Vw}\cdot\MSsy\T + \MCov{\Vn}
	\text{ and }
	\AvgT{\Vw\cdot\Vd\T} = \MCov{\Vw}\cdot\MSsy\T
	\,,
\end{equation}
with $\MCov{\Vw}=\AvgT{\Vw\cdot\Vw\T}$ the wavefront covariance. Noting $\MMpis = \MPis\cdot\MM$ the piston-free influence function matrix, the optimal command estimator now writes\cite{Wallner:83, Fusco:00_PhD, LeRoux:03_PhD, Thibaut:10_FRIM}
\begin{equation}
	\label{eq:MV}
	\MCMV = \underbrace{\Paren{\MMpis\T\cdot\MPup\cdot\MMpis}\Inv\MMpis\T\cdot\MPup}_{\text{Optimal projector }\M{\Pi}}
	\cdot
	\underbrace{\MPis\cdot\MCov{\Vw}\cdot\MSsy\T\Paren{\MSsy\cdot\MCov{\Vw}\cdot\MSsy\T + \MCov{\Vn}}\Inv}_{\text{Optimal reconstructor }\M{R}}
	\text{ and }
	\tilde{\Vc} = \M{\Pi}\cdot\M{R} \cdot \Vd = \MCMV \cdot \Vd
	\,.
\end{equation}
with $\M{\Pi}\in\Reals^{\nact\times\nw}$ and $\M{R}\in\Reals^{\nw\times\ndat}$. Let us notice that both $\M{\Pi}$ and $\M{R}$ are piston-free and can be of infinite size depending on $\nw$. But whatever the size of the reconstructed wavefront, their product $\MCMV=\M{\Pi}\cdot\M{R}\in\Reals^{\nact\times\ndat}$ remains finite.

This minimum variance (MV) method thus combines the optimal projector $\M{\Pi}$ of a wavefront onto the DM that can be computed once for all and an optimal wavefront reconstructor $\M{R}$ at any wanted resolution. As discussed by Thiébaut\&{}Tallon\cite{Thibaut:10_FRIM}, this methods is equivalent to the LS method of \refsec{sec:LS} but by introducing a Tikhonov regularisation that enforces \textit{a priori} covariance on the unknowns. This methods has also been proven to be the most effective to deal with fragmented pupils that induce missing data in the model\cite{Bonnefond:16_FrIM_fragmented_pupil}.

With the notations introduced in \refsec{sec:fitting_error}, the terms of the covariance matrix~$\MCov{\Vw}$ are given by
\begin{equation}
	\Brack{\MCov{\Vw}}_{i,j}
	= \AvgT{\pha_{i}\pha_{j}}
%	= \AvgT{\pha\Paren{\V{p}_{i}}\pha\Paren{\V{p}_{j}}}
	= \frac{1}{2}\AvgT{\pha^2_{i} + \pha^2_{j} - \Paren{\pha_{i} - \pha_{j}}^{2}} = \sigma_{\pha}^2 - \frac{1}{2}\SF{\pha}\Paren{\V{p}_{i}-\V{p}_{j}}
	\,,
\end{equation}
where $\V{p}_{i}$ and $\V{p}_{j}$ are the positions of the wavefront knots $\pha_{i}$ and $\pha_{j}$, $\sigma_{\pha}^2=\AvgT{\pha^2\Paren{\V{0}}}$ is the average variance of the wavefront that is assumed to be stationary (spatially and temporally constant). In principle, this value is unknown. Nonetheless, as $\MCov{\Vw}$ is always multiplied by $\MSsy$ or $\MPis$ that are piston-insensitive, this value is removed and has no impact. Thus all the products implying $\MCov{\Vw}$ can be computed once for all. In the case of Kolmogorov statistics \eq{eq:Kolmogorov}, the Fried's parameter $\rFried$ can be put in factor of all these products and acts as a regularisation hyper-parameter that can be auto-calibrated during the observation.

As for the LS method of \refsec{sec:LS}, the MV method implies to compute an inverse matrix that includes~$\MCov{\V{n}}$ which changes at each iteration of the loop depending on the measurement noise. It has been shown that the problem can be solved in a real time loop with a gradient method via FrIM, a fractal approach that also uses a dedicated pre-conditioner\cite{Thibaut:10_FRIM, Tallon:19_THEMIS}.

In the following, we assess the possibility to use this method to tackle the aliasing error with this purely data science solution based on an inverse problem approach rather than the addition of an optical component. Indeed, the wavefront can be reconstructed (WF reconstruction) at a resolution higher than the actuator grid which prevents the aliasing recovery in the LS method. In the MV method, the missing frequencies that are aliased by the SH-WFS should be partially retrieved by the use of the adequate prior on the turbulence statistics~$\MCov{\Vw}$.

Some numerical methods\cite{Bond:15_anti_aliasing} aiming at reducing the aliasing have already been developed with mitigated results, both in terms of performances and computational costs. They are based on Wiener filters in the Fourier space and use frequency unwrapping algorithm. Here we intend to solve the problem directly in the direct space without any need of further approximation.

\subsection{Overview of the End-to-end Model and its AO Loop}
\label{sec:AO_loop}

As mentioned in the introduction, a dedicated AO bench has been developed at NARIT to test the components and the AO algorithms\cite{Berdeu:22_AO_bench}. Along with this bench, an E2E model has been developed to simulate all its features and predict its performances. Doing so allows us to deeply understand each component and algorithm method and to look for better solutions. The numerical model can help to assess which part of the bench is not working optimally but also to find if there is any experimental feature not taken into account in the numerical model.

\textbf{Wavefront simulation and control} -- A phase plate was designed and procured to emulate the turbulence in the bench\cite{Berdeu:22_AO_bench}. All the following simulations are based on the inner track of this phase plate which corresponds to a seeing of \arcsec{1}. Between each iteration, the phase plate is rotated by an amount $\delta_\theta$ to emulate a wind blowing vertically at \SI{10}{\meter\per\second} and a loop running at \SI{1}{\kilo\hertz}. With the phase plate physical parameters, one complete revolution of the phase plate corresponds to $n_\theta=1834$ iterations. The bench was also used to measure the influence functions of the ALPAO 192DM\cite{Berdeu:22_AO_bench}. These measurements are used to generate the mirror matrix $\MM$. Except if stated otherwise, all the simulations are mono-chromatic at $\lambda = \SI{617}{\nano\meter}$ which is the wavelength of the LED source in the bench.

\begin{figure}[ht!]
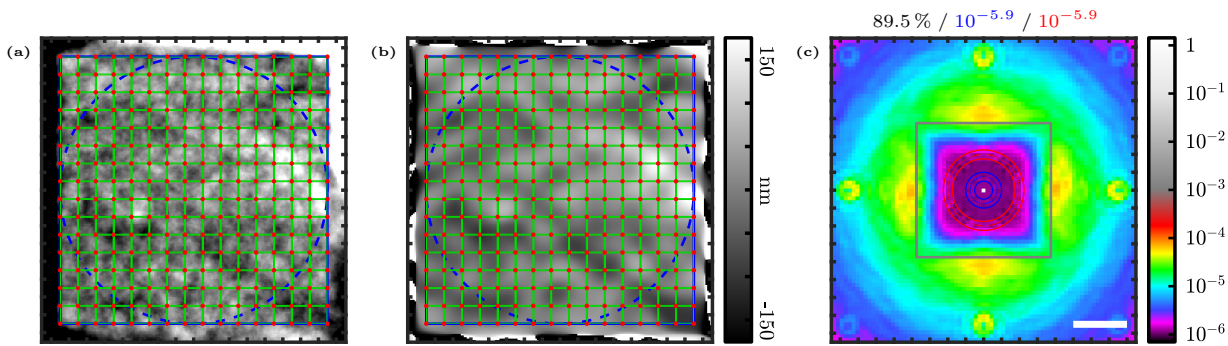
 % fig:SH_fitting
        \centering
        
        % Internal command of the figure for the automatic sizing
        % Line ratio
        \newcommand{\LineRatio}{0.85}
        \newcommand{\PathFig}{figures_SH_}
        
        % First line
        \newcommand{\PathTxt}{./figures/\PathFig fitting_error.txt}
        \newcommand{\FigOne}{\PathFig SH_WF}
        \newcommand{\FigTwo}{\PathFig SH_WF_filt}
        \newcommand{\FigThree}{\PathFig fitting_error}
        
        \newcommand{\subfigColor}{black}        
        
        % Getting the size of the boxes
        \sbox1{\includegraphics{\FigOne}}	% 1st column
        \sbox2{\includegraphics{\FigTwo}}	% 2nd column
        \sbox3{\includegraphics{\FigThree}}	% 3rd column
        % Defining column width command
        \newcommand{\ColumnWidth}[1]
                {\dimexpr \LineRatio \linewidth * \AspectRatio{#1} / (\AspectRatio{1} + \AspectRatio{2} + \AspectRatio{3}) \relax
                }
        \newcommand{\ColumnGap}{\hspace {\dimexpr \linewidth /4 - \LineRatio\linewidth /4 }}

        % Figure table
        \begin{tabular}{
                @{\ColumnGap}
                M{\ColumnWidth{1}}
                @{\ColumnGap}
                M{\ColumnWidth{2}}
                @{\ColumnGap}
                M{\ColumnWidth{3}}
                @{\ColumnGap}
                }
                & & \hspace{-32.5pt}\scriptsize{\input{./\PathFig fitting_error.txt}}
                \\[-2.5pt]
                \subfigimg[width=\linewidth,pos=ul,font=\fontfig{\subfigColor}]{\hspace{-14pt}(a)}{0.0}{\FigOne .pdf} &
                \subfigimg[width=\linewidth,pos=ul,font=\fontfig{\subfigColor}]{\hspace{-14pt}(b)}{0.0}{\FigTwo .pdf} &
                \subfigimg[width=\linewidth,pos=ul,font=\fontfig{\subfigColor}]{\hspace{-14pt}(c)}{0.0}{\FigThree .pdf}
        \end{tabular}
        
        \caption{\label{fig:SH_fitting}
        (a,b) Examples of a wavefront entering the SH-WFS during a closed-loop, without (a) and with (b) filtering pinhole $\Paren{\diam =\lDsub}$. The \SI{95}{\percent} fill factor square sub-apertures are emphasised in green and the actuator positions are in red. The blue square is the aperture diameter on which the loop is closed. The dashed blue circle is the pupil region on which the PSD and SF are computed.
        (c) Long-exposure PSF of the fitting error after one phase plate revolution. The gray square is the cut-off frequency~$\abs{\Vk}<\Paren{2\Vpitch}^{\EWprod-1}$ of the DM. The red and green annuli emphasise the regions on which the contrast is estimated at $1.5\lD$ and $4\lD$. The figure title gives $\Strehl \;/\; \ColorOne{\COne} \;/\; \ColorTwo{\CTwo}$. Scale bar: $6\lD$.}
\end{figure}

\textbf{SH-WFS geometry} -- As shown on \subfigs{fig:SH_fitting}{a,b}, to focus on the aliasing effect and avoid any issue induced by the pupil shape and partially illuminated sub-apertures, the geometry of the SH-WFS is a $15\times15$ square lenslet array of fill factor \SI{95}{\percent} (green squares) that fully covers a square aperture of diameter $D$ (blue square). The diameter of the simulation $\Dsim$ spans over $17\times17$ to benefit from the recovery beyond the field of view allowed by the inverse problem approach.

\textbf{Pseudo-code of the AO loop} -- Algorithm~\ref{alg:AO_loop} sums up the steps of the loop. They are further detailed in the following. Indeed, some lines of this pseudo-code may imply many underlying steps depending on the situation.

\begin{algorithm}
\caption{\label{alg:AO_loop} AO loop algorithm of the end-to-end model.}
\begin{algorithmic}[1]
\small

\State $\theta \gets 0$
	\commentalgo{Initialisation of the phase plate angle.}

\For{$i$ from $1$ to $n_\theta$}
	\commentalgo{Loop iterations.}

	\State $\theta \gets \theta+\delta_\theta$
		\commentalgo{Updating the phase plate angle.}
		
	\State $\Wpp^{i} \gets $ extracting the phase screen of the phase plate rotated by $\theta$ 
		\commentalgo{Incident turbulent wavefront.}
		
	\If{$i<3$}
		\State $\Vc^{i} \gets \M{\Pi}\cdot\Wpp^{i}$
		\label{AO_loop:opti_command}
			\commentalgo{Optimal command.}
	\Else
		\State $\Vc^{i} \gets g_{0}\Vc^{i-2} + g_{\delta}\delta\Vc^{i-2}$
		\label{AO_loop:command}
			\commentalgo{Update of the command with a leaky integrator.}
	\EndIf
		
	\State $\Wdm^{i} \gets \MM\cdot\Vc^{i}$
	\label{AO_loop:DM}
		\commentalgo{Sending the command to the DM.}

	\State $\Wao^{i} \gets \Wpp^{i}-\Wdm^{i}$
	\label{AO_loop:Wao}
		\commentalgo{Applying the correction on the incident wavefront.}
		
	\State $\V{U}_{\Tag{AO}} \gets \MPup\cdot\E{\frac{2\I\pi}{\lambda}\Wao^{i}}$
	\label{AO_loop:Uao}
		\commentalgo{Applying the aperture mask on the complex amplitude.}
		
	\State $\V{U}_{\Tag{WFS}} \gets$ propagation of $\V{U}_{\Tag{AO}}$ towards the WFS
	\label{AO_loop:Uwfs}
		\commentalgo{Wavefront propagation through the AO system.}
		
	\State $\delta\Vd^{i} \gets$ propagation of $\V{U}_{\Tag{WFS}}$ through the SH-WFS and slopes extraction
	\label{AO_loop:slopes}
		\commentalgo{Application of the SH-WFS model.}
		
	\State $\delta\Vc^{i} \gets$ command estimator from $\delta\Vd^{i}$ via $\MC$
	\label{AO_loop:delta_command}
		\commentalgo{Estimation of the optimal command correction.}
		
\EndFor
\end{algorithmic}
\end{algorithm}

\textbf{Applied command} -- At each iteration of the loop, the command applied on the DM is computed with a leaky integrator\cite{Merritt:08_beam_control} with a classical delay of two frames, as shown \refline{AO_loop:command}. Its leak factor is $g_{0}=0.99$ and its gain factor is $g_{\delta}=0.75$. For the two first iterations, the loop is closed using the optimal command (fitting error) by projecting the wavefront on the DM via the optimal projector $\M{\Pi}$, \refline{AO_loop:opti_command}.

\textbf{Propagation towards the SH-WFS} -- The propagation of the wavefront towards the SH-WFS pupil, \refline{AO_loop:Uwfs}, depends on the situation. To test the performance of the MV method, nothing is done: $\V{U}_{\Tag{WFS}} = \V{U}_{\Tag{AO}}$. Nonetheless, we want to compare with the solution that consists in adding a low-pass spatial filter in the beam, as proposed by Poyneer~\etal\cite{Poyneer:04_SFWFS}, to optically block the high spatial frequencies and prevent its aliasing in the command estimation. In this case, $\V{U}_{\Tag{WFS}}$ is first propagated to this pinhole, where its binary transmission, is applied before propagating the resulting wavefront towards the SH-WFS pupil. These propagations are performed via Fourier optics propagation\cite{Goodman:05}, see \refapp{app:sub_CT_SH}. Two square pinholes are tested. The optimal filter\cite{Poyneer:04_SFWFS} of side $\lD$ and a more realistic solution\cite{Sauvage:14_SPHERE_SF} that was proven to work on sky of side $1.2\lD$.

\textbf{Computing the slopes} -- Depending on the situation, \refline{AO_loop:slopes} hides several steps. First the wavefront needs to be propagated through the SH-WFS. Second, the slopes $\delta\Vd^{i}$ must be extracted from the propagation output. These steps are further detailed in the following sections.

\textbf{Computing the optimal command} -- In this work, we want to compare the usual LS approach, that cannot correct the aliasing error, with the proposed MV approach. Depending on the method, LS or MV, the computation of the optimal command correction $\delta\Vc^{i}$ differs, \refline{AO_loop:delta_command}. As mentioned above, in the following equations, $\MG$ can either be $\MGsy$ or $\MGfo$ depending on the context.

For the LS method, as no prior is added on the measurements, it is possible to work with the closed-loop slopes $\delta\Vd^{i}$, directly applying \eq{eq:LS}
\begin{equation}
	\label{eq:command_CL}
	\delta\Vc^{i} = \MCLS\Paren{\nsvd} \cdot \delta\Vd^{i}
	\,,
\end{equation}
where $\MCLS\Paren{\nsvd}$ is the pseudo-inverse of the interaction matrix $\MG$ when removing $\nsvd$ modes in its SVD decomposition, as described in \refsec{sec:LS}. This operation is directly the usual `Matrix$\times$Vector' multiplication.

As the MV method adds some \textit{a priori} regularisation on the reconstructed wavefront by imposing a Kolmogorov statistics, the input slopes must satisfy this statistics. But in a closed-loop, the wavefront entering in the SH-WFS does not follow a Kolmogorov statistics. It is thus necessary to generate the pseudo open-loop slopes via the interaction matrix $\MG$ before applying \eq{eq:MV}
\begin{equation}
	\tilde{\Vc}^{i} = \M{\Pi}\cdot\M{R} \cdot \Paren{\MG \cdot \Vc^{i} + \delta\Vd^{i}} = \MCMV \cdot \Paren{\MG \cdot \Vc^{i} + \delta\Vd^{i}}
	\Rightarrow \delta\Vc^{i} = \tilde{\Vc}^{i} - \Vc^{i}
	\,.
\end{equation}
In practice, this problem is solved in two steps by reconstructing first the super-resolved wavefront $\tilde{\Vw}^{i}$ and then projecting it onto the mirror
\begin{equation}
	\label{eq:command_OL}
	\tilde{\Vw}^{i} = \M{R} \cdot \Paren{\MG \cdot \Vc^{i} + \delta\Vd^{i}}
	\Rightarrow \tilde{\Vc}^{i} = \M{\Pi}\cdot\tilde{\Vw}^{i}
	\Rightarrow \delta\Vc^{i} = \tilde{\Vc}^{i} - \Vc^{i}
	\,.
\end{equation}
The projection step is a pure matrix multiplication. On the other side, the reconstruction step implies to compute the inverse of the combination of the noise covariance matrices in \eq{eq:MV}. This is done iteratively via a conjugate gradient algorithm\cite{Barrett:94_ConjGrad, Thibaut:10_FRIM}.

\textbf{Fitting error} -- Finally, let us mention that to avoid any issue with the Fourier transform of the PSD on a square grid, the PSD of the wavefronts and the associated SF are computed on the circular aperture emphasised by the dashed blue circle in \subfigs{fig:SH_fitting}{a,b}. As discussed in \refsec{sec:fitting_error}, the long-exposure PSF of the turbulence residuals limited by the fitting error
\begin{equation}
	\Vw_{\Tag{fitting}}^{i} = \Wpp^{i} - \MM\cdot\M{\Pi}\cdot\Wpp^{i}
\end{equation}
gives the optimal performances expected for a perfect AO loop and a perfect coronagraph. After one phase plate revolution, \subfig{fig:SH_fitting}{c} shows that the maximal achievable Strehl ratio is \SI{89.5}{\percent}. The average value on the blue (resp. red) annulus of width $1\lD$ gives the raw contrast at $1.5\lD$ (resp. $4\lD$) to assess the performance of the algorithm at very close inner working angle (IWA) (resp. at approximately the middle of the AO corrected area). When limited by the fitting error, the raw contrast is $10^{-5.9}$.

\textbf{Important note: frozen correction} -- As we focus on the aliasing error, all the results in the following sections are based on the statistics of the frozen correction
\begin{equation}
	\label{eq:frozen}
	\Vw_{\Tag{frozen}}^{i} = \Wao^{i} - \MM\cdot\delta\Vc^{i} = \Wpp^{i} - \MM\cdot\Paren{\Vc^{i} + \delta\Vc^{i}}
\end{equation}
where $\Wao^{i}$ is the AO residuals of the current incident wavefront $\Wpp^{i}$ corrected with the current command $\Wdm^{i} = \MM\cdot\Vc^{i}$. In other words, the estimated command correction is applied directly on the same measured wavefront that produced the slopes to compute it. Note that the servo-lag error still applies on the closed loop described in Algorithm~\ref{alg:AO_loop} due to \refline{AO_loop:command}. As the wind blows vertically in the simulation, this explains the vertical hourglass shape in \fig{fig:aliasing}: the spots laying at the edge of the aperture always see a very distorted wavefront that has not been properly corrected yet due to the lag, impacting the slope measurements that in return impact the frozen correction. The wavefront defined in \eqref{eq:frozen} is never seen by the AO loop nor the SH-WFS. It is purely used to remove the servo-lag error and study only the correction of the aliasing error.

\subsection{Proof of Concept in an Inverse Crime Simulation}
\label{sec:POC}

To test the hypothesis that the MV method is sensitive to the aliasing via the super-resolution of the wavefront reconstruction and thus can be used to reduce it, we first run an AO loop in an inverse crime simulation\cite{Bonnefond:16_FrIM_fragmented_pupil}. This means that the model that we inverse to close the loop is exactly the model used to run the loop. Namely, the propagation through the SH-WFS %in \eq{eq:data}, and in 
\refline{AO_loop:slopes} of Algorithm~\ref{alg:AO_loop}, is purely based on the synthetic model $\MSsy$ by applying \eq{eq:data}. It is a linear operator that computes the average gradient on each sub-aperture, the gradient being computed by finite differences.

\begin{figure}[ht!]
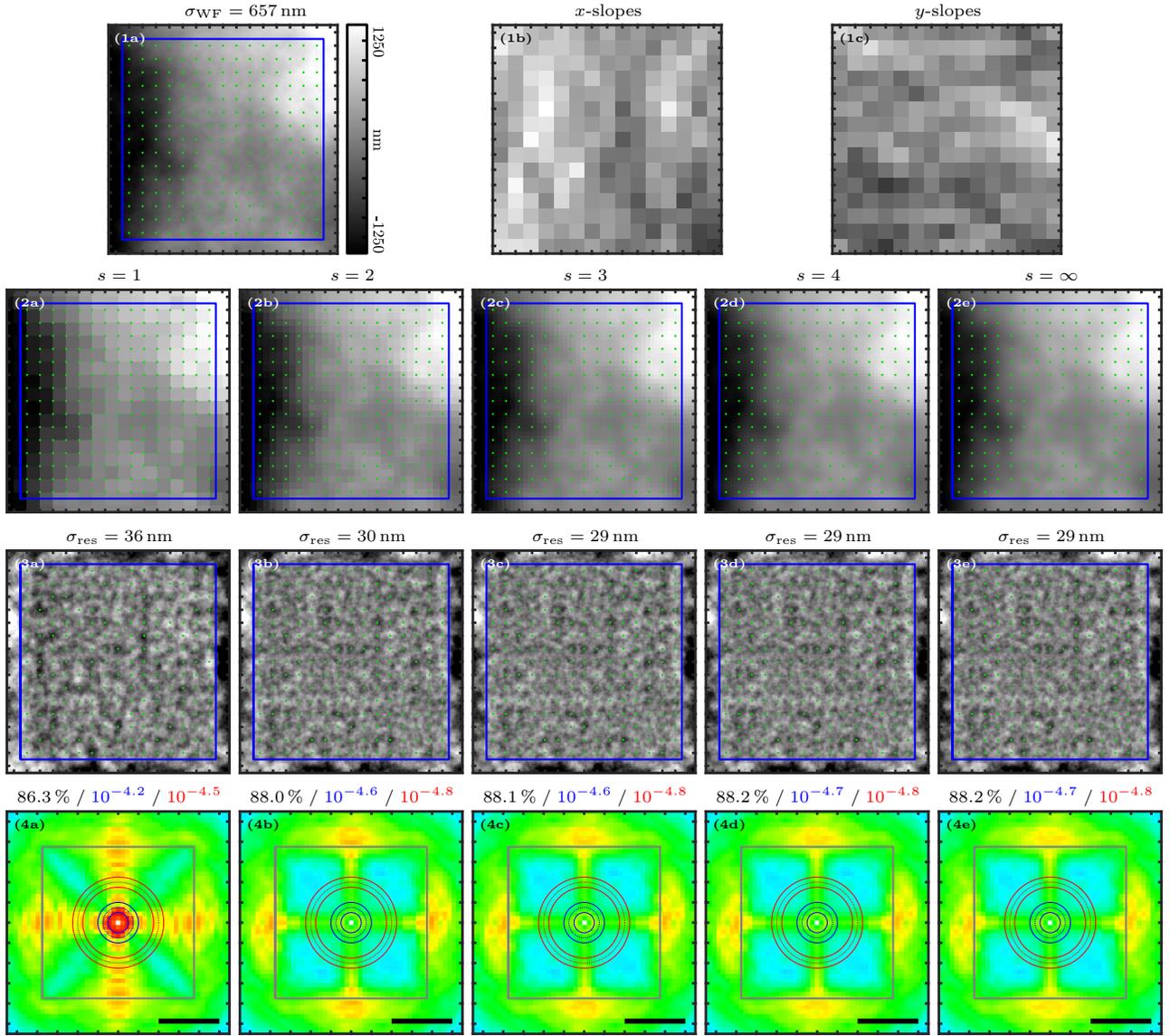
 % fig:WF_recons
        \centering
        
        % Internal command of the figure for the automatic sizing
        \newcommand{\PathFig}{figures_WF_recons_}

        % First table
        \newcommand{\LineRatio}{0.65}
        \newcommand{\FigOne}{\PathFig WF_OL}
        \newcommand{\FigTwo}{\PathFig WF_OL_slopes_x}
        \newcommand{\FigThree}{\PathFig WF_OL_slopes_y}
        
        \newcommand{\subfigColor}{black}        
        
        % Getting the size of the boxes
        \sbox1{\includegraphics{\FigOne}}               % 1st column
        \sbox2{\includegraphics{\FigTwo}}               % 2nd column
        \sbox3{\includegraphics{\FigThree}}     % 3rd column
        % Defining column width command
        \newcommand{\ColumnWidth}[1]
                {\dimexpr \LineRatio \linewidth * \AspectRatio{#1} / (\AspectRatio{1} + \AspectRatio{2} + \AspectRatio{3}) \relax
                }
        \newcommand{\ColumnGap}{\hspace {\dimexpr \linewidth /4 - \LineRatio\linewidth /4 }}

        % Figure table
        \begin{tabular}{
                @{\ColumnGap}
                M{\ColumnWidth{1}}
                @{\ColumnGap}
                M{\ColumnWidth{2}}
                @{\ColumnGap}
                M{\ColumnWidth{3}}
                @{\ColumnGap}
                }
                \\
                \scriptsize{\input{./\FigOne.txt}}
                &
                \scriptsize{$x$-slopes}
                &
                \scriptsize{$y$-slopes}
                \\[-2.5pt]
                \subfigimg[width=\linewidth,pos=ul,font=\fontfig{white}]{$\;$(1a)}{0.0}{\FigOne .pdf} &
                \subfigimg[width=\linewidth,pos=ul,font=\fontfig{\subfigColor}]{$\;$(1b)}{0.0}{\FigTwo .pdf} &
                \subfigimg[width=\linewidth,pos=ul,font=\fontfig{\subfigColor}]{$\;$(1c)}{0.0}{\FigThree .pdf}
        \end{tabular}        
        
        \vspace{-22pt}
        
        % Second table
        \renewcommand{\LineRatio}{0.99}
        \newcommand{\FlagOne}{WF_recons}
        \newcommand{\FlagTwo}{WF_res}
        \newcommand{\FlagThree}{PSF_frozen}
        \newcommand{\FigOneOne}{\PathFig \FlagOne _S1}
        \newcommand{\FigOneTwo}{\PathFig \FlagOne _S2}
        \newcommand{\FigOneThree}{\PathFig \FlagOne _S3}
        \newcommand{\FigOneFour}{\PathFig \FlagOne _S4}
        \newcommand{\FigOneFive}{\PathFig \FlagOne }
        
        \newcommand{\FigTwoOne}{\PathFig \FlagTwo _S1}
        \newcommand{\FigTwoTwo}{\PathFig \FlagTwo _S2}
        \newcommand{\FigTwoThree}{\PathFig \FlagTwo _S3}
        \newcommand{\FigTwoFour}{\PathFig \FlagTwo _S4}
        \newcommand{\FigTwoFive}{\PathFig \FlagTwo }
        
        \newcommand{\FigThreeOne}{\PathFig \FlagThree _S1}
        \newcommand{\FigThreeTwo}{\PathFig \FlagThree _S2}
        \newcommand{\FigThreeThree}{\PathFig \FlagThree _S3}
        \newcommand{\FigThreeFour}{\PathFig \FlagThree _S4}
        \newcommand{\FigThreeFive}{\PathFig \FlagThree }
        
        \renewcommand{\subfigColor}{white}        
        
        % Getting the size of the boxes
        \sbox1{\includegraphics{\FigOneOne}}               % 1st column
        \sbox2{\includegraphics{\FigOneTwo}}               % 2nd column
        \sbox3{\includegraphics{\FigOneThree}}     % 3rd column
        \sbox4{\includegraphics{\FigOneFour}}     % 4th column
        \sbox5{\includegraphics{\FigOneFive}}     % 4th column
        % Defining column width command
        \renewcommand{\ColumnWidth}[1]
                {\dimexpr \LineRatio \linewidth * \AspectRatio{#1} / (\AspectRatio{1} + \AspectRatio{2} + \AspectRatio{3} + \AspectRatio{4}+ \AspectRatio{5}) \relax
                }
        \renewcommand{\ColumnGap}{\hspace {\dimexpr \linewidth /6 - \LineRatio\linewidth /6 }}

        % Figure table
        \begin{tabular}{
                @{\ColumnGap}
                M{\ColumnWidth{1}}
                @{\ColumnGap}
                M{\ColumnWidth{2}}
                @{\ColumnGap}
                M{\ColumnWidth{3}}
                @{\ColumnGap}
                M{\ColumnWidth{4}}
                @{\ColumnGap}
                M{\ColumnWidth{5}}
                @{\ColumnGap}
                }
                \\
                \scriptsize{$s=1$}
                &
                \scriptsize{$s=2$}
                &
                \scriptsize{$s=3$}
                &
                \scriptsize{$s=4$}
                &
                \scriptsize{$s=\infty$}
                \\[-2.5pt]
                \subfigimg[width=\linewidth,pos=ul,font=\fontfig{\subfigColor}]{$\;$(2a)}{0.0}{\FigOneOne .pdf} &
                \subfigimg[width=\linewidth,pos=ul,font=\fontfig{\subfigColor}]{$\;$(2b)}{0.0}{\FigOneTwo .pdf} &
                \subfigimg[width=\linewidth,pos=ul,font=\fontfig{\subfigColor}]{$\;$(2c)}{0.0}{\FigOneThree .pdf} &
                \subfigimg[width=\linewidth,pos=ul,font=\fontfig{\subfigColor}]{$\;$(2d)}{0.0}{\FigOneFour .pdf}&
                \subfigimg[width=\linewidth,pos=ul,font=\fontfig{\subfigColor}]{$\;$(2e)}{0.0}{\FigOneFive .pdf}
                \\[-5pt]
                \scriptsize{\input{./\FigTwoOne.txt}}
                &
                \scriptsize{\input{./\FigTwoTwo.txt}}
                &
                \scriptsize{\input{./\FigTwoThree.txt}}
                &
                \scriptsize{\input{./\FigTwoFour.txt}}
                &
                \scriptsize{\input{./\FigTwoFive.txt}}
                \\[-2.5pt]
                \subfigimg[width=\linewidth,pos=ul,font=\fontfig{\subfigColor}]{$\;$(3a)}{0.0}{\FigTwoOne .pdf} &
                \subfigimg[width=\linewidth,pos=ul,font=\fontfig{\subfigColor}]{$\;$(3b)}{0.0}{\FigTwoTwo .pdf} &
                \subfigimg[width=\linewidth,pos=ul,font=\fontfig{\subfigColor}]{$\;$(3c)}{0.0}{\FigTwoThree .pdf} &
                \subfigimg[width=\linewidth,pos=ul,font=\fontfig{\subfigColor}]{$\;$(3d)}{0.0}{\FigTwoFour .pdf}&
                \subfigimg[width=\linewidth,pos=ul,font=\fontfig{\subfigColor}]{$\;$(3e)}{0.0}{\FigTwoFive .pdf}
                \\[-5pt]
                \scriptsize{\input{./\FigThreeOne.txt}}
                &
                \scriptsize{\input{./\FigThreeTwo.txt}}
                &
                \scriptsize{\input{./\FigThreeThree.txt}}
                &
                \scriptsize{\input{./\FigThreeFour.txt}}
                &
                \scriptsize{\input{./\FigThreeFive.txt}}
                \\[-2.5pt]
                \subfigimg[width=\linewidth,pos=ul,font=\fontfig{black}]{$\;$(4a)}{0.0}{\FigThreeOne .pdf} &
                \subfigimg[width=\linewidth,pos=ul,font=\fontfig{black}]{$\;$(4b)}{0.0}{\FigThreeTwo .pdf} &
                \subfigimg[width=\linewidth,pos=ul,font=\fontfig{black}]{$\;$(4c)}{0.0}{\FigThreeThree .pdf} &
                \subfigimg[width=\linewidth,pos=ul,font=\fontfig{black}]{$\;$(4d)}{0.0}{\FigThreeFour .pdf}&
                \subfigimg[width=\linewidth,pos=ul,font=\fontfig{black}]{$\;$(4e)}{0.0}{\FigThreeFive .pdf}
        \end{tabular}

        \caption{\label{fig:WF_recons} Effect of the sampling parameter~$s$ on the WF reconstruction with the minimum variance estimator and inverse crime simulation. (1a)~Phase screen on the phase plate. $\sigWF$ is the RMS value on the pupil defined by the blue square. The green dots emphasise the position of the SH-WFS sub-apertures. (1b, 1c)~Generated open-loop $xy$-slopes. (2)~WF reconstructed for different values of the sampling parameter~$s$. (3)~Residuals of the synthesised high resolution WF with the theoretical wavefront. The RMS value on the pupil~$\sigRes$ is given for each sampling situation. The color scale is one-tenth of (1a). (4)~Long-exposure PSF of the turbulence residuals alone, $\PSFres$, after one phase plate revolution. See legend and color bar of \subfig{fig:SH_fitting}{c}.%The gray square is the cut-off frequency~$\abs{\Vk}<\Paren{2\Vpitch}^{\EWprod-1}$ of the DM. The red and green annuli emphasise the regions on which the contrast is estimated at $1.5\lD$ and $4\lD$. The sub-figure titles give $\Strehl \;/\; \COne \;/\; \CTwo$. Scale bar: $6\lD$.
        }
\end{figure}

The output of this operator is directly the slopes of the wavefront, as shown in \subfigs{fig:WF_recons}{1b,1c}. In other words, no WFS raw image with spots is produced and there is no need to use a centroiding algorithm.

This operator must be directly applied on the wavefront phase $\Wwfs$, not its complex amplitude $\V{U}_{\Tag{WFS}}$. Without any filtering pinhole, we already mentioned that $\V{U}_{\Tag{WFS}} = \V{U}_{\Tag{AO}}$, and $\MSsy$ is applied on $\Wao$, as for example in \subfig{fig:SH_fitting}{a}. When a spatial filter is inserted in the path, the beam is propagated via Fourier optics, using its complex amplitude. $\MSsy$ is thus applied on $\Wwfs$, the phase of the complex amplitude $\V{U}_{\Tag{WFS}}$, as for example in \subfig{fig:SH_fitting}{b}. This method could lead to phase ambiguity if the absolute value phase goes beyond $\pi$, as seen on the edges of \subfig{fig:SH_fitting}{b}. In practice, in a closed loop we never experienced a phase ambiguity on the pupil and we did not have to implement phase unwrapping algorithm\cite{Zhao:18_unwrapping}. As a side remark, let us notice that the pinhole filters the spatial features as expected: $\Wwfs$ in \subfig{fig:SH_fitting}{b} is a smoother version of $\Wao$ in \subfig{fig:SH_fitting}{a}.

Similarly, the interaction matrix used for the pseudo-inverse in \eq{eq:command_CL} or to generate the open-loop data in \eq{eq:command_OL} is based on this synthetic model $\MGsy = \MSsy \cdot \MM$. No noise is added in the simulation.

First, we focus on the MV method by assessing the impact of the wavefront resolution on the aliasing correction. The lowest resolution corresponds to a sampling $s=1$ wavefront pixel per sub-aperture, placed at the actuator locations (corners of the sub-apertures), as shown in \subfig{fig:WF_recons}{2a}. We then test a sampling $s\in\Brace{2,3,4}$ pixels per sub-aperture, \subfigs{fig:WF_recons}{2b-2d}. These tests are compared with the high resolution wavefront $s=\infty$ corresponding to a sampling of $s=12$, \subfigs{fig:WF_recons}{2e}.

\subfigsfull{fig:WF_recons}{2,3} compare the different sampling parameters $s$. \subfigsfull{fig:WF_recons}{2a-2e} are the reconstructed wavefronts based on the slopes in \subfigs{fig:WF_recons}{1b,1c} of the wavefront of \subfig{fig:WF_recons}{1a}. \subfigsfull{fig:WF_recons}{3a-3e} show the reconstruction residuals. Looking at the reconstruction for $s\geq2$, it can be seen that the regularised wavefront reconstruction successfully retrieves details that are smaller than the sub-aperture size. These details are lost for $s=1$.

This is also confirmed by the root mean square (RMS) error on the pupil given in \subfigs{fig:WF_recons}{3a-3e}. Most of the gain (\SI{15}{\percent}) happens between $s=1$ and $s=2$, with higher residuals for $s=1$. There is a negligible extra gain at $s=3$ that implies that the MV can recover frequencies at the third of the sub-aperture size, despite the quick drop of the PSD power law at high spatial frequencies.

Let us also notice that the inverse problem approach can reconstruct the wavefront beyond the measured aperture, as seen on the region outside the blue square in \subfigs{fig:WF_recons}{2a-3e}. This could be useful in reducing the servo-lag error by having a predictive control loop more evolved than the simple leaky integrator implemented in this study. One could use the information that is retrieved beyond the pupil and that is about to enter in the pupil. It can also be used to slave actuators that lie beyond the pupil aperture to improve the AO residuals on the aperture edges. This is nonetheless beyond the scope of this work.

The improvements in the wavefront reconstruction similarly impact the long-exposure PSF, in \subfigs{fig:WF_recons}{4a-3e}. If the sampling is enough, the MV method can produce Strehl contrasts up to \SI{88.2}{\percent}, only \SI{1.3}{\percent} below the limit of the fitting error, in \subfig{fig:SH_fitting}{c}. As expected, $s=1$ is not sufficient to recover the aliased frequencies that are wrapped inside the AO corrected area, reducing the Strehl ratio and the contrast performances.

Contrary to the Strehl ratio, the performances in terms of raw contrast are less impressive. In the middle of the AO corrected area, the MV method cannot produce contrast below $10^{-4.8}$, slightly more than one magnitude worse than the $10^{-5.9}$ ultimate raw contrast of the fitting error, in \subfig{fig:SH_fitting}{c}. 

\begin{figure}[ht!]
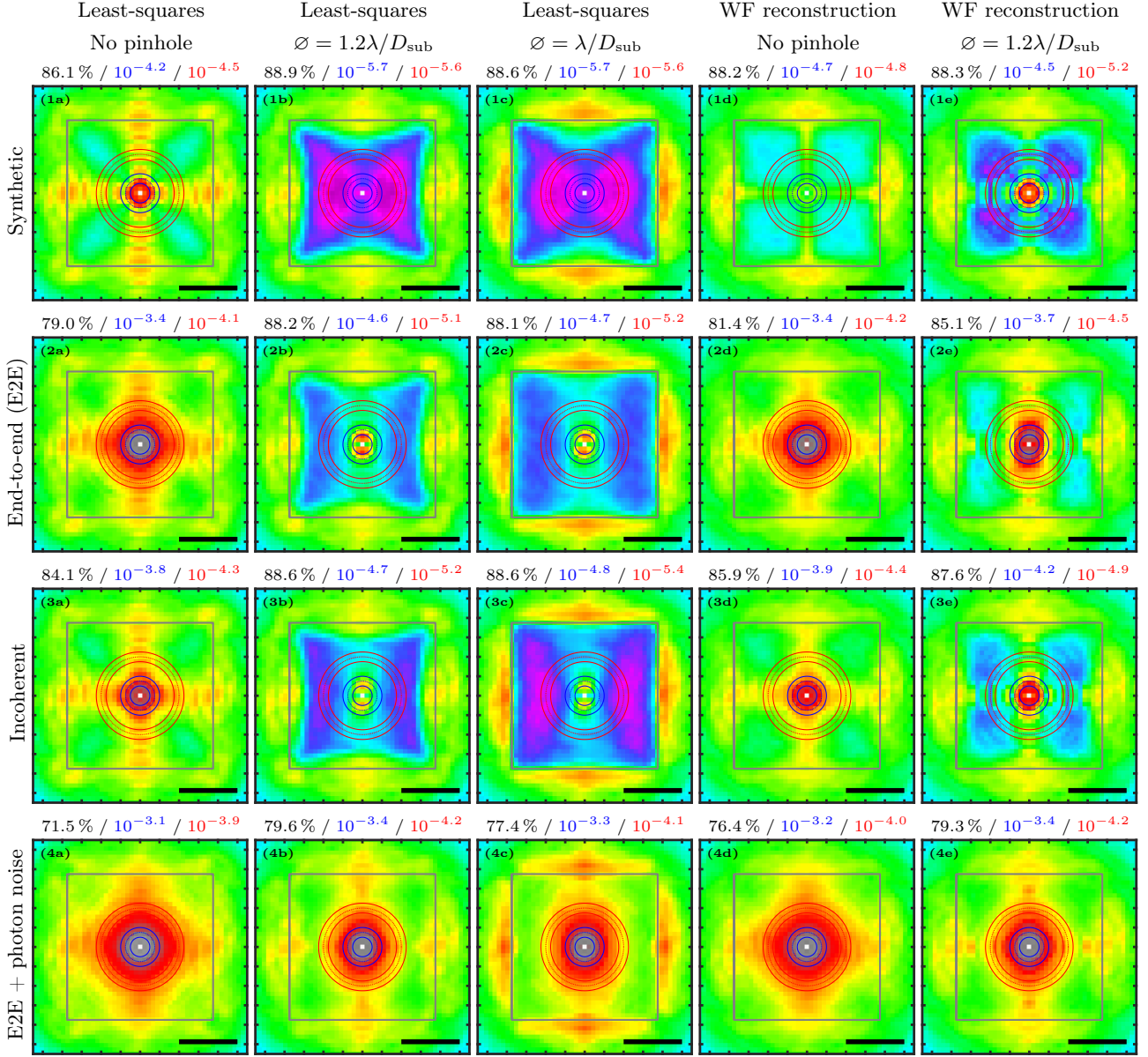
 % fig:aliasing
        \centering
        
        % Internal command of the figure for the automatic sizing
        \newcommand{\PathFig}{figures_aliasing_}
        \newcommand{\LineRatio}{1}
        \newcommand{\FirstCol}{11pt}
        \newcommand{\subfigColor}{black}
        
        % Flags for the propagation
        \newcommand{\PropOne}{PSF_Synth}
        \newcommand{\PropTwo}{PSF_FO}
        \newcommand{\PropThree}{PSF_Inc}
        \newcommand{\PropFour}{PSF_FO_noise}
        
        % Flags for the loop
        \newcommand{\LoopOne}{_pinv}
        \newcommand{\LoopTwo}{_FS1p2_pinv}
        \newcommand{\LoopThree}{_FS_pinv}
        \newcommand{\LoopFour}{_WF_recons}
        \newcommand{\LoopFive}{_FS1p2_WF_recons}
        
		% Flags for the table        
        \newcommand{\FigOneOne}{\PathFig \PropOne \LoopOne}
        \newcommand{\FigOneTwo}{\PathFig \PropOne \LoopTwo}
        \newcommand{\FigOneThree}{\PathFig \PropOne \LoopThree}
        \newcommand{\FigOneFour}{\PathFig \PropOne \LoopFour}
        \newcommand{\FigOneFive}{\PathFig \PropOne \LoopFive}   
          
        \newcommand{\FigTwoOne}{\PathFig \PropTwo \LoopOne}
        \newcommand{\FigTwoTwo}{\PathFig \PropTwo \LoopTwo}
        \newcommand{\FigTwoThree}{\PathFig \PropTwo \LoopThree}
        \newcommand{\FigTwoFour}{\PathFig \PropTwo \LoopFour}
        \newcommand{\FigTwoFive}{\PathFig \PropTwo \LoopFive}
        
        \newcommand{\FigThreeOne}{\PathFig \PropThree \LoopOne}
        \newcommand{\FigThreeTwo}{\PathFig \PropThree \LoopTwo}
        \newcommand{\FigThreeThree}{\PathFig \PropThree \LoopThree}
        \newcommand{\FigThreeFour}{\PathFig \PropThree \LoopFour}
        \newcommand{\FigThreeFive}{\PathFig \PropThree \LoopFive}
        
        \newcommand{\FigFourOne}{\PathFig \PropFour \LoopOne}
        \newcommand{\FigFourTwo}{\PathFig \PropFour \LoopTwo}
        \newcommand{\FigFourThree}{\PathFig \PropFour \LoopThree}
        \newcommand{\FigFourFour}{\PathFig \PropFour \LoopFour}
        \newcommand{\FigFourFive}{\PathFig \PropFour \LoopFive}

        % Getting the size of the boxes
        \sbox1{\includegraphics{\FigOneOne}}               % 1st column
        \sbox2{\includegraphics{\FigOneTwo}}               % 2nd column
        \sbox3{\includegraphics{\FigOneThree}}     % 3rd column
        \sbox4{\includegraphics{\FigOneFour}}     % 4th column
        \sbox5{\includegraphics{\FigOneFive}}     % 4th column
        % Defining column width command
        \newcommand{\ColumnWidth}[1]
                {\the \dimexpr (\linewidth-\FirstCol) * \LineRatio / 5 \relax}

        % Figure table
        \begin{tabular}{
                @{}
                M{\FirstCol}
                @{}
                M{\ColumnWidth{1}}
                @{}
                M{\ColumnWidth{2}}
                @{}
                M{\ColumnWidth{3}}
                @{}
                M{\ColumnWidth{4}}
                @{}
                M{\ColumnWidth{5}}
                @{}
                }
                &
                \small{
                	\begin{tabular}{@{}c@{}}
	                	Least-squares
    	            	\\
        	        	No pinhole
                	\end{tabular}
                }
                &
                \small{
                	\begin{tabular}{@{}c@{}}
	                	Least-squares
    	            	\\
        	        	$\diam =1.2\lDsub$
                	\end{tabular}
                }
                &
                \small{
                	\begin{tabular}{@{}c@{}}
	                	Least-squares
    	            	\\
        	        	$\diam =\lDsub$
                	\end{tabular}
                }
                &
                \small{
                	\begin{tabular}{@{}c@{}}
	                	WF reconstruction
    	            	\\
        	        	No pinhole
                	\end{tabular}
                }
                &
                \small{
                	\begin{tabular}{@{}c@{}}
	                	WF reconstruction
    	            	\\
        	        	$\diam =1.2\lDsub$
                	\end{tabular}
                }
                \\[-2.5pt]
                &
                \scriptsize{\input{./\FigOneOne.txt}}
                &
                \scriptsize{\input{./\FigOneTwo.txt}}
                &
                \scriptsize{\input{./\FigOneThree.txt}}
                &
                \scriptsize{\input{./\FigOneFour.txt}}
                &
                \scriptsize{\input{./\FigOneFive.txt}}
                \\[-2.5pt]
                \rotatebox[origin=l]{90}{\small Synthetic}
                &
                \subfigimg[width=\linewidth,pos=ul,font=\fontfig{\subfigColor}]{$\;$(1a)}{0.0}{\FigOneOne .pdf} &
                \subfigimg[width=\linewidth,pos=ul,font=\fontfig{\subfigColor}]{$\;$(1b)}{0.0}{\FigOneTwo .pdf} &
                \subfigimg[width=\linewidth,pos=ul,font=\fontfig{\subfigColor}]{$\;$(1c)}{0.0}{\FigOneThree .pdf} &
                \subfigimg[width=\linewidth,pos=ul,font=\fontfig{\subfigColor}]{$\;$(1d)}{0.0}{\FigOneFour .pdf}&
                \subfigimg[width=\linewidth,pos=ul,font=\fontfig{\subfigColor}]{$\;$(1e)}{0.0}{\FigOneFive .pdf}
                \\[-5pt]
                &
                \scriptsize{\input{./\FigTwoOne.txt}}
                &
                \scriptsize{\input{./\FigTwoTwo.txt}}
                &
                \scriptsize{\input{./\FigTwoThree.txt}}
                &
                \scriptsize{\input{./\FigTwoFour.txt}}
                &
                \scriptsize{\input{./\FigTwoFive.txt}}
                \\[-2.5pt]
                \rotatebox[origin=l]{90}{\small End-to-end (E2E)}
                &
                \subfigimg[width=\linewidth,pos=ul,font=\fontfig{\subfigColor}]{$\;$(2a)}{0.0}{\FigTwoOne .pdf} &
                \subfigimg[width=\linewidth,pos=ul,font=\fontfig{\subfigColor}]{$\;$(2b)}{0.0}{\FigTwoTwo .pdf} &
                \subfigimg[width=\linewidth,pos=ul,font=\fontfig{\subfigColor}]{$\;$(2c)}{0.0}{\FigTwoThree .pdf} &
                \subfigimg[width=\linewidth,pos=ul,font=\fontfig{\subfigColor}]{$\;$(2d)}{0.0}{\FigTwoFour .pdf}&
                \subfigimg[width=\linewidth,pos=ul,font=\fontfig{\subfigColor}]{$\;$(2e)}{0.0}{\FigTwoFive .pdf}
                \\[-5pt]
                &
                \scriptsize{\input{./\FigThreeOne.txt}}
                &
                \scriptsize{\input{./\FigThreeTwo.txt}}
                &
                \scriptsize{\input{./\FigThreeThree.txt}}
                &
                \scriptsize{\input{./\FigThreeFour.txt}}
                &
                \scriptsize{\input{./\FigThreeFive.txt}}
                \\[-2.5pt]
                \rotatebox[origin=l]{90}{\small Incoherent}
                &
                \subfigimg[width=\linewidth,pos=ul,font=\fontfig{\subfigColor}]{$\;$(3a)}{0.0}{\FigThreeOne .pdf} &
                \subfigimg[width=\linewidth,pos=ul,font=\fontfig{\subfigColor}]{$\;$(3b)}{0.0}{\FigThreeTwo .pdf} &
                \subfigimg[width=\linewidth,pos=ul,font=\fontfig{\subfigColor}]{$\;$(3c)}{0.0}{\FigThreeThree .pdf} &
                \subfigimg[width=\linewidth,pos=ul,font=\fontfig{\subfigColor}]{$\;$(3d)}{0.0}{\FigThreeFour .pdf}&
                \subfigimg[width=\linewidth,pos=ul,font=\fontfig{\subfigColor}]{$\;$(3e)}{0.0}{\FigThreeFive .pdf}
                \\[-5pt]
                &
                \scriptsize{\input{./\FigFourOne.txt}}
                &
                \scriptsize{\input{./\FigFourTwo.txt}}
                &
                \scriptsize{\input{./\FigFourThree.txt}}
                &
                \scriptsize{\input{./\FigFourFour.txt}}
                &
                \scriptsize{\input{./\FigFourFive.txt}}
                \\[-2.5pt]
                \rotatebox[origin=l]{90}{\small E2E + photon noise}
                &
                \subfigimg[width=\linewidth,pos=ul,font=\fontfig{\subfigColor}]{$\;$(4a)}{0.0}{\FigFourOne .pdf} &
                \subfigimg[width=\linewidth,pos=ul,font=\fontfig{\subfigColor}]{$\;$(4b)}{0.0}{\FigFourTwo .pdf} &
                \subfigimg[width=\linewidth,pos=ul,font=\fontfig{\subfigColor}]{$\;$(4c)}{0.0}{\FigFourThree .pdf} &
                \subfigimg[width=\linewidth,pos=ul,font=\fontfig{\subfigColor}]{$\;$(4d)}{0.0}{\FigFourFour .pdf}&
                \subfigimg[width=\linewidth,pos=ul,font=\fontfig{\subfigColor}]{$\;$(4e)}{0.0}{\FigFourFive .pdf}
        \end{tabular}

        \caption{\label{fig:aliasing} Long-exposure PSF of the turbulence residuals after one phase plate revolution for different propagation situations and command estimators. See legend and color bar of \subfig{fig:SH_fitting}{c}.
        (a)~Standard least-squares method without any filtering pinhole.
        (b)~Standard least-squares method with a filtering pinhole of diameter  $\diam =1.2\lDsub$.
        (c)~Standard least-squares method with a filtering pinhole of diameter  $\diam =\lDsub$.
        (d)~Minimum variance method with super-resolved WF reconstruction without any filtering pinhole.
        (e)~Minimum variance method with super-resolved WF reconstruction with a filtering pinhole of diameter $\diam =1.2\lDsub$.
        (1)~The WF is propagated through the synthetic model of the SH-WFS.
        (2)~Full end-to-end noiseless simulation with coherent Fourier optics propagation.
        (3)~Incoherent propagation where each sub-aperture is propagated independently.
        (4)~Full end-to-end noisy simulation with coherent Fourier optics propagation: 200 photons per sub-aperture.
        }
\end{figure}

In this synthetic propagation framework, the MV method with a super-resolved ($s=\infty$) WF reconstruction is further compared with the LS method without pinhole or with the square pinholes mentioned earlier. To avoid the numerical noise amplification of poorly constrained modes, $\nsvd=5$ modes are removed. For the same reason, a small value of $\sigma_{\Vn}=\SI{0.05}{pixel}$ is given to $\MCov{\Vn}$ for the MV method for $s=1$. The results are gathered in \subfigs{fig:aliasing}{1a-1d}.

The first noticeable thing is the similarity between \subfig{fig:WF_recons}{4a} and \subfig{fig:aliasing}{1a}. In the absence of noise, the LS method and MV method with $s=1$ are equivalent. As the WF reconstruction is performed at the same resolution than the SH-WFS sampling, the aliasing cannot be corrected.

Then, it appears that the use of a filtering pinhole combined with the LS method is extremely effective, as shown in \subfigs{fig:aliasing}{1b,1c}. As expected, the use of a bigger pinhole of $\diam =1.2\lDsub$ induces some aliasing with some frequencies wrapping on the edges of the AO corrected area compared to $\diam = 1\lDsub$. Nonetheless the impact on the contrast in the inner regions of the AO corrected area is negligible with raw contrasts at $\COne = 10^{-5.7}$ and $\CTwo = 10^{-5.6}$ that are close to the limit of $10^{-5.9}$ for both pinholes. Surprisingly, the best Strehl ratio is obtained with $\diam =1.2\lDsub$: \SI{88.9}{\percent}, only \SI{0.6}{\percent} below the limit of the fitting error. We interpret this result as follows: using a pinhole limits the maximal slope that can be detected by the SH-WFS, around one pixel according to its design. Thus, the model saturates and is not linear any more. The bigger the pinhole, the larger is the dynamics on which the model is linear. 

Comparing the MV method, \subfig{fig:aliasing}{1d}, with the usual LS method \subfig{fig:aliasing}{1a}, it seems that it can recover $\nicefrac{3}{4}$ of the Strehl (\SI{88.6}{\percent}) lost in the aliasing (\SI{86.1}{\percent}) compared to the used of pinhole (\SI{88.9}{\percent}), without the need of adding any optical component in the beam. This is a success for the MV method. But once again, the results on the raw contrast are less exciting, with a bit less than half an order of magnitude between the LS and the MV methods, the latter being still one order of magnitude lower than the LS method with a pinhole.

\subsection{Noiseless End-to-end Simulation}
\label{sec:E2E}

The next step is to perform a full E2E simulation. This is done by replacing  \eq{eq:data} and the \refline{AO_loop:slopes} of Algorithm~\ref{alg:AO_loop} by a Fourier optics propagation of the wavefront through the SH-WFS and fitting the resulting spots position to extract the slopes.

In the Fourier optics framework, the SH-WFS is modelled by an equivalent complex transmittance as explained in \refapp{app:sub_CT_SH}. The optimal sampling of the simulation is set as described in \refapp{app:sub_samp}. The centroiding of the spot is done via a center of gravity\cite{Fusco:04_SH_noise} on the boxes of $8 \times 8$ pixels of the EvWaCo SH-WFS, keeping the pixels brighter than \SI{0.1}{\percent} of the maximal pixel in each box. This implies that the E2E model $\MSfo$ is not strictly a linear operator and thus not exactly equal to $\MSsy$.

For the LS method, to apply \eq{eq:command_CL}, the interaction matrix $\MGfo$ is not synthetic but generated as in usual experimental AO systems. All the actuators are poked one by one and their influence function is propagated via the end-to-end model. The produced slopes are measured to build the interaction matrix. For the MV method, to apply the WF reconstructor in \eq{eq:command_OL}, $\MGsy$ and $\MSsy$ are used, as defined in the previous \refsec{sec:POC}.

At first, no noise is added to the simulation. The results are gathered in \subfigs{fig:aliasing}{2a-2d}.

It first appears that the LS method coupled with a pinhole has a Strehl ratio almost as good as with the synthetic model ($\strehl > \SI{88}{\percent}$). Nonetheless, the contrasts worsen, especially at close IWA with a drop of one magnitude for $\COne$.

The results of the MV method are disappointing. The raw contrast are now the same than with the LS method without pinhole. The Strehl ratio is slightly better, but not in the $\nicefrac{3}{4}$ extend of the synthetic loop.

This under-performance could be explained by a discrepancy between the synthetic model $\MGsy$ and the Fourier optics propagation $\MGfo$. To test this hypothesis, \subfigs{fig:spot_noise}{1a-1d} show the count maps of the residuals between the closed-loop slopes $\xfo$ and $\yfo$ measured on the E2E data after Fourier optics propagation and the closed-loop slopes $\xsy$ and $\ysy$ predicted by the synthetic model. As we work in a closed loop, most of the slopes are confined within $\pm 0.25$ pixel.

\begin{figure}[ht!] % fig:spot_noise
        \centering
        
        % Internal command of the figure for the automatic sizing
        \newcommand{\PathFig}{figures_spot_noise_}
        \newcommand{\LineRatio}{1}
        \newcommand{\FirstCol}{11pt}
        \newcommand{\subfigColor}{black}
        
        % Flags for the propagation
        \newcommand{\PropOne}{Spot_noise_FO}
        \newcommand{\PropTwo}{Spot_noise_Inc}
        \newcommand{\PropThree}{Spot_noise_FO_noise}
        
        % Flags for the loop
        \newcommand{\LoopOne}{_pinv}
        \newcommand{\LoopTwo}{_FS1p2_pinv}
        \newcommand{\LoopThree}{_FS_pinv}
        \newcommand{\LoopFour}{_WF_recons}
        \newcommand{\LoopFive}{_FS1p2_WF_recons}
        
		% Flags for the table        
        \newcommand{\FigOneOne}{\PathFig \PropOne \LoopOne}
        \newcommand{\FigOneTwo}{\PathFig \PropOne \LoopTwo}
        \newcommand{\FigOneThree}{\PathFig \PropOne \LoopThree}
        \newcommand{\FigOneFour}{\PathFig \PropOne \LoopFour}
        \newcommand{\FigOneFive}{\PathFig \PropOne \LoopFive}   
          
        \newcommand{\FigTwoOne}{\PathFig \PropTwo \LoopOne}
        \newcommand{\FigTwoTwo}{\PathFig \PropTwo \LoopTwo}
        \newcommand{\FigTwoThree}{\PathFig \PropTwo \LoopThree}
        \newcommand{\FigTwoFour}{\PathFig \PropTwo \LoopFour}
        \newcommand{\FigTwoFive}{\PathFig \PropTwo \LoopFive}
        
        \newcommand{\FigThreeOne}{\PathFig \PropThree \LoopOne}
        \newcommand{\FigThreeTwo}{\PathFig \PropThree \LoopTwo}
        \newcommand{\FigThreeThree}{\PathFig \PropThree \LoopThree}
        \newcommand{\FigThreeFour}{\PathFig \PropThree \LoopFour}
        \newcommand{\FigThreeFive}{\PathFig \PropThree \LoopFive}

        % Getting the size of the boxes
        \sbox1{\includegraphics{\FigOneOne}}               % 1st column
        \sbox2{\includegraphics{\FigOneTwo}}               % 2nd column
        \sbox3{\includegraphics{\FigOneThree}}     % 3rd column
        \sbox4{\includegraphics{\FigOneFour}}     % 4th column
        \sbox5{\includegraphics{\FigOneFive}}     % 5th column
        % Defining column width command
        \newcommand{\ColumnWidth}[1]
                {\the \dimexpr (\linewidth-\FirstCol) * \LineRatio / 5 \relax}

        % Figure table
        \begin{tabular}{
                @{}
                M{\FirstCol}
                @{}
                M{\ColumnWidth{1}}
                @{}
                M{\ColumnWidth{2}}
                @{}
                M{\ColumnWidth{3}}
                @{}
                M{\ColumnWidth{4}}
                @{}
                M{\ColumnWidth{5}}
                @{}
                }
                &
                \small{
                	\begin{tabular}{@{}c@{}}
	                	Least-squares
    	            	\\
        	        	No pinhole
                	\end{tabular}
                }
                &
                \small{
                	\begin{tabular}{@{}c@{}}
	                	Least-squares
    	            	\\
        	        	$\diam =1.2\lDsub$
                	\end{tabular}
                }
                &
                \small{
                	\begin{tabular}{@{}c@{}}
	                	Least-squares
    	            	\\
        	        	$\diam =\lDsub$
                	\end{tabular}
                }
                &
                \small{
                	\begin{tabular}{@{}c@{}}
	                	WF reconstruction
    	            	\\
        	        	No pinhole
                	\end{tabular}
                }
                &
                \small{
                	\begin{tabular}{@{}c@{}}
	                	WF reconstruction
    	            	\\
        	        	$\diam =1.2\lDsub$
                	\end{tabular}
                }
                \\[-1pt]
                &
                \scriptsize{\input{./\FigOneOne.txt}}
                &
                \scriptsize{\input{./\FigOneTwo.txt}}
                &
                \scriptsize{\input{./\FigOneThree.txt}}
                &
                \scriptsize{\input{./\FigOneFour.txt}}
                &
                \scriptsize{\input{./\FigOneFive.txt}}
                \\[-1pt]
                \rotatebox[origin=l]{90}{\small E2E}
                &
                \subfigimg[width=\linewidth,pos=ul,font=\fontfig{\subfigColor}]{$\;$(1a)}{0.0}{\FigOneOne .pdf} &
                \subfigimg[width=\linewidth,pos=ul,font=\fontfig{\subfigColor}]{$\;$(1b)}{0.0}{\FigOneTwo .pdf} &
                \subfigimg[width=\linewidth,pos=ul,font=\fontfig{\subfigColor}]{$\;$(1c)}{0.0}{\FigOneThree .pdf} &
                \subfigimg[width=\linewidth,pos=ul,font=\fontfig{\subfigColor}]{$\;$(1d)}{0.0}{\FigOneFour .pdf}&
                \subfigimg[width=\linewidth,pos=ul,font=\fontfig{\subfigColor}]{$\;$(1e)}{0.0}{\FigOneFive .pdf}
                \\[-5pt]
                &
                \scriptsize{\input{./\FigTwoOne.txt}}
                &
                \scriptsize{\input{./\FigTwoTwo.txt}}
                &
                \scriptsize{\input{./\FigTwoThree.txt}}
                &
                \scriptsize{\input{./\FigTwoFour.txt}}
                &
                \scriptsize{\input{./\FigTwoFive.txt}}
                \\[-1pt]
                \rotatebox[origin=l]{90}{\small Incoherent}
                &
                \subfigimg[width=\linewidth,pos=ul,font=\fontfig{\subfigColor}]{$\;$(2a)}{0.0}{\FigTwoOne .pdf} &
                \subfigimg[width=\linewidth,pos=ul,font=\fontfig{\subfigColor}]{$\;$(2b)}{0.0}{\FigTwoTwo .pdf} &
                \subfigimg[width=\linewidth,pos=ul,font=\fontfig{\subfigColor}]{$\;$(2c)}{0.0}{\FigTwoThree .pdf} &
                \subfigimg[width=\linewidth,pos=ul,font=\fontfig{\subfigColor}]{$\;$(2d)}{0.0}{\FigTwoFour .pdf}&
                \subfigimg[width=\linewidth,pos=ul,font=\fontfig{\subfigColor}]{$\;$(2e)}{0.0}{\FigTwoFive .pdf}
                \\[-5pt]
                &
                \scriptsize{\input{./\FigThreeOne.txt}}
                &
                \scriptsize{\input{./\FigThreeTwo.txt}}
                &
                \scriptsize{\input{./\FigThreeThree.txt}}
                &
                \scriptsize{\input{./\FigThreeFour.txt}}
                &
                \scriptsize{\input{./\FigThreeFive.txt}}
                \\[-1pt]
                \rotatebox[origin=l]{90}{\small E2E + noise}
                &
                \subfigimg[width=\linewidth,pos=ul,font=\fontfig{\subfigColor}]{$\;$(3a)}{0.0}{\FigThreeOne .pdf} &
                \subfigimg[width=\linewidth,pos=ul,font=\fontfig{\subfigColor}]{$\;$(3b)}{0.0}{\FigThreeTwo .pdf} &
                \subfigimg[width=\linewidth,pos=ul,font=\fontfig{\subfigColor}]{$\;$(3c)}{0.0}{\FigThreeThree .pdf} &
                \subfigimg[width=\linewidth,pos=ul,font=\fontfig{\subfigColor}]{$\;$(3d)}{0.0}{\FigThreeFour .pdf}&
                \subfigimg[width=\linewidth,pos=ul,font=\fontfig{\subfigColor}]{$\;$(3e)}{0.0}{\FigThreeFive .pdf}
        \end{tabular}

        \caption{\label{fig:spot_noise} Count maps combining the residuals $\xfo-\xsy$ \vs $\xfo$ and $\yfo-\ysy$ \vs $\yfo$.
        (a)~Standard least-squares method without any filtering pinhole.
        (b)~Standard least-squares method with a filtering pinhole of diameter  $\diam =1.2\lDsub$.
        (c)~Standard least-squares method with a filtering pinhole of diameter  $\diam =\lDsub$.
        (d)~Minimum variance method with super-resolved WF reconstruction without any filtering pinhole.
        (e)~Minimum variance method with super-resolved WF reconstruction with a filtering pinhole of diameter  $\diam =1.2\lDsub$.
        (1)~Full end-to-end noiseless simulation with coherent Fourier optics propagation.
        (2)~Incoherent propagation where each sub-aperture is propagated independently.
        (3)~Full end-to-end noisy simulation with coherent Fourier optics propagation: 200 photons per sub-aperture.
        Scale bar: 0.2 pixel.
        }
\end{figure}

The main feature is a global slope, shared by all the graphs. That suggests that the gain of the synthetic interaction matrix does not match the gain of the E2E model. As all the LS method loops are closed using the simulated interaction matrix $\MGfo$, this gain is learnt in the loop calibration step. On the opposite, this gain could impact the MV method that relies on the synthetic matrix to generate the pseudo-open loop.

The second noticeable feature is the noise around this gain bias. Using the spatial filter pinhole strongly reduces the noise on the spot positioning. This noise is quite surprising as the simulations are noiseless in the sense that no noise is added to the SH-WFS raw data simulation. This noise is consequently coming from modelling errors, induced by a non-modelled discrepancy between $\MSsy$ and $\MSfo$ or a bias induced by the centroiding strategy.

Let us mention that this noise in the slope measurements was accounted for in the MV method by adapting $\MCov{\Vn}$ accordingly to $\sigma_{\Vn}=\SI{0.085}{pixel}$, used in this section and in the next \refsecs{sec:E2E_inc}{sec:E2E_noise}.

\subsection{Studying the Impact of the Interferences Between Sub-apertures}
\label{sec:E2E_inc}

\begin{figure}[ht!]
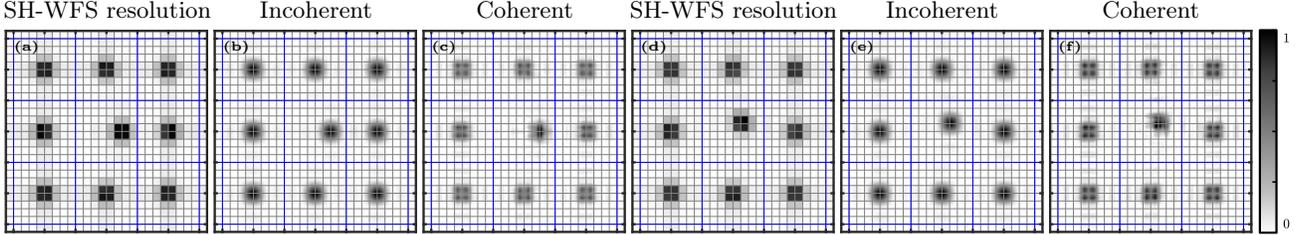
 % fig:IncVSCoh
        \centering
        
        % Internal command of the figure for the automatic sizing
        % Line ratio
        \newcommand{\LineRatio}{1}
        \newcommand{\PathFig}{figures_IncVSCoh_}
        
        % First line
        \newcommand{\FigOne}{\PathFig I_coh_mono_7x7_EvWaCo_617_21}
        \newcommand{\FigTwo}{\PathFig I_inc_mono_7x7_EvWaCo_617_Shannon_21}
        \newcommand{\FigThree}{\PathFig I_coh_mono_7x7_EvWaCo_617_Shannon_21}
        \newcommand{\FigFour}{\PathFig I_coh_mono_7x7_EvWaCo_617_110}
        \newcommand{\FigFive}{\PathFig I_inc_mono_7x7_EvWaCo_617_Shannon_110}
        \newcommand{\FigSix}{\PathFig I_coh_mono_7x7_EvWaCo_617_Shannon_110}
        \newcommand{\FigSeven}{\PathFig bar}
        
        \newcommand{\subfigColor}{black}        
        
        % Getting the size of the boxes
        \sbox1{\includegraphics{\FigOne}}	% 1st column
        \sbox2{\includegraphics{\FigTwo}}	% 2nd column
        \sbox3{\includegraphics{\FigThree}}	% 3rd column
        \sbox4{\includegraphics{\FigFour}}	% 4th column
        \sbox5{\includegraphics{\FigFive}}	% 5th column
        \sbox6{\includegraphics{\FigSix}}	% 6th column
        \sbox7{\includegraphics{\FigSeven}}	% 7th column
        % Defining column width command
        \newcommand{\ColumnWidth}[1]
                {\dimexpr \LineRatio \linewidth * \AspectRatio{#1} / (\AspectRatio{1} + \AspectRatio{2} + \AspectRatio{3} + \AspectRatio{4} + \AspectRatio{5} + \AspectRatio{6} + \AspectRatio{7}) \relax
                }
        \newcommand{\ColumnGap}{\hspace {\dimexpr \linewidth /8 - \LineRatio\linewidth /8 }}

        % Figure table
        \begin{tabular}{
                @{\ColumnGap}
                M{\ColumnWidth{1}}
                @{\ColumnGap}
                M{\ColumnWidth{2}}
                @{\ColumnGap}
                M{\ColumnWidth{3}}
                @{\ColumnGap}
                M{\ColumnWidth{4}}
                @{\ColumnGap}
                M{\ColumnWidth{5}}
                @{\ColumnGap}
                M{\ColumnWidth{6}}
                @{\ColumnGap}
                M{\ColumnWidth{7}}
                @{\ColumnGap}
                }
                \small{SH-WFS resolution} &
                \small{Incoherent} &
                \small{Coherent} &
                \small{SH-WFS resolution} &
                \small{Incoherent} &
                \small{Coherent} &       
                \\[-1pt]
                \subfigimg[width=\linewidth,pos=ul,font=\fontfig{\subfigColor}]{$\;$(a)}{0.0}{\FigOne .pdf} &
                \subfigimg[width=\linewidth,pos=ul,font=\fontfig{\subfigColor}]{$\;$(b)}{0.0}{\FigTwo .pdf} &
                \subfigimg[width=\linewidth,pos=ul,font=\fontfig{\subfigColor}]{$\;$(c)}{0.0}{\FigThree .pdf} &
                \subfigimg[width=\linewidth,pos=ul,font=\fontfig{\subfigColor}]{$\;$(d)}{0.0}{\FigFour .pdf} &
                \subfigimg[width=\linewidth,pos=ul,font=\fontfig{\subfigColor}]{$\;$(e)}{0.0}{\FigFive .pdf} &
                \subfigimg[width=\linewidth,pos=ul,font=\fontfig{\subfigColor}]{$\;$(f)}{0.0}{\FigSix .pdf} &
                \subfigimg[width=\linewidth,pos=ul,font=\fontfig{\subfigColor}]{}{0.0}{\FigSeven}
        \end{tabular}

        \caption{\label{fig:IncVSCoh} Impact of the interferences on the shape and location of the spots for a displacement on the $x$-axis (a,b,c) and on the diagonal (d,e,f). The blue (resp. gray) squares emphasise the sub-aperture (resp. WFS pixel) edges.
        (a,d)~Low resolution simulated data of the SH-WFS.
        (b,e) Generally assumed situation of an incoherent and independent propagation for each sub-aperture. (c,f) Global end-to-end simulation accounting for the interferences between the sub-apertures at the Shannon resolution.}
\end{figure}

Our first assumption is that the noise discussed in the previous \refsec{sec:E2E} is induced by the interferences between the spots. As further detailed in \refapp{app:inter}, this effect is usually not implemented in end-to-end simulations that propagate each sub-aperture independently. Our Fourier optics formalism for the SH-WFS, presented in \refapp{app:sub_CT_SH}, correctly models this effect by coherently propagating the wavefront through the SH-WFS sub-apertures, as shown in \fig{fig:IncVSCoh}. Behind the low resolution of the SH-WFS sensor, as in \subfigs{fig:IncVSCoh}{a,d}, the high resolution spots are distorted by the interferences with their neighbours, as in \subfigs{fig:IncVSCoh}{c,f}. The real spot shape is then quite different from the usually assumed incoherent propagation where each sub-aperture is propagated independently from its neighbours, as in \subfigs{fig:IncVSCoh}{b,e}.

In this section, we implement a incoherent version of our E2E model. Each sub-aperture is individually propagated to simulate its $8 \times 8$ pixels box. All the boxes are then stitched together to mimic real raw SH-WFS data. All cross-talk between the spots in thus suppressed.

The results on the long-exposure PSF are gathered in \subfigs{fig:aliasing}{3a-3d}. Compared to the full E2E simulation of \refsec{sec:E2E}, the impact on the LS method combined with a pinhole is negligible, the Strehl ratio and the raw contrasts being slightly improved. The effect in the absence of pinhole, for both the LS and MV methods, is more pronounced, with a gain of a few percent in Strehl and almost half a magnitude at close IWA. But once again, despite its slightly higher Strehl ratio, the MV method is similar to the LS method without pinhole in terms of contrast and is still clearly outperformed by the use of a filtering pinhole.

The count maps of the slope residuals of \subfigs{fig:spot_noise}{2a-2d} show that suppressing the interferences between the sub-aperture removes the unexpected gain in the slope measurements observed in \refsec{sec:E2E}. This is consistent with the more in depth study performed in \refapp{app:inter} where we show that the interferences can produce a bias in the slope extraction that behaves as a gain factor. One could expect that a poly-chromatic illumination would cancel out this effect. We nonetheless prove that broadband illuminations also present this artefact. For on-sky application, this can have an impact if the calibrations are performed with an internal source that has an average wavelength noticeably different from the average working wavelength during the observations.

As discussed in \refapp{app:inter}, this gain is a problem due to local interferences for slopes within a few tenths of a pixel around their equilibrium position in a closed loop. It is not a global gain on the open-loop slopes. Thus it is better to use the synthetic matrix $\MGsy$ than the E2E matrix $\MGfo$ to close the loop with the MV method, as done in \refsec{sec:E2E}. Indeed, the bias in the gain of the E2E matrix would produce corrupted open-loop slopes at large scale and make the loop diverge. Some simulations were done (not shown here) which corroborated this assumption.

\begin{figure}[ht!]
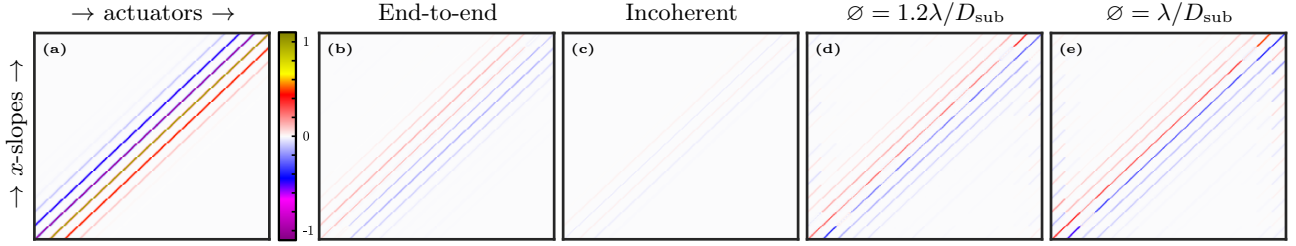
 % fig:IntMat
        \centering
        
        % Internal command of the figure for the automatic sizing
        % Line ratio
        \newcommand{\FirstCol}{11pt}
        \newcommand{\LineRatio}{1}
        \newcommand{\PathFig}{figures_IntMat_}
        
        \newcommand{\FigOne}{\PathFig IntMat_syn}
        \newcommand{\FigTwo}{\PathFig bar}
        \newcommand{\FigThree}{\PathFig res_IntMat_FO_pinv}
        \newcommand{\FigFour}{\PathFig res_IntMat_Inc_pinv}
        \newcommand{\FigFive}{\PathFig res_IntMat_FO_FS1p2_pinv}
        \newcommand{\FigSix}{\PathFig res_IntMat_FO_FS_pinv}
        
        \newcommand{\subfigColor}{black}        
        
        % Getting the size of the boxes
        \sbox1{\includegraphics{\FigOne}}               % 1st column
        \sbox2{\includegraphics{\FigTwo}}               % 2nd column
        \sbox3{\includegraphics{\FigThree}}     		% 3rd column
        \sbox4{\includegraphics{\FigFour}}     			% 4th column
        \sbox5{\includegraphics{\FigFive}}     			% 5th column
        \sbox6{\includegraphics{\FigSix}}     			% 6th column
        
        % Defining column width command
        \newcommand{\ColumnWidth}[1]
                {\the \dimexpr (\linewidth - \FirstCol) * \LineRatio *  \AspectRatio{#1} / (\AspectRatio{1} + \AspectRatio{2} + \AspectRatio{3} + \AspectRatio{4}+ \AspectRatio{5} + \AspectRatio{6}) \relax
                }
        \newcommand{\ColumnGap}{\hspace {\the \dimexpr (\linewidth-\FirstCol) * (1-\LineRatio) / 7 \relax}}

        % Figure table
        \begin{tabular}{
                @{}
                M{\FirstCol}
                @{\ColumnGap}
                M{\ColumnWidth{1}}
                @{\ColumnGap}
                M{\ColumnWidth{2}}
                @{\ColumnGap}
                M{\ColumnWidth{3}}
                @{\ColumnGap}
                M{\ColumnWidth{4}}
                @{\ColumnGap}
                M{\ColumnWidth{5}}
                @{\ColumnGap}
                M{\ColumnWidth{6}}
                @{\ColumnGap}
                }
                &
                {\small $\rightarrow$ actuators $\rightarrow$} &
                &
                {\small End-to-end} &
                {\small Incoherent} &
                {\small $\diam =1.2\lDsub$} &
                {\small $\diam =\lDsub$}
                \\[-1pt]
                \rotatebox[origin=l]{90}{\small $\rightarrow$ $x$-slopes $\rightarrow$}
                &
                \subfigimg[width=\linewidth,pos=ul,font=\fontfig{\subfigColor}]{$\;$(a)}{0.0}{\FigOne .pdf} &
                \subfigimg[width=\linewidth,pos=ul,font=\fontfig{\subfigColor}]{}{0.0}{\FigTwo .pdf} &
                \subfigimg[width=\linewidth,pos=ul,font=\fontfig{\subfigColor}]{$\;$(b)}{0.0}{\FigThree .pdf} &
                \subfigimg[width=\linewidth,pos=ul,font=\fontfig{\subfigColor}]{$\;$(c)}{0.0}{\FigFour .pdf} &
                \subfigimg[width=\linewidth,pos=ul,font=\fontfig{\subfigColor}]{$\;$(d)}{0.0}{\FigFive .pdf} &
                \subfigimg[width=\linewidth,pos=ul,font=\fontfig{\subfigColor}]{$\;$(e)}{0.0}{\FigSix .pdf}
        \end{tabular}

        \caption{\label{fig:IntMat} Impact of the interferences between the sub-apertures on the interaction matrix.
        (a)~Normalised interaction matrix $\MGsy$ of the synthetic model of the SH-WFS for the $x$-slopes.
        (b-e) Residuals $\MGfo - \MGsy$ of the fitted interaction matrix in the E2E model and the synthetic interaction matrix for different scenarios: the full end-to-end noiseless simulation with coherent Fourier optics propagation without pinhole (b) and with a pinhole of diameter $\diam =1.2\lDsub$ (d) and $\diam =\lDsub$ (e) or for an incoherent propagation where each sub-aperture is propagated independently (c).}
\end{figure}

The impact of the interferences in the E2E interaction matrix fit is also seen in \fig{fig:IntMat} that compares the residuals of the different $\MGfo$ with the synthetic matric $\MGsy$. With the interferences between the sub-apertures, in \subfig{fig:IntMat}{b}, in $\MGfo$ is noticeable different from $\MGsy$, in \subfig{fig:IntMat}{a}. For incoherent propagation, in \subfig{fig:IntMat}{c}, the matrices are almost identical. For information, the situations where a pinhole is inserted in the beam are given in \subfigs{fig:IntMat}{d,e}. The resulting interaction matrices are less sparse because the filtering impact the pupil edges, spreading the influence function of the actuators on more sub-apertures.

If the incoherent propagation successfully removes the unexpected gain previously observed in \subfigs{fig:spot_noise}{1a-1d}, it does not solve the problem of the noise on the positions. This noise consequently has to come from the spot fitting strategy. We tested different threshold values for the center of gravity algorithm and we tried to increase the resolution of the SH-WFS sensor or to increase the field of view of each sub-aperture (not shown here). None of this test had an impact on this noise. Our best assumption is that this residual noise is induced by the spot deformation due to the turbulence inside each sub-aperture pupil, as mentioned by Fusco~\etal\cite{Fusco:04_SH_noise}. As using a pinhole filters out the high frequencies and smooths the wavefront, this would explain why the noise is smaller in \subfigs{fig:spot_noise}{2b-2c} than in \subfigs{fig:spot_noise}{2a-2d}.

\subsection{Noisy Simulations}
\label{sec:E2E_noise}

So far we ran noiseless simulations and we saw in \refsecs{sec:E2E}{sec:E2E_inc} that without a filtering pinhole, the LS and MV methods are dominated by model noise on the spot fitting. In this section, we simulate a more realistic case with our coherent E2E model.

In the optical path, they are four mirrors in the TNT prior the EvWaCo instrument exposed to the ambient air\cite{Buisset:18_EvWaCo_spec}. Inside the instrument, they are four additional reflective surfaces (tip-tilt mirror, DM and off-axis parabolas), the beam splitter, the collimating achromatic doublet and the lenslet array. With usual transmittance ranging from \SI{80}{\percent} for the external optics to \SI{97}{\percent} for internal optics and accounting for the camera efficiency, the global throughput is roughly \SI{30}{\percent}. The surface of the sub-apertures\cite{Berdeu:22_AO_bench} on the primary mirror are $7.8\times\SI{7.8}{\square\cm}$. As mentioned in \refsec{sec:intro}, the spectral range of the EvWaCo WFS spans over \SI{2000}{\angstrom}. In the V-band, a 0-magnitude star has a flux\cite{Lena:12_ObsAstro} of \SI{1005}{photons\per\second\per\square\cm\per\angstrom}. Everything combined gives approximately \SI{200}{photons \per \text{sub-aperture}} for an exposure time of \SI{1}{\milli\second} and a star of magnitude of $5.5$.

When the electron-multiplication mode is enabled, the readout noise of the Nüvü~$128^\Tag{AO}$ EMCCD detector can be neglected. In the following we just add the photon noise in the simulations for a flux of \SI{200}{photons} in each sub-aperture. To account for this extra noise, the number of removed modes for the LS method is increased to $\nsvd = 10$ and the threshold value in the spot centroiding is raised to \SI{5}{\percent} to the value of the maximal pixel in each box. For the MV method, $\MCov{\Vn}$ is adapted accordingly.

The results of the E2E simulations, in \subfigs{fig:aliasing}{4a-4d}, show that once again, the optically filtered LS method gives the best results ($\strehl = \SI{79.6}{\percent}$ for $\diam =1.2\lDsub$). Nonetheless, the gap with the MV method is reduced. With a Strehl ratio of \SI{76.4}{\percent}, it manages to retrieve more than half what is lost by the unfiltered LS method ($\strehl = \SI{71.5}{\percent}$). Interestingly, the difference between $\diam =1.2\lDsub$ and $\diam =1\lDsub$ becomes noticeable with more than \SI{2}{\percent} difference in the Strehl ratio. For $\diam =1\lDsub$, the performances of the filtered LS and the MV methods even become comparable. On the tested annuli, all the situation are comparable, with a slight advantage when inserting a pinhole. For regions closer to the edges of the AO corrected area, the optically filtered LS method gives the best results (green \vs orange and yellow).

The count maps of slopes residuals, in \subfigs{fig:spot_noise}{3a-3d}, show that now the simulations are dominated by the photon noise. The noise is slightly more compact in \subfigs{fig:spot_noise}{3b,3c} for filtered cases. The unfiltered LS and MV methods have similar level of noise, in \subfigs{fig:spot_noise}{3a,3d} but different Strehl ratios in \subfigs{fig:aliasing}{4a,4d}. This shows the superiority of the MV command estimator in a noisy situation to increase the AO loop performances.
% This discrepancy comes from the command estimator. Using the correct one is essential to increase the performances.\CommentMichel{I didn't catch the meaning of these two last sentences (sorry!)}

\subsection{Combining Spatial Filtering and Wavefront Reconstruction: a Bad Good Idea?}

We saw in the previous sections that using a pinhole is an effective method to reduce the artefacts impacting the spot positioning. One could wonder if this optical filter could be combined with the MV method to gather their advantages. The results of such simulations are given in the last column (e) of \figs{fig:aliasing}{fig:spot_noise}.

If all the results appear to be better than the unfiltered MV method, they are all worse than the usual filtered LS method. This result is expected: the MV method is a regularised method that assumes a Kolmogorov spectrum as an input. If a spatial filter is used, the wavefront entering the SH-WFS does not follow the expected statistics any more. As a consequence, the WF reconstructor in \eq{eq:command_OL} interprets some low frequencies as high frequencies and alias the reconstructed wavefront. This degrades the performances for small IWA as seen in \subfigs{fig:aliasing}{1e-4e}, compared to \subfigs{fig:aliasing}{1b-4b} and \subfigs{fig:aliasing}{1c-4c}.

A solution to this problem would be to update the synthetic model $\MSsy$ to make it more similar to the `truth' by modelling the smoothing effect of the spatial filter. If not, $\MSsy$ wrongly says that the SH-WFS sees aliased signal. This could be done by including a convolution in $\MS$ that would act as a low-pass filter on the wavefront seen by the WFS. This is nonetheless beyond the scope of this paper.

% A solution to this problem would be to update the covariance matrix $\MCov{\Vw}$ to account for the new statistics of the filtered wavefront. This is nonetheless beyond the scope of this paper. \CommentMichel{Using a spatial filter does not change the priors $\MCov{\Vw}$. Anyway, we are looking for the full $\Vw$ seen by the DM. The correct path is to model the smoothing effect of the spatial filter, for instance by including a convolution in $\MS$, that will act as a low-pass filter on the wavefront seen by the WFS (the kernel of this convolution is small, same size as the subaperture). If not, $\MS$ wrongly says that the WFS sees aliased signal. We need a better modeling of the truth.}

\section{CONCLUSION}

In this work, we introduced a novel analytical model\cite{Berdeu:22_IF} to compute the PSD of the fitting error. This model accounts for the profile of its influence functions of the DM. Contrary to the usual binary mask applied on the PSD, this model can predict the features of long-exposure PSF in terms of achievable contrast, especially close to the edge of the AO corrected area. This model was applied to a real case: the DM192 from ALPAO, based on influence function measurements performed with our AO characterisation bench at NARIT. The model predictions accurately match Monte Carlo simulations.

We also presented a new method to numerically reduce the aliasing error induced by the SH-WFS without the need to change the optical design, for example by adding a filtering pinhole. Based on a minimum variance estimator, this method adds \textit{a priori} knowledge of the statistics of the turbulence to reconstruct the incident wavefront at a resolution higher than the sampling of the SH-WFS. This approach was compared with the usual least-squares method. The results on a loop closed in an inverse crime problem with a synthetic model are promising.

A full end-to-end model was then implemented to test this method on a more realistic framework. To do so, we presented the modelling of a SH-WFS in Fourier optics formalism. We saw that in a noiseless context, the interferences between neighbouring sub-apertures play a major role in the discrepancies between the synthetic model and the end-to-end model. The performances of the MV algorithm are limited by the model noise induced on the spot fitting, likely coming from their deformation due to the turbulence at the scale of the sub-aperture. We also showed that increasing the Strehl ratio is not necessary linked with an improvement of the raw contrast which are then two different quality metrics to assess the effectiveness of an AO loop.

When adding realistic photon noise, the proposed MV method outperforms the usual unfiltered LS method. This is a promising result as using a pinhole is well adapted for XAO with unresolved sources, but cannot be used for more conventional AO systems with resolved sources such as a laser guide star\cite{Fusco:04_SH_noise} or in solar astronomy\cite{Tallon:19_THEMIS}. An additional gain is awaited for noisy simulation by implementing a method extracting the associated covariance of the measurement noise to properly inverse the problem, for example by propagating the sensor noise model or with matching filter techniques\cite{Thiebaut:18_SPIE_THEMIS}.

Finally, the MV method is a natural framework for super-resolution wavefront reconstruction\cite{Correia:21_super_resolution}. The fact that it can retrieve the wavefront beyond the pupil field could be use for predictive algorithm.

All the presented results were obtained in simulations via an end-to-end model emulating the bench already aligned in the NARIT cleanroom\cite{Berdeu:22_AO_bench}. These results will be experimentally tested in this bench once the hardware control is fully operational.

\appendix    %>>>> this command starts appendixes

\section{SH-WFS MODELLING IN FOURIER OPTICS FRAMEWORK}

In this appendix, we present the equivalent complex transmittance plane of a SH-WFS in Fourier optics framework and discuss about its numerical implementation in terms of sampling, especially for a poly-chromatic illumination. This section is mainly a short overview of the main steps that we followed to assess the best simulation strategy. A more detailed study can be provided by the first author upon request, the complete analysis being beyond the scope of this appendix.

\subsection{Complex Transmittance of a SH-WFS Pupil}
\label{app:sub_CT_SH}

\renewcommand{\FT}[2]{\FTnot\left[#1\right]\left(#2\right)}

\begin{figure}[ht!]
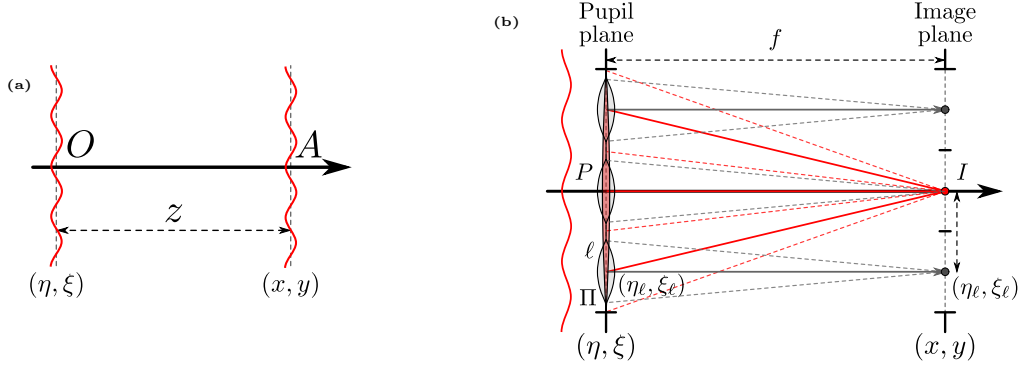
 % fig:FO_SH
        \centering
        
        % Internal command of the figure for the automatic sizing
        % Line ratio
        \newcommand{\LineRatio}{0.75}
        \newcommand{\PathFig}{figures_FO_SH_}
        
        % First line
        \newcommand{\FigOne}{\PathFig propagation_z}
        \newcommand{\FigTwo}{\PathFig WFS_scheme}
        
        \newcommand{\subfigColor}{black}        
        
        % Getting the size of the boxes
        \sbox1{\includegraphics{\FigOne}}               % 1st column
        \sbox2{\includegraphics{\FigTwo}}               % 2nd column
        % Defining column width command
        \newcommand{\ColumnWidth}[1]
                {\dimexpr \LineRatio \linewidth * \AspectRatio{#1} / (\AspectRatio{1} + \AspectRatio{2}) \relax
                }
        \newcommand{\ColumnGap}{\hspace {\dimexpr \linewidth /3 - \LineRatio\linewidth /3 }}

        % Figure table
        \begin{tabular}{
                @{\ColumnGap}
                M{\ColumnWidth{1}}
                @{\ColumnGap}
                M{\ColumnWidth{2}}
                @{\ColumnGap}
                }
                \subfigimg[width=0.65\linewidth,pos=ul,font=\fontfig{\subfigColor}]{\hspace{-11pt}(a)}{0.0}{\FigOne .pdf} &
                \subfigimg[width=\linewidth,pos=ul,font=\fontfig{\subfigColor}]{\hspace{-22pt}(b)}{0.0}{\FigTwo .pdf}
        \end{tabular}

        \caption{\label{fig:FO_SH} (a)~Scheme of the propagation of a wavefront between plane located at~$O$ and~$A$ and separated by a distance~$z$. (b)~Optical scheme of the SH-WFS. Each lenslet~$\ell$ of the array of axis~$\left(\eta_{\ell}, \xi_{\ell}\right)$ focuses its incident light around the position~$\left(\eta_{\ell}, \xi_{\ell}\right)$ on the sensor plane.}
\end{figure}

\textbf{Fourier optics propagation} --  The propagation between two planes separated by a distance~$z$ is depicted in \subfig{fig:FO_SH}{a}. Using the notation of the figure, and noting $\FT{h}{u,v} = \iint_{-\infty}^{+\infty} h\left(x,y\right)e^{-2i\pi\left(xu+yv\right)}\D{x} \D{y}$ the 2D Fourier transform~$\FT{h}{u,v}$ of~$h\left(x,y\right)$, the wavefront~$U_{A}$, in the plane located at~$A$, can be expressed in terms of the wavefront~$U_{O}$, in the plane located at~$O$ under the Fresnel approximation\cite{Goodman:05}
\begin{equation}
	\label{eq:prop_Fresnel}
	U_{A}\left(x,y\right) = \frac{e^{ikz}}{i\lambda z} e^{i\frac{k}{2z}\left(x^{2}+y^{2}\right)} \FT{U_{O}\left(\eta,\xi\right)e^{i\frac{k}{2z}\left(\eta^{2} + \xi^{2}\right)}}{\frac{x}{\lambda z},\frac{y}{\lambda z}} \,,
\end{equation}
or
\begin{equation}
	\label{eq:prop_Fourier}
	\FT{U_{A}\left(x,y\right)}{f_{x}, f_{y}} = e^{ikz} e^{-i\pi\lambda z \left(f_{x}^{2}+f_{y}^{2}\right)} \FT{U_{O}\left(\eta, \xi\right)}{f_{x}, f_{y}} \,,
\end{equation}
with~$\lambda$ the wavelength and~$k=\frac{2\pi}{\lambda}$ the wavenumber with a refractive index assumed to be equal to 1.

We also remind here that the complex transmittance~$m_{\ell}$ of a infinite thin lens of focal length~$f$ in the Fresnel approximation\cite{Goodman:05} is
\begin{equation} % eq:t_lens
	\label{eq:t_lens}
	m_{\ell}\left(x,y\right) = e^{-i\frac{k}{2f}\left(x^{2} + y^{2}\right)}e^{i\varphi_{0}} \,,
\end{equation}
where~$\varphi_{0}$ is the phase introduced at the center of the lens (due to its thickness). In the following, all the lenses will be considered to be identical. As a consequence this term will be the same for all the lenses, introducing a constant phase term that has no incidence on the simulated intensities.  Thus, it will be neglected in the following.

\textbf{Modelling a microlens array} -- Combining \eq{eq:t_lens} with \eqs{eq:prop_Fresnel}{eq:prop_Fourier} allows using only Fourier transforms to alternate between image and pupil planes in a numerical simulation\cite{Schmidt:10_FO}. In these peculiar planes, complex transmittance masks are applied to emulate the different optical elements. The aim of this section is to find a similar formalism when the lens is replaced by an array of microlenses of focal length~$f$, so called lenslets, as in a SH-WFS placed in a pupil plane conjugated with the entrance pupil of the instrument, as shown in \subfig{fig:FO_SH}{b}.

Simply applying a binary amplitude mask on the aperture, composed of the juxtaposition of all the subapertures of the lenslets, is equivalent to the red lens in the figure: the transmitted light is focused at a unique point on the optical axis. To correctly deflect the light of each lenslet~$\ell$, as shown in gray, an adequate phase ramp must be applied in each subaperture.

Noting,~$\left(\eta_{\ell}, \xi_{\ell}\right)$ the position of the axis of each lenslet~$\ell$ and~$\Pi$ their gate function assumed to be identical for all the lenslets\footnote{If not, the gate function of each lenslet must be indexed by~$\ell$ in the following~$\Pi \rightarrow \Pi_{\ell}$}
\begin{equation}
	\Pi\left(\eta, \xi\right) = 
	\begin{cases}
	1 \text{ if } \left(\eta, \xi\right) \text{ is in the subaperture}
	\\
	0 \text{ otherwise}
	\end{cases}
	\,,
\end{equation}
and according to~\eq{eq:t_lens}, the complex transmittance of each lenslet is
\begin{equation}
	\Pi\left(\eta-\eta_{\ell}, \xi-\xi_{\ell}\right)e^{-i\frac{k}{2f}\left(\left(\eta-\eta_{\ell}\right)^{2} + \left(\xi-\xi_{\ell}\right)^{2}\right)}
	\,.
\end{equation}

Introducing $\kappa\left(x,y\right) = \frac{e^{ikf}}{i\lambda f}e^{i\frac{k}{2f}\left(x^{2}+y^{2}\right)}$ and using~\eq{eq:prop_Fresnel}, the propagation between the pupil plane~$P$ and the image plane~$I$ is thus given by
\begin{equation}
	U_{I}\left(x,y\right)
		= \kappa\left(x,y\right)\FT{U_{P}\left(\eta,\xi\right)\sum_{\ell}
		\underbrace{\Pi\left(\eta-\eta_{\ell}, \xi-\xi_{\ell}\right)e^{i\frac{k}{f}\left(\eta_{\ell}\left(\eta-\eta_{\ell}\right) + \xi_{\ell}\left(\xi-\xi_{\ell}\right)\right)}e^{i\frac{k}{2f}\left(\eta_{\ell}^{2} + \xi_{\ell}^{2}\right)}}
		_{m_{\ell}\left(\eta, \xi\right)}
		}{\frac{x}{\lambda f},\frac{y}{\lambda f}}
	\,.
\end{equation}
And thus, the light propagation through a given lenslet~$\ell$ is given by the Fourier transform of the product of the incident wavefront~$U_{P}$ with the complex mask
\begin{equation} % eq:t_lenslet
	\label{eq:t_lenslet}
	\newcommand{\textbox}[2]{\parbox{#1}{\centering \scriptsize #2}}
	m_{\ell}\left(\eta, \xi\right)
	{}={}\underbrace{e^{i\frac{k}{2f}\left(\eta_{\ell}^{2} + \xi_{\ell}^{2}\right)}}_{\textbox{3.5em}{Spherical \\ phase}}
	\underbrace{\Pi\left(\eta-\eta_{\ell}, \xi-\xi_{\ell}\right)}_{\textbox{5em}{Subaperture of the \\ lenslet~$\ell$}}
	\underbrace{e^{\frac{2i\pi}{\lambda f}\left(\eta_{\ell}\left(\eta-\eta_{\ell}\right) + \xi_{\ell}\left(\xi-\xi_{\ell}\right)\right)}}_{\textbox{6em}{Phase ramp of wave vector \\ $\frac{2\pi}{\lambda f}\left(\eta_{\ell}, \xi_{\ell}\right)$}}
	\,.
\end{equation}
As a conclusion, the framework of Fourier optics consisting in propagating the wavefronts via Fourier transforms still holds, the whole lenslet array being modelled by a single complex plane~$m_{\mu}$
\begin{equation} % eq:t_array
	\label{eq:t_array}
	m_{\mu}\left(\eta, \xi\right) = \sum_{\ell}m_{\ell}\left(\eta, \xi\right)
	\Rightarrow
	U_{I}
	=
	\kappa\left(x,y\right)\FT{U_{P}\left(\eta,\xi\right)m_{\mu}\left(\eta, \xi\right)}{\frac{x}{\lambda f},\frac{y}{\lambda f}}
	\,.
\end{equation}

\eqfull{eq:t_lenslet} carries different physical properties of the lens array. Firstly,~$\Pi$ corresponds to the lenslet area. Secondly, for a given lenslet~$\ell$, the phase ramp corresponds to a mono-chromatic plane wave of wave vector~$\frac{2\pi}{\lambda f}\left(\eta_{\ell}, \xi_{\ell}\right)$ whose origin is the center~$\left(\eta_{\ell}, \xi_{\ell}\right)$ of the lenslet\footnote{For numerical simulations, in terms of angular position $\left(x,y\right) = f\left(\alpha_{x},\alpha_{y}\right)$, this can also be interpreted as a tilt introduced in the wavefront, deflecting the light in the direction of the wave vector~$\left(\eta_{\ell}/f, \xi_{\ell}/f\right) = (\alpha_{x_{\ell}},\alpha_{y_{\ell}})$.}. Finally, the quadratic term~$e^{i\frac{k}{2f}\left(\eta_{\ell}^{2} + \xi_{\ell}^{2}\right)}$ corresponds to spherical phase introduced by the position of the lenslet~$\ell$, not laying on the main optical axis\footnote{Similarly, for a numerical implementation where the positions on the image plane are expressed in terms of angular positions, the quadratic phase term is applied as follows: $e^{i\frac{k}{2f}\left(\eta_{\ell}^{2} + \xi_{\ell}^{2}\right)}
	\leftrightarrow
	e^{\frac{i\pi}{\lambda}\left(\alpha_{x_{\ell}}\eta_{\ell} + \alpha_{y_{\ell}}\xi_{\ell}\right)}$.}. This is equivalent to an offset in the phase ramp that is generally accounted for\cite{Primot:03_SH_grating}, but that is sometimes forgotten in numerical simulations\cite{Henault:19}. Implementing this term is critical to correctly model the interferences between the different sub-apertures as it carries their phase piston.

In the general case and in a numerical implementation, \eq{eq:t_array} will be applied. But as a final side remark, it is possible to further investigate this equation in the case of a mono-chromatic plane wave illumination~$U_{P}\left(\eta,\xi\right) = 1$. It comes
\begin{equation}
	U_{I}\left(x,y\right) = \sum_{\ell}\kappa\left(x-\eta_{\ell},y-\xi_{\ell}\right)\FT{\Pi}{\frac{x-\eta_{\ell}}{\lambda f},\frac{y-\xi_{\ell}}{\lambda f}}
	\,.
\end{equation}
As expected, the final wavefront is the interference of the translated Fourier transforms of lenslet subapertures~$\Pi$, but weighted by the quadratic phase term~$\kappa$. If some theoretical works, where analytical formulas are used, account for this term\cite{Dai:07_FO_SH}, it is often forgotten\cite{Roblin:93_SH_th}.

We cross-checked the numerical formula for the complex transmittance and the analytical solution in a simulation, as shown in \refapp{app:sub_samp}.

\subsection{Numerical Apodisation of the Aperture Masks}
\label{app:sub_apod}

In the following, we assume a square grid of $N \times N$ pixels of size $\pitch{I} = \angunit/s$ in the image plane where $\angunit$ is the angular unit of the simulation (generally in terms of $\lambda_{0}/D$ where $\lambda_{0}$ is a reference wavelength) and $s$ the sampling parameters giving the number of pixels in~$1\angunit$. For an illumination of wavelength~$\lambda$, the Fourier optics propagation scales the pupil plane sampling of resolution~$\pitch{P}$ according to \eq{eq:prop_Fresnel} as
\begin{equation}
	\label{eq:sampling}
	\pitch{I} = \lambda / \Paren{N\pitch{P}}
	\,.
\end{equation}
This leads to the famous equivalence
\begin{equation}
	\label{eq:shannon}
	N\pitch{P} \geq 2 D \Leftrightarrow s \geq 2\frac{D}{\lambda}\angunit
	\,,
\end{equation}
that states that a pupil at least padded twice is required to sample one angular resolution element $\lD$ with at least two pixels.

Nonetheless, even if the aperture is correctly zero-padded, as it is spatially limited, its diffraction pattern has an infinite support. As performing a numerical Fourier transform implies to work on a periodic signal, this infinite diffraction pattern produces self-interference patterns well seen on the edges of the simulated domain. To reduce this numerical artefact, a technique can be to perform the simulation on a domain~$p$ times larger than the wanted field of view of~$n$ angular elements~$\angunit$. To obtain the final results, this extra field of view, assuming to contain most of this artefact, is discarded. With these notations, the total number of pixels of the simulation is $N = n \times s \times p$.

The discretisation on a Cartesian grid raises another issue when it comes to describe well defined sharp edges compared to the pixel size $\pitch{P}$. It is straightforward to set the pixel value to 1 inside the shape and 0 outside. But what is the correct value to set to a pixel overlapping the shape's edge? Noting~$\D{p}$ the pixel pitch and~$d_{\Vk}$ the algebraic distance of the pixel~$\Vk$ to the shape's edge (positive if the pixel is outside the shape and negative if it is inside), a pragmatic solution is to set the value of that pixel to
\begin{equation}
	\label{eq:W_pupil}
	m_{\Vk} =
	\begin{cases}
	1/2-d_{\Vk}/\D{p} \text{ if } \left| d_{\Vk} \right| \leq \D{p}/2
	\\
	1 \text{ if } d_{\Vk} \leq -\D{p}/2
	\\
	0 \text{ if } d_{\Vk} \geq \D{p}/2
	\end{cases}
	\,.
\end{equation}
Thus, if the shape's edge passes exactly at the center of the pixel~$\Vk$ (so~$d_{\Vk}=0$), the value of that pixel is~$m_{\Vk}=1/2$.

\begin{figure}[ht!]
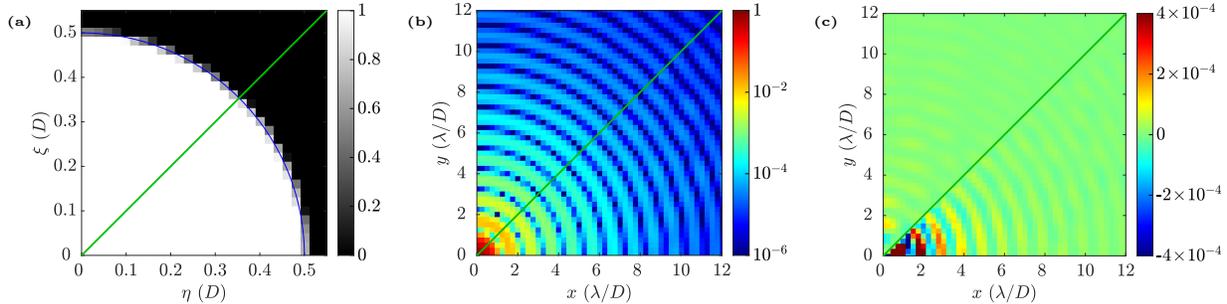
 % fig:FO_apod
        \centering
        
        % Internal command of the figure for the automatic sizing
        % Line ratio
        \newcommand{\LineRatio}{0.85}
        \newcommand{\PathFig}{figures_FO_apod_}
        
        % First line
        \newcommand{\FigOne}{\PathFig Aperture_amplitude_disc_samp_4_pad_2}
        \newcommand{\FigTwo}{\PathFig I_S_comp_disc_samp_4_pad_2}
        \newcommand{\FigThree}{\PathFig Res_comp_disc_samp_4_pad_2}
        
        \newcommand{\subfigColor}{black}        
        
        % Getting the size of the boxes
        \sbox1{\includegraphics{\FigOne}}               % 1st column
        \sbox2{\includegraphics{\FigTwo}}               % 2nd column
        \sbox3{\includegraphics{\FigThree}}           	% 3rd column
        
        % Defining column width command
        \newcommand{\ColumnWidth}[1]
                {\dimexpr \LineRatio \linewidth * \AspectRatio{#1} / (\AspectRatio{1} + \AspectRatio{2} + \AspectRatio{3}) \relax
                }
        \newcommand{\ColumnGap}{\hspace {\dimexpr \linewidth /4 - \LineRatio\linewidth /4 }}

        % Figure table
        \begin{tabular}{
                @{\ColumnGap}
                M{\ColumnWidth{1}}
                @{\ColumnGap}
                M{\ColumnWidth{2}}
                @{\ColumnGap}
                M{\ColumnWidth{3}}
                @{\ColumnGap}
                }
                \subfigimg[width=\linewidth,pos=ul,font=\fontfig{\subfigColor}]{\hspace{-11pt}(a)}{0.0}{\FigOne .pdf} &
                \subfigimg[width=\linewidth,pos=ul,font=\fontfig{\subfigColor}]{\hspace{-11pt}(b)}{0.0}{\FigTwo .pdf} &
                \subfigimg[width=\linewidth,pos=ul,font=\fontfig{\subfigColor}]{\hspace{-11pt}(c)}{0.0}{\FigThree .pdf}
        \end{tabular}

        \caption{\label{fig:FO_apod} Comparison of the numerical simulation with the analytical solution for a circular aperture on $n=25$ angular units $\angunit=\lD$, padding twice the image plane $p=2$ and for a sampling parameter~$s=10$. The numerical solution obtained with the case~$c=1$ (resp.~$c=2$) is above (resp. under) the green diagonal. (a) Zoom on the aperture~$m_{A}^{c}$. (b) Zoom on the numerical diffraction pattern~$I^{c}/\Sap$. (c) Zoom on the residuals~$I^{c} = I^{\Tag{d}}-k_{c}I^{c}$.}
\end{figure}

The next question concerns the energy conservation. The model propagates the complex amplitude of the wavefront but the energy is carried by its intensity, that is to say its squared modulus. So, should~$m_{\Vk}$ be applied on the amplitude of the discretised shape or its squared modulus? To answer this question, \fig{fig:FO_apod} compares the simulation to a theoretical prediction for a circular aperture of size~$D$ where the analytical solution is known for $s=10$ and $p=2$. For an illumination of wavelength~$\lambda$, the theoretical irradiance is $I^{\Tag{d}}_{\Vk} = \sqrt{2/\pi}\Sap \left(2 J_{1}\left( r_{\Vk} \right)/r_{\Vk}\right)^{2}$ with $\Sap = \pi D^{2}/4$ and $ r_{\Vk} = \frac{\pi D}{\lambda} \sin \left(\frac{\lambda}{D}\sqrt{\alpha^{2}_{x,\Vk} + \alpha^{2}_{y,\Vk}}\right)$. The angular positions are given in terms of~$\lD$ and~$J_{1}$ is the Bessel function of the first kind of order one.

The two cases~$c$ are tested. For $c=1$, the amplitude of the aperture mask~$m^{1}_{A}$ at pixel~$\Vk$ is equal to~$m_{\Vk}$. For $c=2$, the amplitude of the aperture mask~$m^{2}_{A}$ at pixel~$\Vk$ is equal to~$m_{\Vk}^{0.5}$ ($m_{\Vk}$ is applied on the squared modulus of the aperture mask). To determine which solution best matches the analytical formula, the residuals are defined as $I^{c} = I^{\Tag{d}}-k_{c}I^{c}$ where~$k_{\Tag{c}}$ is the numerical ratio to achieve the best match between the~$I^{\Tag{d}}$ and~$I^{c}$. Indeed, they can differ by a numerical factor that is not physical but introduced by the numerical modelling of the problem. From a least-squares point of view, $k_{\Tag{c}} = \sum_{\Vk} I^{c}_{\Vk}\times I^{\Tag{d}}_{\Vk} / \sum_{\Vk} I^{c}_{\Vk}\times I^{c}_{\Vk}$.

We get $k_1 = \SI{101.7}{\percent}$ and $k_2 = \SI{99.7}{\percent}$ that implies a better energy conservation for the second case as it could be expected. Nonetheless, it appears that the similarity in the morphology of the diffraction patterns is better with $c=1$ according to \subfig{fig:FO_apod}{c} by more than one order of magnitude. This shows that $m_{\Vk}$ must be applied on the complex amplitude of the simulated shape\footnote[1]{More situations where tested and compared. They can be provided on request.}.

\subsection{Numerical Sampling and Monochromatic \vs Polychromatic Simulations}
\label{app:sub_samp}

Depending on the simulation parameters, the pixel pitch~$\pitch{S}$ of the sensor may be different from the sampling pitch~$\pitch{I}$ of the simulation in an image plane. For example if the illumination is poly-chromatic, the final result is the incoherent summation of mono-chromatic simulations and \eq{eq:sampling} shows that the sampling depends on the simulated wavelength~$\lambda$ whereas the sensor pitch~$\pitch{S}$ is fixed whatever the wavelength. Another example is when the sensor pixel pitch may provide a sampling that is too coarse to correctly insure the zero-padding of the aperture as presented in \eq{eq:shannon}.

\begin{figure}[ht!]
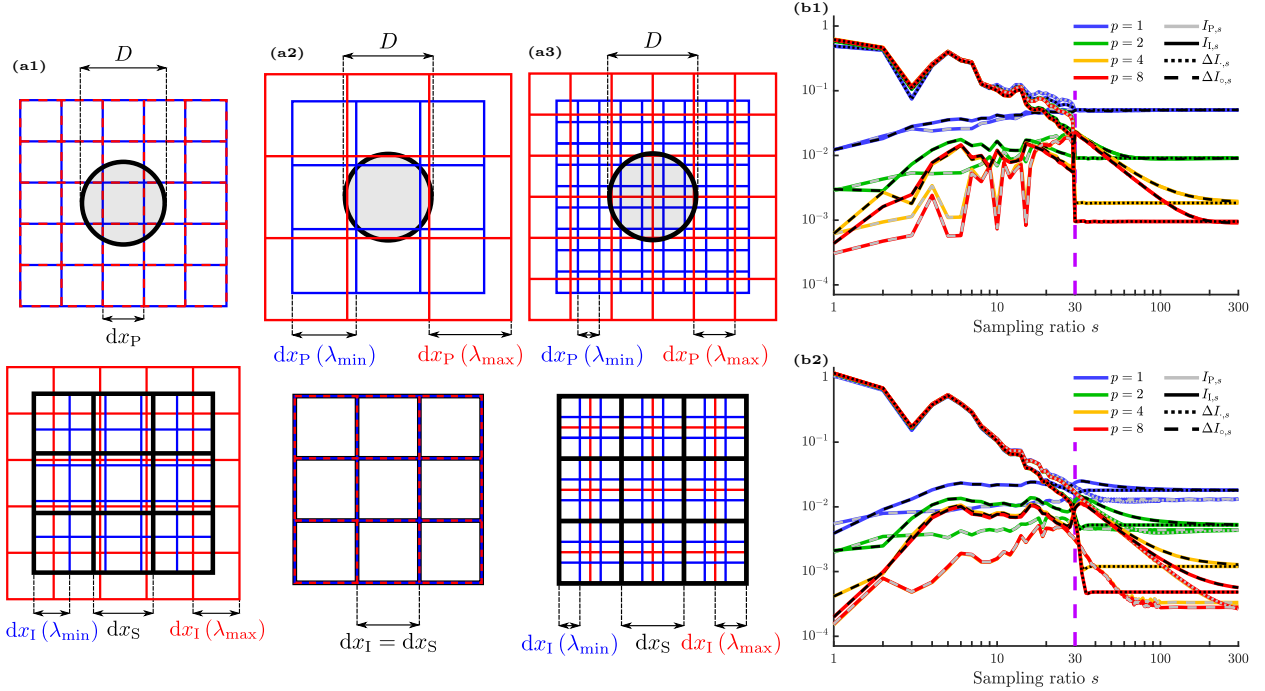
 % fig:FO_samp
        \centering
        
        % Internal command of the figure for the automatic sizing
        % Line ratio
        \newcommand{\LineRatio}{1}
        \newcommand{\PathFig}{figures_FO_samp_}
        
        % First line
        \newcommand{\FigOne}{\PathFig Pupil_sampling}
        \newcommand{\FigTwo}{\PathFig Image_sampling}
        \newcommand{\FigThree}{\PathFig Image_sampling_US}
        \newcommand{\FigFour}{\PathFig MAD_mono_0}
        \newcommand{\FigFive}{\PathFig MAD_poly_0}
        
        \newcommand{\subfigColor}{black}

        % Figure table
        \begin{tabular}{
                @{\hspace{0.016\linewidth}}
                M{0.2\linewidth}
                @{}
                M{0.2\linewidth}
                @{}
                M{0.2\linewidth}
                @{\hspace{0.016\linewidth}}
                C{0.35\linewidth}
                @{\hspace{0.016\linewidth}}
                }
                \subfigimg[width=\linewidth,pos=ul,font=\fontfig{\subfigColor}]{(a1)}{0.0}{\FigOne .pdf} &
                \subfigimg[width=\linewidth,pos=ul,font=\fontfig{\subfigColor}]{(a2)}{0.0}{\FigTwo .pdf} &
                \subfigimg[width=\linewidth,pos=ul,font=\fontfig{\subfigColor}]{(a3)}{0.0}{\FigThree .pdf} &
                \begin{tabular}{@{}c@{}}
                	\subfigimg[width=\linewidth,pos=ul,font=\fontfig{\subfigColor}]{\hspace{-5pt}(b1)}{0.0}{\FigFour .pdf} \\
                	\subfigimg[width=\linewidth,pos=ul,font=\fontfig{\subfigColor}]{\hspace{-5pt}(b2)}{0.0}{\FigFive .pdf}
       			\end{tabular}  
        \end{tabular}
        
%
%        % Figure table
%        \begin{tabular}{
%                @{}
%                M{0.2\linewidth}
%                @{}
%                M{0.2\linewidth}
%                @{}
%                M{0.2\linewidth}
%                @{}
%                }
%                \subfigimg[width=\linewidth,pos=ul,font=\fontfig{\subfigColor}]{(a1)}{0.0}{\FigOne} &
%                \subfigimg[width=\linewidth,pos=ul,font=\fontfig{\subfigColor}]{(a2)}{0.0}{\FigTwo} &
%                \subfigimg[width=\linewidth,pos=ul,font=\fontfig{\subfigColor}]{(a3)}{0.0}{\FigThree}
%        \end{tabular}
%        
%        
%        \begin{tabular}{
%                @{}
%                M{0.33\linewidth}
%                @{\hspace{0.1\linewidth}}
%                M{0.33\linewidth}
%                @{}
%                }
%                \subfigimg[width=\linewidth,pos=ul,font=\fontfig{\subfigColor}]{\hspace{-5pt}(b)}{0.0}{\FigFour} &
%                \subfigimg[width=\linewidth,pos=ul,font=\fontfig{\subfigColor}]{\hspace{-5pt}(b)}{0.0}{\FigFive}
%        \end{tabular}

        \caption{\label{fig:FO_samp} (a)~Sampling strategies in the pupil (top) and image (bottom) planes to simulate the intensity received by each pixel (black squares) of pitch~$\pitch{S}$ of the simulated sensor. The sampling of the simulation scales according to the wavelength (coloured squares) and must be correctly adapted. (a1)~The sampling~$\pitch{P}$ is fixed in a pupil plane. The sampling~$\pitch{I}$ in an image plane scales according to the wavelength. (a2)~The sampling~$\pitch{I}=\pitch{S}$ is fixed in an image plane. The sampling~$\pitch{P}$ in a pupil plane scales according to the wavelength. (a3)~If, as presented on the figure, the sampling~$\pitch{I}$ is fixed in an image plane but the sensor pixel pitch~$\pitch{S}$ is too large to ensure a correct sampling of the simulation, the pixel pitch of the simulation is a subdivision of the sensor pixel pitch. In this situation, this subdivision factor depends on the wavelength.
        (b)~Comparing the NAD (see text) of the two strategies with respect to their convergence solution. The black curves (resp. gray) correspond to a sampling fixed in an image plane (resp. a pupil plane). The dotted curves correspond to the NAD between the simulation~$\Is{s}$ at sampling~$s$ and the sampled analytical solution~$\Iss{s}$. The dashed curves correspond to the NAD between the simulation~$\Is{s}$ at sampling~$s$ and the integrated analytical solution~$\Iis{s}$. The color of the curve indicates its corresponding padding parameter~$p \in \left\lbrace 1,2,4,8 \right\rbrace$. (b1)~Monochromatic illumination. (b2)~Polychromatic illumination.}
\end{figure}

For poly-chromatic simulations, the chromatic scaling can be either carried by~$\pitch{I}\left(\lambda\right)$ with a fixed sampling~$\pitch{P}$ in the pupil plane, as presented in \subfig{fig:FO_samp}{a1} or by~$\pitch{P}\left(\lambda\right)$ with a fixed sampling~$\pitch{I}$ in the image plane, as presented in \subfig{fig:FO_samp}{a2}. This last situation occurs if the pixel pitch~$\pitch{S}$ is small enough to insure a correct sampling at all wavelength: $\forall \lambda,\pitch{I}\left(\lambda\right) = \pitch{S}$. Otherwise, as presented in \subfig{fig:FO_samp}{a3}, $\pitch{I}\left(\lambda\right)$ must be adapted to insure a sufficient sampling.

The final step of the simulation is to account for the sensor pixel integration, that is to say, to obtain the intensity measured by each pixel of the sensor (in black in \subfig{fig:FO_samp}{a}) from the knowledge of the values of the pixels of the simulation (in blue and red in \subfig{fig:FO_samp}{a}). The sensor pixel response is assumed to be a gate function whose size is the size of the pixel.

These two strategies are further detailed in the following. In the case of a poly-chromatic simulation, let's define~$\lmin$ and~$\lmax$, the extremal wavelengths in the illumination spectrum.

\subsubsection{Strategy 1: The sampling is fixed in a pupil plane.}

In this case, the sampling in a pupil plane is identical for all the simulated wavelengths:~$\forall \lambda, \pitch{P}\left(\lambda\right) = \pitch{P}$ and the sampling of an image plane scales as
\begin{equation} % eq:Samp_P_scale_I
	\label{eq:Samp_P_scale_I}
	\frac{\pitch{I}\left(\lambda_{1}\right)}{\lambda_{1}}
	= 
	\frac{\pitch{I}\left(\lambda_{2}\right)}{\lambda_{2}}
	\,.
\end{equation}
For a given sensor of~$\npix{S}$ pixels of pitch~$\pitch{S}$, it is possible to determine the optimal number of pixels~$\npix{sim}$ in the simulation and their size~$\pitch{P}$ in a pupil plane that insure the sampling requirements. As presented in \subfig{fig:FO_samp}{a1}, the number of pixels is given by the total simulated field of view, constrained by~
$\npix{S}\pitch{S}\leq \npix{sim}\pitch{I}\left(\lmin\right)$
. In addition, the sampling of the simulation for a given wavelength must be smaller than the fixed sampling of the simulated sensor constrained by~$\pitch{S}\geq \pitch{I}\left(\lmax\right)$. Finally, the sampling must satisfied a sufficient 0-padding of the aperture
$\npix{sim}\pitch{P}\geq2D$
. Noting~$\ceil*{\cdot}$ the operator rounding up to the closest integer, all these conditions give the optimal number of pixels of the simulation and the associated optimal sampling of a pupil plane
\begin{equation}
	\label{eq:Samp_P_optim}
	\begin{cases}
		\npix{sim}
		=
		\ceil*{\frac{\npix{S}\pitch{S}\angunit}{\lmin}\Tag{max}\left(2D, \frac{\lmax}{\pitch{S}\angunit}\right)}
	\\
		\pitch{P}
		=
		\frac{1}{\npix{sim}}\Tag{max}\left(2D, \frac{\lmax}{\pitch{S}\angunit}\right)
	\end{cases}
	\,.
\end{equation}
These parameters are identical for all the wavelengths of the simulation.

Each mono-chromatic simulation is sampled on a grid that has no reason to correspond to the sensor grid. To obtain the intensity in each pixel of the sensor each mono-chromatic simulation is convolved by the pixel gate function to obtain a blurred picture that accounts for the pixel spatial integration\footnote{As the Fourier transform of the gate function is analytical, this convolution is carried out by product in the Fourier space.}. This blurred intensity is then interpolated on the position of the sensor pixels to obtain the final mono-chromatic intensity on the sensor grid.

\subsubsection{Strategy 2: The sampling is fixed in an image plane.}

In this case, the sampling in an image plane is identical for all the simulated wavelengths:~$\forall \lambda, \pitch{I}\left(\lambda\right) = \pitch{I} = \pitch{S}$ and the sampling of a pupil plane scales with~$\lambda$
\begin{equation}
	\pitch{P}\left(\lambda\right)
	\underset{\text{\eq{eq:sampling}}}{=}
	\frac{\angunit^{-1}\lambda}{\npix{sim}\pitch{I}}
	\,,
\end{equation}
with~$\npix{sim} = \npix{S}$. In this situation, as presented on \subfig{fig:FO_samp}{a2}, the intensity on a sensor pixel is directly the intensity of the corresponding pixel in the simulation, without any further need of integration nor interpolation.

The above paragraph is true at all wavelengths, only if the sensor sampling~$\pitch{S}$ is small enough to ensure a sufficient 0-padding of the aperture plane, \eq{eq:shannon}, a constraint that is strongest for the shortest wavelength. If this condition is not met, in the present strategy, the sensor pixel is subdivided by a sampling ratio $s \in \PosIntegers^{\star}$ into smaller pixels for the simulation as presented on \subfig{fig:FO_samp}{a3} whose optimal value is given by
\begin{equation}
	\label{eq:Samp_I_optim}
	\frac{s}{\pitch{S}}
	\underset{\text{\eq{eq:shannon}}}{\geq}
	\frac{2D}{\lambda}\angunit
	\Rightarrow
	s
	=
	\ceil*{\pitch{S}\frac{2D}{\lambda}\angunit}
	\,,
\end{equation}
changing the number of pixels in the simulation to~$\npix{sim} = s\npix{S}$. In this situation, the integrated intensity in a given pixel of the sensor is directly the summation of all the intensities of its corresponding~$s^{2}$ subpixels of the simulation\footnote{This subdivision of the pixel in the simulation hides a positioning issue as the position of the zero on the sensor plane may not match the zero of the simulation. More details can be provided on request.}.

\subsubsection{Comparing the strategies}

Both of the proposed strategies have their own pros and cons that are not discussed here. In the following, the quality of the convergence of each strategy towards a ground truth gives a criterion on which is the best one that should be used. The simulation is run according to \eq{eq:t_array} and has the following parameters.
\begin{itemize}
	\item The element of resolution is $\lambda / D$.
	\item A wavefront sensor of $3\times3$ square sub-apertures separated by $D_{\Tag{sub}}=D/3$ in the pupil plane of diameter~$95 \% D_{\Tag{sub}}$ and by~$15\;\lambda / D$ ($= 5\;\lambda / D_{\Tag{sub}}$) in the image plane. This square situation provides a known analytical model with the famous sinc function.
	\item The sensor field of view corresponds to $5\times5$ subapertures, that is to say $75 \times 75 \; \lambda / D$.
	\item The field of view of the simulation is extended by a factor of~$p \in \left\lbrace 1,2,4,8 \right\rbrace$ (as defined in \refapp{app:sub_apod}).
	\item The base for the sensor resolution is one pixel per subaperture, that is to say, $\pitch{S}\left(1\right)=15\;\lambda / D$ and $\npix{S}\left(1\right)=5$ on each axis.
	\item The illumination can be mono-chromatic (at $\lambda = \SI{0.5}{\micro\meter}$) or poly-chromatic (with a Gaussian profile centred at $\lambda = \SI{0.5}{\micro\meter}$ with a spectral ratio (full width at half-maximum) of $\Delta \lambda / \lambda = 15 \%$, simulated from $\lmin = \SI{0.4}{\micro\meter}$ to $\lmax = \SI{0.6}{\micro\meter}$ by step of $\delta \lambda = \SI{5}{\nano\meter}$).
\end{itemize}

The two strategies are compared for different sampling ratios~$s$, that is to say, for sensor grids of $\pitch{S}\left(s\right)=\pitch{S}\left(1\right)/s$ and $\npix{S}\left(s\right)=s\npix{S}\left(1\right)$ on each axis. Due to the computational burden of the poly-chromatic simulation, the tested ratios are $s\in \left\lbrace \left\llbracket 1,100 \right\rrbracket, 25\times \left\llbracket 5,12 \right\rrbracket \right\rbrace$. In the following, $\Iips{P}{s}$ (resp. $\Iips{I}{s}$) stands for the simulation with the sampling fixed in a pupil plane (resp. in an image plane) with a sampling ratio~$s$.

In practise, most of the simulations usually done in instrument modelling needs to be performed with a small sampling ratio~$s$ around the optimal sampling of \eq{eq:sampling}. Studying the convergence of each method towards the analytical solution gives a hint on which is the most suitable strategy. Two analytical ground truths are used.
\begin{itemize}
	\item $\Iss{s}$ corresponds to the analytical solution computed at the pixel position. This solution is equivalent to say that a pixel samples the field at its exact position, probing on a point (Dirac function).
	\item $\Iis{s}$ accounts for the pixel integration with the same technique than when the sampling is fixed in an image plane: the pixel is subdivided in sub-pixels on which the analytical field is computed and that are then summed up. The pixels are subdivided so that the high resolution analytical solution is sampled with~$s\geq 1024$ (for instance, for~$s=300$, a pixel is subdivided in~$4\times4$ sub-pixels and the analytical solution is estimated on a~$1200\times1200$ grid). This solution approximately accounts for the fact that the pixel integrates the flux on its surface and provided a more realistic estimate than~$\Iss{s}$.
\end{itemize}

In addition, as mentioned in \refapp{app:sub_apod}, the numerical implementation~$I^{\Tag{n}}$ can present a constant factor~$k_{\Tag{n}}$ with the analytical solution~$I^{\Tag{a}}$. The similarity between a numerical solution and an analytical solution is measured by the normalised absolute deviation (NAD) that integrates (linearly) the discrepancies between the two
\begin{equation}
	\frac{\sum_{\Vk} \left| I^{\Tag{a}}_{\Vk} - k_{\Tag{n}}I^{\Tag{n}}_{\Vk} \right|}{\sum_{\Vk} I^{\Tag{a}}_{\Vk}}
	\text{ with } k_{\Tag{n}}
	=
	\frac{\sum_{\Vk} I^{\Tag{n}}_{\Vk}\times I^{\Tag{a}}_{\Vk}}{\sum_{\Vk} I^{\Tag{n}}_{\Vk}\times I^{\Tag{n}}_{\Vk}}	
	\,.
\end{equation}
The evolution of the NAD for the two strategies, under a mono-chromatic and poly-chromatic illumination and for the different padding parameters is given in \subfig{fig:FO_samp}{b}.

From \subfig{fig:FO_samp}{b1}, it appears that for~$s\geq30$ (vertical purple dashed line in the figure), the two strategies are equivalent for a mono-chromatic illumination. This was expected as for~$s\geq30$, the two strategies run exactly the same simulation. Indeed,
\begin{equation}
	\forall s\geq30,
	\quad
	\Tag{max}\left(2D, \frac{\lambda}{\pitch{S}\left(s\right)\angunit}\right)
	=
	\frac{s}{15}D
	\underset{\text{\eqs{eq:sampling}{eq:Samp_P_optim}}}{\Rightarrow}
	\begin{cases}
		\npix{sim}
		=
		\npix{S}\left(s\right)
	\\
		\pitch{I}
		=
		\pitch{S}\left(s\right)
	\end{cases}
	\text{and}
	\quad
	1 
	\underset{\text{\eq{eq:Samp_I_optim}}}{=}
	\ceil*{\frac{\pitch{S}}{s}\frac{2D}{\lambda}\angunit=\frac{30}{s}}
	\,.
\end{equation}
Under this threshold, whatever the illumination spectrum (mono-chromatic or poly-chromatic) the comparison with the pseudo-integrated analytical solution~$\Iis{s}$ (dashed curves) gives a better NAD than with the sampled analytical solution~$\Iss{s}$ (dotted curves). This is expected as by definition, under this threshold, the sensor sampling is not sufficient to correctly sample the signal and grasp its variation. Accounting for the pixel integration is essential.

Beyond this threshold, under mono-chromatic illumination, the NAD reaches a plateau if compared with the sampled solution~$\Iss{s}$ whereas the convergence is slower when compared with the integrated solution~$\Iis{s}$. This is expected as the sensor sampling is directly the simulation sampling and this sampling is fine enough to grasps the signal as discussed in \refapp{app:sub_apod}. This corresponds to a Dirac sampling and consequently compares well with~$\Iss{s}$. On the opposite it assumes that the value integrated on the pixel is equal to the value at its center, that is not correct, and consequently does not compared well with~$\Iis{s}$.

The plateau behaviour and its dependence according to the padding parameters~$p$ shows that the error that dominates the simulation is the self-interference due to the limited size of the simulated domain. In other words, the sampling is sufficient enough according to the criterion obtained in \refapp{app:sub_apod} (corresponding to~$s\geq30$ in the present situation), further refining the sampling~$s$ does not provides a more accurate simulation. Similar conclusions are obtained from \subfig{fig:FO_samp}{b2} for a poly-chromatic simulation by fixing the sampling in an image plane ($\Iips{I}{s}$, black curves), but with a smoother transition around~$s=30$, the transition threshold being different for each simulated wavelength.

The noticeable feature of the poly-chromatic illumination in \subfig{fig:FO_samp}{b2} is the discrepancy between the two sampling strategies (gray versus black curves), the simulation in this situation being different if an image plane (black curves) or a pupil plane (gray curves) is chosen to fix the sampling. When a pupil is chosen to fix the sampling, $\forall \lambda > \lmin$, the simulation is oversampled for all the wavelengths and the integration on the pixel is accounted for in the down-sampling operation. Thus this strategy provides a better estimate of the flux integrated by a pixel than fixing the sampling in an image plane (that assumes a Dirac probing of the field), as shown by the better convergence of the dashed gray curves. As a consequence, this solution is not adapted when compared with the sampled analytical solution (dotted gray curves).

As previously, increasing~$p$ first leads to a better NAD. But for~$p\in\left\lbrace 4,8 \right\rbrace$ the curves almost perfectly overlaps: the simulation is not dominated by the self-interferences but by the pixel integration with the convolution by a gate function that introduces errors that dominate at high samplings.

For a given~$p$, comparing the gray and black curves shows that as noticed above, the strategy 1 provides a better solution at high samplings than the strategy 2. This is expected as this strategy implies to increase the simulated field with~$\lambda$ where the self-interferences can occur, thus diminishing its effects.

\begin{figure}[p!]
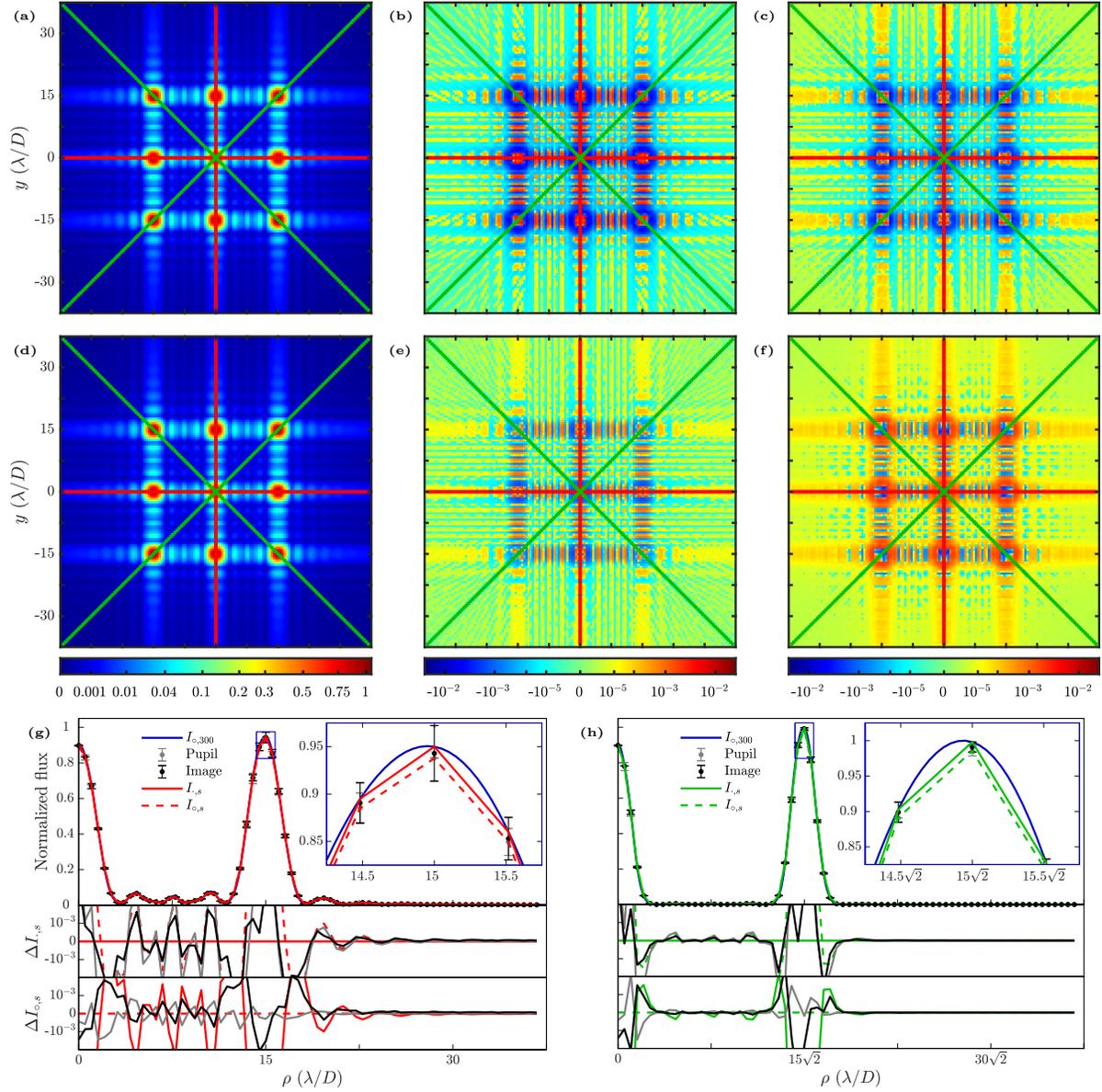
 % fig:PvsI_poly_s_29
	\centering
	% Path of the subfigures
	\newcommand{\PathFig}{figures_FO_samp_Convergence_poly_}
	\newcommand{\SampFig}{29}
	
	% Line ratio
	\newcommand{\LineRatio}{0.85}
	
	% Analytical solution: sampled
	\newcommand{\SampType}{samp}
	\newcommand{\FigOne}{\PathFig sol_\SampType _I_S_tot_\SampFig}
	\newcommand{\FigTwo}{\PathFig res_\SampType _P_I_S_tot_\SampFig}
	\newcommand{\FigThree}{\PathFig res_\SampType _I_I_S_tot_\SampFig}
	
	% Internal command of the figure for the automatic sizing
	% Getting the size of the boxes
	\sbox1{\includegraphics{\FigOne .pdf}} % 1st column
	\sbox2{\includegraphics{\FigTwo .pdf}} % 2nd column
	\sbox3{\includegraphics{\FigThree .pdf}} % 3rd column
	% Defining column width command
	\newcommand{\ColumnWidth}[1]
		{\dimexpr \LineRatio \linewidth * \AspectRatio{#1} / (\AspectRatio{1} + \AspectRatio{2} + \AspectRatio{3}) \relax
		}
	\newcommand{\ColumnGap}{\hspace {\dimexpr \linewidth /4 - \LineRatio\linewidth /4}}

	% Figure table
	\begin{tabular}{
		@{\ColumnGap}
		m{\ColumnWidth{1}}
		@{\ColumnGap}
		m{\ColumnWidth{2}}
		@{\ColumnGap}
		m{\ColumnWidth{3}}
		@{\ColumnGap}
		}
		\subfigimg[width=\linewidth,pos=ul,font=\fontfig{black}]{(a)}{0.0}{\FigOne .pdf}
		&
		\subfigimg[width=\linewidth,pos=ul,font=\fontfig{black}]{\hspace{-15pt}(b)}{0.0}{\FigTwo .pdf}
		&
		\subfigimg[width=\linewidth,pos=ul,font=\fontfig{black}]{\hspace{-15pt}(c)}{0.0}{\FigThree .pdf}
	\end{tabular}
	
	% Analytical solution: integrated
	\renewcommand{\SampType}{int}
	\renewcommand{\FigOne}{\PathFig sol_\SampType _I_S_tot_\SampFig}
	\renewcommand{\FigTwo}{\PathFig res_\SampType _P_I_S_tot_\SampFig}
	\renewcommand{\FigThree}{\PathFig res_\SampType _I_I_S_tot_\SampFig}
	
	% Internal command of the figure for the automatic sizing
	% Getting the size of the boxes
	\sbox1{\includegraphics{\FigOne .pdf}} % 1st column
	\sbox2{\includegraphics{\FigTwo .pdf}} % 2nd column
	\sbox3{\includegraphics{\FigThree .pdf}} % 3rd column

	% Figure table
	\begin{tabular}{
		@{\ColumnGap}
		m{\ColumnWidth{1}}
		@{\ColumnGap}
		m{\ColumnWidth{2}}
		@{\ColumnGap}
		m{\ColumnWidth{3}}
		@{\ColumnGap}
		}
		\subfigimg[width=\linewidth,pos=ul,font=\fontfig{black}]{(d)}{0.0}{\FigOne}
		&
		\subfigimg[width=\linewidth,pos=ul,font=\fontfig{black}]{\hspace{-15pt}(e)}{0.0}{\FigTwo}
		&
		\subfigimg[width=\linewidth,pos=ul,font=\fontfig{black}]{\hspace{-15pt}(f)}{0.0}{\FigThree}
	\end{tabular}

	% Internal command of the figure for the automatic sizing
	% Line ratio
	% Subfigures
	\renewcommand{\FigOne}{\PathFig prof_axis_I_S_tot_\SampFig}
	\renewcommand{\FigTwo}{\PathFig prof_diag_I_S_tot_\SampFig}
	% Getting the size of the boxes
	\sbox1{\includegraphics{\FigOne}} % 1st column
	\sbox2{\includegraphics{\FigTwo}} % 2nd column
	% Defining column width command
	\renewcommand{\ColumnWidth}[1]
		{\dimexpr \LineRatio \linewidth * \AspectRatio{#1} / (\AspectRatio{1} + \AspectRatio{2}) \relax
		}
	\renewcommand{\ColumnGap}{\hspace {\dimexpr \linewidth / 3 - \LineRatio\linewidth / 3}}

	% Figure table
	\begin{tabular}{
		@{\ColumnGap}
		m{\ColumnWidth{1}}
		@{\ColumnGap}
		m{\ColumnWidth{2}}
		@{\ColumnGap}
		}
		\subfigimg[width=\linewidth,pos=ul,font=\fontfig{black}]{(g)}{0.0}{\FigOne .pdf}
		&
		\subfigimg[width=\linewidth,pos=ul,font=\fontfig{black}]{\hspace{-15pt}(h)}{0.0}{\FigTwo .pdf}
	\end{tabular}
	
	\caption{\label{fig:PvsI_poly_s_29}Polychromatic simulation with a sampling ratio~$s=29$ and a padding parameter~$p=4$. (a)~Analytical solution sampled at the pixel positions~$\Iss{s}$. (b)~Residuals between the analytical solution sampled at the pixel positions and the simulation with a sampling fixed in a pupil plane~$\Iss{s} - k_{\Tag{n}}\Iips{P}{s}$. (c)~Residuals between the analytical solution sampled at the pixel positions and the simulation with a sampling fixed in an image plane~$\Iss{s} - k_{\Tag{n}}\Iips{I}{s}$ (d)~Analytical solution integrated on the pixels~$\Iis{s}$. (e)~Residuals between the analytical solution integrated on the pixels and the simulation with a sampling fixed in a pupil plane~$\Iis{s} - k_{\Tag{n}}\Iips{P}{s}$. (f)~Residuals between the analytical solution integrated on the pixels and the simulation with a sampling fixed in an image plane~$\Iis{s} - k_{\Tag{n}}\Iips{I}{s}$. (g-h)~Comparison of the four extracted profiles along the $x$ and $y$-axes (red) (e) or the diagonals (green) (f). The black (resp. gray) curve corresponds to a sampling fixed in an image plane (resp. a pupil plane). The solid curves correspond to the solution at~$s=300$. $\Delta\Iss{s}$ shows the difference with the analytical solution sampled at the pixel positions~$\Iss{s}$ (red and green solid curves). $\Delta\Iis{s}$ shows the difference with the analytical solution integrated on the pixels~$\Iis{s}$ (red and green dashed curves). The error bars indicate the standard deviation of the four symmetric points on the profiles, scaled with a factor~$10^{14}$.}
\end{figure}

\figfull{fig:PvsI_poly_s_29} presents the details of a specific sampling ratio~$s=29$ for a padding parameter of~$p=4$. First, the symmetry of the simulation is tested by comparing the simulated intensity obtained on four symmetric profiles, the four axes (red) and the four diagonals (green). The standard deviations are below~$10^{-15}$ corresponding to numerical rounding errors. Even for even number of pixels, the extra negative position due to the definition of the discrete Fourier frequencies does not impact the symmetry of the simulation.

The final conclusion is that the best option will depend on the needs. If one wants to simulate the field at specific locations (Dirac probing) with a very fine sampling, fixing the sampling in an image plane is the fastest solution with a homogeneous convergence. The precision of the simulation will be limited by the padding of the simulated field, or equivalently the resolution in the pupil plane. On the opposite, if one is looking for a simulation accounting for the pixel integration and around or below the optimal sampling criterion given by \eq{eq:shannon}, fixing the sampling in the pupil plane is more adapted. In this situation, the padding parameter is not the limiting factor, as seen in \subfig{fig:FO_samp}{b2}. A better model for the pixel integration must be quested as the core of the proposed method is based on a optimal sampling of the simulation that assumes the equivalence between a Dirac probing and what will be measured by the sensor. In other words, a better modelling of the integration by the pixel is needed, possibly meaning higher resolution simulation before a down-sampling operator.

\section{EFFECTS OF THE INTERFERENCES BETWEEN THE SUB-APERTURES of A SH-WFS}
\label{app:inter}

In standard simulation tools such as YAO\cite{Rigaut:13_YAO} (Yorick Adaptive Optics), OOMAO\cite{Conan:14_OOMAO} (Object-Oriented Matlab Adaptive Optics), COMPASS\cite{Gratadour:14_COMPASS} (COMputing Platform for Adaptive opticS Systems) or SOAPY\cite{Reeves_16:Soapy} (Simulation AO PYthon), due to computational burden, it is common to propagate each sub-aperture independently. As a consequence, the interferences between the spots are not accounted for. Let us mention HCIPy\cite{Por:18_HCIPy} (High Contrast Imaging for Python) that performs Fourier optics propagation through the full system and thus accounts for the interferences.

Nonetheless, the impact of the interferences between the sub-apertures of a SH-WFS has already been studied. Let us for example cite Roblin\&{}Horville\cite{Roblin:93_SH_th} who worked on an analytical model. But as discussed in \refapp{app:sub_CT_SH}, the quadratic term to correctly account for the interference was forgotten, invaliding the conclusions.  Dai~\etal\cite{Dai:07_FO_SH} performed a deeper analytical and numerical study on the error in the spot positioning. Dai~\etal mainly focused on the disturbance of a single spots centroiding due to its neighbours. In the following, we focus also on the impact of a tilted-spot on its neighbours and also study the poly-chromatic case.

The parameters of the simulated SH-WFS are identical to the EvWaCo WFS: sub-apertures of $8\times 8$ pixels of FOV $6.4\times\arcsec{6.4}$. To limit the numerical burden of the poly-chromatic illumination, an array of only $7\times7$ sub-apertures is simulated. The central spot is shifted by introducing a phase ramp (that is to say a tip-tilt) on its sub-aperture. The central spot is displaced that way over all the FOV of its sub-aperture and the positions of the spots are obtained by fitting the best 2D Gaussian pattern on each spot, a method similar to a weighted center of gravity or a matched filter\cite{Fusco:04_SH_noise}.

\begin{figure}[ht!]
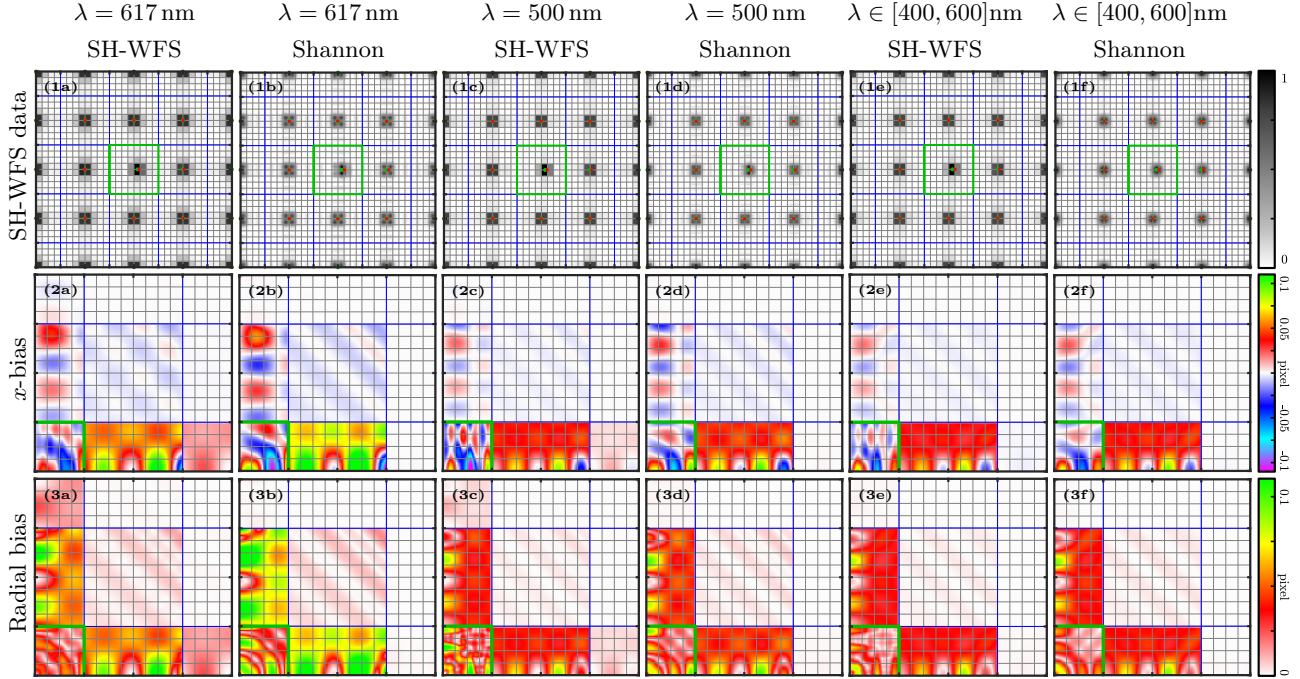
 % fig:app_inter
        \centering
        
        % Internal command of the figure for the automatic sizing
        \newcommand{\PathFig}{figures_Inter_}
        \newcommand{\LineRatio}{1}
        \newcommand{\FirstCol}{11pt}
        \newcommand{\subfigColor}{black}
        
        % Flags for the rows
        \newcommand{\RowOne}{I_coh_}
        \newcommand{\RowTwo}{Shift_coh_x_}
        \newcommand{\RowThree}{Shift_coh_rad_}
        
        % Flags for the columns
        \newcommand{\ColOne}{mono_7x7_EvWaCo_617}
        \newcommand{\ColTwo}{mono_7x7_EvWaCo_617_Shannon}
        \newcommand{\ColThree}{mono_7x7_EvWaCo}
        \newcommand{\ColFour}{mono_7x7_EvWaCo_Shannon}
        \newcommand{\ColFive}{poly_7x7_EvWaCo}
        \newcommand{\ColSix}{poly_7x7_EvWaCo_Shannon}
        
		% Flags for the table        
        \newcommand{\FigOneOne}{\PathFig \RowOne \ColOne _7}
        \newcommand{\FigOneTwo}{\PathFig \RowOne \ColTwo _7}
        \newcommand{\FigOneThree}{\PathFig \RowOne \ColThree _7}
        \newcommand{\FigOneFour}{\PathFig \RowOne \ColFour _7}
        \newcommand{\FigOneFive}{\PathFig \RowOne \ColFive _7}
        \newcommand{\FigOneSix}{\PathFig \RowOne \ColSix _7}
        \newcommand{\FigOneSeven}{\PathFig bar}   
          
        \newcommand{\FigTwoOne}{\PathFig \RowTwo \ColOne}
        \newcommand{\FigTwoTwo}{\PathFig \RowTwo \ColTwo}
        \newcommand{\FigTwoThree}{\PathFig \RowTwo \ColThree}
        \newcommand{\FigTwoFour}{\PathFig \RowTwo \ColFour}
        \newcommand{\FigTwoFive}{\PathFig \RowTwo \ColFive}
        \newcommand{\FigTwoSix}{\PathFig \RowTwo \ColSix}
        \newcommand{\FigTwoSeven}{\PathFig bar_shift_y}
        
        \newcommand{\FigThreeOne}{\PathFig \RowThree \ColOne}
        \newcommand{\FigThreeTwo}{\PathFig \RowThree \ColTwo}
        \newcommand{\FigThreeThree}{\PathFig \RowThree \ColThree}
        \newcommand{\FigThreeFour}{\PathFig \RowThree \ColFour}
        \newcommand{\FigThreeFive}{\PathFig \RowThree \ColFive}
        \newcommand{\FigThreeSix}{\PathFig \RowThree \ColSix}
        \newcommand{\FigThreeSeven}{\PathFig bar_shift_rad}

        % Getting the size of the boxes
        \sbox1{\includegraphics{\FigOneOne}}               % 1st column
        \sbox2{\includegraphics{\FigOneTwo}}               % 2nd column
        \sbox3{\includegraphics{\FigOneThree}}     % 3rd column
        \sbox4{\includegraphics{\FigOneFour}}     % 4th column
        \sbox5{\includegraphics{\FigOneFive}}     % 5th column
        \sbox6{\includegraphics{\FigOneSix}}     % 6th column
        \sbox7{\includegraphics{\FigTwoSeven}}     % 7th column
        \sbox8{\includegraphics{\FigOneSeven}}     % Auxiliary
        % Defining column width command
        \newcommand{\ColumnWidth}[1]
                {\the \dimexpr (\linewidth - \FirstCol) * \LineRatio *  \AspectRatio{#1} / (\AspectRatio{1} + \AspectRatio{2} + \AspectRatio{3} + \AspectRatio{4} + \AspectRatio{5} + \AspectRatio{6} + \AspectRatio{7}) \relax
                }
        \newcommand{\ColumnGap}{\hspace {\the \dimexpr (\linewidth-\FirstCol) * (1-\LineRatio) / 8 \relax}}

        % Figure table
        \begin{tabular}{
                @{}
                M{\FirstCol}
                @{\ColumnGap}
                M{\ColumnWidth{1}}
                @{\ColumnGap}
                M{\ColumnWidth{2}}
                @{\ColumnGap}
                M{\ColumnWidth{3}}
                @{\ColumnGap}
                M{\ColumnWidth{4}}
                @{\ColumnGap}
                L{\ColumnWidth{5}}
                @{\ColumnGap}
                L{\ColumnWidth{6}}
                @{\ColumnGap}
                L{\ColumnWidth{7}}
                @{\ColumnGap}
                }
                &
                \small{
                	\begin{tabular}{@{}c@{}}
	                	$\lambda = \SI{617}{\nano\meter}$
    	            	\\
        	        	SH-WFS
                	\end{tabular}
                	}
                &
                \small{
                	\begin{tabular}{@{}c@{}}
	                	$\lambda = \SI{617}{\nano\meter}$
    	            	\\
        	        	Shannon
                	\end{tabular}
                	}
                &
                \small{
                	\begin{tabular}{@{}c@{}}
	                	$\lambda = \SI{500}{\nano\meter}$
    	            	\\
        	        	SH-WFS
                	\end{tabular}
                	}
                &
                \small{
                	\begin{tabular}{@{}c@{}}
	                	$\lambda = \SI{500}{\nano\meter}$
    	            	\\
        	        	Shannon
                	\end{tabular}
                	}
                &
                \small{
                	\begin{tabular}{@{}c@{}}
	                	$\lambda \in \Brack{400,600}\SI{}{\nano\meter}$
    	            	\\
        	        	SH-WFS
                	\end{tabular}
                	}
                &
                \small{
                	\begin{tabular}{@{}c@{}}
	                	$\lambda \in \Brack{400,600}\SI{}{\nano\meter}$
    	            	\\
        	        	Shannon
                	\end{tabular}
                	}
                &
                \\
                \rotatebox[origin=l]{90}{\small SH-WFS data}
                &
                \subfigimg[width=\linewidth,pos=ul,font=\fontfig{\subfigColor}]{$\;$(1a)}{0.0}{\FigOneOne .pdf} &
                \subfigimg[width=\linewidth,pos=ul,font=\fontfig{\subfigColor}]{$\;$(1b)}{0.0}{\FigOneTwo .pdf} &
                \subfigimg[width=\linewidth,pos=ul,font=\fontfig{\subfigColor}]{$\;$(1c)}{0.0}{\FigOneThree .pdf} &
                \subfigimg[width=\linewidth,pos=ul,font=\fontfig{\subfigColor}]{$\;$(1d)}{0.0}{\FigOneFour .pdf}&
                \subfigimg[width=\linewidth,pos=ul,font=\fontfig{\subfigColor}]{$\;$(1e)}{0.0}{\FigOneFive .pdf}&
                \subfigimg[width=\linewidth,pos=ul,font=\fontfig{\subfigColor}]{$\;$(1f)}{0.0}{\FigOneSix .pdf}&
                \subfigimg[width=\the \dimexpr \linewidth * \AspectRatio{8} / \AspectRatio{7},pos=ul,font=\fontfig{\subfigColor}]{}{0.0}{\FigOneSeven .pdf}
                \\[-5pt]
                \rotatebox[origin=l]{90}{\small $x$-bias}
                &
                \subfigimg[width=\linewidth,pos=ul,font=\fontfig{\subfigColor}]{$\;$(2a)}{0.0}{\FigTwoOne .pdf} &
                \subfigimg[width=\linewidth,pos=ul,font=\fontfig{\subfigColor}]{$\;$(2b)}{0.0}{\FigTwoTwo .pdf} &
                \subfigimg[width=\linewidth,pos=ul,font=\fontfig{\subfigColor}]{$\;$(2c)}{0.0}{\FigTwoThree .pdf} &
                \subfigimg[width=\linewidth,pos=ul,font=\fontfig{\subfigColor}]{$\;$(2d)}{0.0}{\FigTwoFour .pdf}&
                \subfigimg[width=\linewidth,pos=ul,font=\fontfig{\subfigColor}]{$\;$(2e)}{0.0}{\FigTwoFive .pdf} &
                \subfigimg[width=\linewidth,pos=ul,font=\fontfig{\subfigColor}]{$\;$(2f)}{0.0}{\FigTwoSix .pdf} &
                \subfigimg[width=\linewidth,pos=ul,font=\fontfig{\subfigColor}]{}{0.0}{\FigTwoSeven .pdf}
                \\[-5pt]
                \rotatebox[origin=l]{90}{\small Radial bias}
                &
                \subfigimg[width=\linewidth,pos=ul,font=\fontfig{\subfigColor}]{$\;$(3a)}{0.0}{\FigThreeOne .pdf} &
                \subfigimg[width=\linewidth,pos=ul,font=\fontfig{\subfigColor}]{$\;$(3b)}{0.0}{\FigThreeTwo .pdf} &
                \subfigimg[width=\linewidth,pos=ul,font=\fontfig{\subfigColor}]{$\;$(3c)}{0.0}{\FigThreeThree .pdf} &
                \subfigimg[width=\linewidth,pos=ul,font=\fontfig{\subfigColor}]{$\;$(3d)}{0.0}{\FigThreeFour .pdf} &
                \subfigimg[width=\linewidth,pos=ul,font=\fontfig{\subfigColor}]{$\;$(3e)}{0.0}{\FigThreeFive .pdf} &
                \subfigimg[width=\linewidth,pos=ul,font=\fontfig{\subfigColor}]{$\;$(3f)}{0.0}{\FigThreeSix .pdf} &
                \subfigimg[width=\linewidth,pos=ul,font=\fontfig{\subfigColor}]{}{0.0}{\FigThreeSeven .pdf}
        \end{tabular}

        \caption{\label{fig:app_inter} Bias on the position of the spots due to the interferences for different situations with a SH-WFS with $7 \times 7$ sub-apertures (blue squares). The central sub-aperture is framed in green. The pixels at the resolution of the EvWaCo WFS are framed in gray. Simulations are performed at the resolution of the EvWaCo WFS (a,c,e) or at the the Shannon resolution of the numerical propagation (b,d,f).
        (1)~Zoom of the simulated noiseless raw data. The green spots indicate where the spots are supposed to be according to the input tilt. The red spots indicate where the spots are found by the centroiding algorithm. For display purpose, the relative shift compared to the expected position has been multiplied by~5.
        (2,3)~$x$-bias (2) and radial bias (3) on the position of the different spots induced by a pure tip-tilt on the central spot. See the main text for the explanation on how to read the figure. For symmetry reason, only the upper right quadrant is displayed.
        (a,b)~Monochromatic simulation with $\lambda=\SI{617}{\nano\meter}$.
        (c,d)~Monochromatic simulation with $\lambda=\SI{500}{\nano\meter}$.
        (e,f)~Polychromatic simulation with $\lambda\in \Brack{400,600}\SI{}{\nano\meter}$.
        }
\end{figure}

Three different illuminations are simulated and gathered in \fig{fig:app_inter}:
\begin{itemize}
	\item A mono-chromatic illumination of $\lambda=\SI{617}{\nano\meter}$ similar to the one we have in the AO bench and that has been used for all the end-to-end simulations of this work.
	\item A mono-chromatic illumination of $\lambda=\SI{500}{\nano\meter}$ at the centre of the bandwidth of the EvWaCo WFS.
	\item A poly-chromatic illumination of $\lambda\in\Brack{400, 600}\SI{}{\nano\meter}$ corresponding to the bandwidth of the EvWaCo WFS. It is done by summing mono-chromatic simulations performed every step of $\delta \lambda = \SI{10}{\nano\meter}$. As discussed in \refapp{app:sub_samp}, the sampling is fixed in the pupil image.
\end{itemize}

\subfigfull{fig:app_inter}{1} presents the simulated raw data of the SH-WFS for different samplings. \subfigsfull{fig:app_inter}{1a,1c,1e} are done at the resolution of the EvWaCo WFS. This also corresponds to the resolution performed by simulation tools that propagate each lenslet independently: the resolution is at the Shannon sampling at the scale of a sub-aperture. The simulation at the Shannon sampling of the full aperture of the SH-WFS, \subfigs{fig:app_inter}{1b,1d,1f} shows that at the scale of an EvWaCo pixel, the spots are highly distorted by the high frequency interference fringes. This introduces a bias compared to the expected location of the spots (red points \vs green points). Thus, inducing a tilt on the central spot may displace its neighbouring spots.

This unexpected shifts are quantified in \subfigs{fig:app_inter}{2,3}. \subfigfull{fig:app_inter}{2} shows the bias $\delta_x$ along the $x$-axis.  \subfigfull{fig:app_inter}{2} shows the total radial bias $\sqrt{\delta^2_x + \delta^2_y}$.

For the central sub-aperture, framed in green, the color of figure encodes for the amount of bias found at the location of the supposedly induced tilt. For example, it is possible to read on \subfig{fig:app_inter}{2a} that for a tilt purely along $x$ of 1 pixel of the central spot, the bias on its position will be a positive extra 0.03 pixel along $x$ (red color). And for a tilt purely along $x$ of 3 pixels, the bias will be a negative 0.04 pixel along $x$ (dark blue color).

For the other sub-apertures, the color of figure encodes for the amount of bias found on the spot for a corresponding tilt induced on the central spot. For example, it is possible to read on \subfig{fig:app_inter}{2a} that for a tilt purely along $x$ of 1 pixel of the central spot, the bias introduced on its neighbouring spot on the right will be 0.075 pixel along $x$ (yellow color). And for a tilt purely along $y$ of 2 pixels, the bias will be a positive 0.03 pixel  along $x$ (red color).

Only the upper right quadrant is given for symmetry reasons. Indeed, the $x$-bias is symmetric around the $x$-axis and anti-symmetric around the $y$-axis. And the $y$-bias map is the transpose of the $x$-bias map.

Comparing the simulation at the different samplings \subfigs{fig:app_inter}{a,c,e} \vs \subfigs{fig:app_inter}{b,d,f}, the main difference is the already known phenomenon of the bias induced by the coarse pixel sampling (see the vertical patterns at the pixel pitch frenquency on \subfigs{fig:app_inter}{a,c,e}). These structures are smoothed in the over-sampled simulations of \subfigs{fig:app_inter}{b,d,f}. Otherwise, the effect on the bias is negligible: its order of magnitude and structure are similar. This means that despite their resolution lower than the Shannon frequency of the full pupil, the standard SH-WFS are sensitive to the interferences that distort the spot shapes.

Comparing the mono-chromatic simulations at $\lambda=\SI{617}{\nano\meter}$ in \subfigs{fig:app_inter}{a,b}, \vs $\lambda=\SI{500}{\nano\meter}$ in \subfigs{fig:app_inter}{c,d} shows that the pattern of the bias naturally scales with the wavelength. The intensity of its peaks nonetheless remains identical as these peaks are mainly due to the secondary maxima of the spot diffraction patterns whose intensities does not depend on the wavelength. For a radial displacement of less than a pixel, the bias on the central pixel can peak higher than 0.075 pixel and around 0.05 pixel for its closest neighbours.

As the bias pattern depends on the wavelength, it is generally assumed that the use of a poly-chromatic source would cancel out these biases. As seen on \subfig{fig:app_inter}{1f}, the structure of the poly-chromatic spot indeed appears smoother than its equivalent on \subfig{fig:app_inter}{1b,1d} without strong corruption by high frequency fringes. Nonetheless, the bias maps in \subfig{fig:app_inter}{2e,2f,3e,3f} show that the biases are extremely similar to the mono-chromatic case of $\lambda=\SI{500}{\nano\meter}$, in \subfig{fig:app_inter}{2c,2d,3c,3d}, especially for small induced tilt (corresponding to the working region of a closed loop). We can conclude that a poly-chromatic illumination does not average out the effects of the interferences and these effects remain similar to the ones induced by a mono-chromatic illumination centred in the wavelength bandwidth of the SH-WFS.

How can these biases be interpreted in the context of a closed AO loop?
\begin{itemize}
	\item The bias on the central pixel is equivalent to a gain in its displacement different than the one predicted by the synthetic model. The gain is accounted for in the interaction matrix for the least-squares method as it learned during the measurement of the interaction matrix if the poke on the actuator is small enough for the spot to stay close to its equilibrium position. As discussed above about the wavelength dependency, this implies to be careful for on-sky loop if the calibration of the loop is done with a laser or an internal source not centred on the SH-WFS bandwidth.
	\item For the bias on the neighbouring pixels, it appears that only the four closest neighbours are really impacted by the displacement of the central spot. The interferences thus remain mainly a local effect. In a closed-loop where all the spots are supposed to wander around their equilibrium position, the averaged bias induced on the neighbours will remain null. The effect of the interferences thus consists in adding an extra noise on their positions. This explains why the noise on the count maps in \fig{fig:spot_noise} is larger for \subfigs{fig:spot_noise}{a} (coherent) than for \subfigs{fig:spot_noise}{b} (incoherent).
\end{itemize}

\acknowledgments % equivalent to \section*{ACKNOWLEDGMENTS}       

The authors thank Pierre Chavel (Institut d'Optique Graduate School) for the fruitful discussions on the modelling of a Shack-Hartmann wavefront sensor in a Fourier optics framework.
\\
The authors acknowledge the supports from Chulalongkorn University’s CUniverse (CUAASC) grant and from the Program Management Unit for Human Resources \& Institutional Development, Research and Innovation, NXPO (grant number B16F630069). 

% References
\bibliography{AO_loop} % bibliography data in report.bib
\bibliographystyle{spiebib} % makes bibtex use spiebib.bst

\end{document}